\documentstyle[epsf,psfig,12pt]{article}

\renewcommand{\baselinestretch}{1.23}

\begin{document}

%\draft

\renewcommand{\theequation}{\thesection.\arabic{equation}}
\def\be{\begin{eqnarray}}
\def\en{\end{eqnarray}}
\def\non{\nonumber}
\def\la{\langle}
\def\ra{\rangle}
\def\up{\uparrow}
\def\down{\downarrow}
\def\ep{\varepsilon}
\def\ums{{\mu}_{_{\overline{\rm MS}}}}
\def\u{\mu_{\rm fact}}
\def\gg{\Delta\sigma^{\gamma G}}
\def\lsim{ {\ \lower-1.2pt\vbox{\hbox{\rlap{$<$}\lower5pt\vbox{\hbox{$\sim$}
}}}\ } }
\def\gsim{ {\ \lower-1.2pt\vbox{\hbox{\rlap{$>$}\lower5pt\vbox{\hbox{$\sim$}
}}}\ } }
\def\getao{g_{_{\eta_0NN}}}
\def\gghost{g_{_{GNN}}}
\def\geta{g_{_{\eta NN}}}
\def\gpi{g_{_{\pi NN}}}
\def\getap{g_{_{\eta'NN}}}
\def\getao{g_{_{\eta_0NN}}}
\def\gghost{g_{_{GNN}}}
\def\geta{g_{_{\eta NN}}}
\def\gpi{g_{_{\pi NN}}}
\def\getap{g_{_{\eta'NN}}}
\def\dk{\partial\!\cdot\!K}\def\dk{\partial\!\cdot\!K}
\def\pr{{\sl Phys. Rev.}~}
\def\prl{{\sl Phys. Rev. Lett.}~}
\def\pl{{\sl Phys. Lett.}~}
\def\np{{\sl Nucl. Phys.}~}
\def\zp{{\sl Z. Phys.}~}

\title{
\begin{flushright}
{\normalsize IP-ASTP-03-96}
%{\normalsize IP-ASTP-03-96\\
%June, 1996}
\end{flushright}
\vskip 10mm
\Large\bf
Status of the Proton Spin Problem$^{*}$
}
\author{{\bf Hai-Yang Cheng}\\
{\it Institute of Physics, Academia Sinica}\\
{\it Taipei, Taiwan 115, Republic of China}\\
}
\date{(August, 1996)}

\maketitle

\begin{abstract}
    The proton spin problem triggered by the EMC experiment and its present
status are closely examined. Recent experimental and theoretical progresses
and their implications are reviewed. It is pointed out that the sign of
the sea-quark polarization generated perturbatively by hard gluons via the 
anomaly mechanism is predictable: It is negative if the gluon spin 
component is positive. We
stress that the polarized nucleon structure function $g_1(x)$ is
independent of the $k_\perp$-factorization scheme chosen in defining
the quark spin density and the hard photon-gluon scattering cross section.
Consequently, the anomalous gluon and sea-quark interpretations for 
$\Gamma_1$, the first moment of $g_1(x)$, are equivalent. It is the axial
anomaly that accounts for the observed suppression of $\Gamma_1^p$.

%\vskip 3 cm
%\tableofcontents

\vskip 4 cm
\centerline{ $^*$ Lecture presented at the}
\centerline{ {\it Xth Spring School on Particles and Fields}}
\centerline{National Cheng-Kung University, Taiwan, ROC} 
\centerline{March 20-22, 1996}
\end{abstract}
\pagebreak

\tableofcontents
\pagebreak

%\narrowtext
\section{Introduction}
   Experiments on polarized deep inelastic lepton-nucleon scattering 
started in the middle 70s \cite{E80}. Measurements of cross section
differences with the longitudinally polarized lepton beam and nucleon
target determine the polarized nucleon structure function $g_1(x)$.
In 1983 the first moment of the proton spin structure function, 
$\Gamma_1^p\equiv\int^1_0g_1^p(x)dx$,
was obtained by the SLAC-Yale group to be $0.17\pm 0.05$ \cite{E130}, which
is in nice agreement with the prediction $0.171\pm 0.006$ expected from the
Ellis-Jaffe sum rule \cite{EJ} based on the assumption of vanishing 
strange-sea polarization, i.e., $\Delta s=0$. Therefore, the polarized 
DIS data can be explained simply by the valence quark spin components.
Around the same time, two theoretical analyses by Lam and Li \cite{Lam}
and by Ratcliffe \cite{Rat} were devoted
to studying the gluon effects on the polarized proton structure function and
its first moment. It appears that there is an anomalous gluon 
contribution to $\Gamma_1^p$ in the sense that, though the gluon effect
is formally a leading-order QCD correction, it does {\it not}
vanish in the asymptotic limit $Q^2\to\infty$ \cite{Lam}. However, the
implication of this observation was not clear at that time.

    The 1987 EMC experiment \cite{EMC} then came to a surprise. The 
result published in 1988 and later indicated that $\Gamma_1^p=0.126\pm 
0.018$, substantially lower than the expectation from the Ellis-Jaffe
conjecture. This led to the stunning implication that very little 
($\lsim 15\%$) of
the proton spin is carried by the quarks, contrary to the naive
quark-model picture. While the proton spin arises entirely from the quarks
in the non-relativistic constituent quark model, the sum of the 
$z$-component of quark spins, $\Delta\Sigma$, accounts for 3/4 of the 
proton spin in the relativistic quark model. The EMC data implied a
substantial sea-quark polarization in the region $x<0.1$\,, a range not 
probed by earlier SLAC experiments. The question of what is the
proton spin content triggered by the EMC experiment has stimulated
a great deal of interest in understanding the origin of the so-called 
(although not pertinently) ``proton spin crisis". Up to date, there are
over a thousand published papers connected to this and related topics.

   During the period of 1988-1993, theorists tried hard to resolve the
proton spin enigma and seek explanations for the EMC measurement of
$\Gamma_1^p$, assuming the validity of the EMC data at small $x$ ($0.01<x
<0.1$) and of the extrapolation procedure to the unmeasured small $x$ 
region $(x<0.01)$. One of the main theoretical problems is that hard 
gluons cannot induce sea polarization perturbatively for massless
quarks due to helicity conservation. Hence, it is difficult to
accommodate a large strange-sea polarization $\Delta s\approx -0.10$ 
in the naive parton model. Right after the EMC experiment, the effect of
anomalous gluon contributions to $\Gamma_1^p$ was revived by Efremov and
Teryaev \cite{Efr}, Altarelli and Ross \cite{AR}, Carlitz, Collins and
Mueller \cite{CCM} (see also Leader and Anselmino \cite{Lead}). Roughly
speaking, the anomalous gluon effect originating from the axial anomaly
mimics the role of sea polarization if the gluon spin component $\Delta G$
has a sign opposite to $\Delta s$. Consequently, $\Delta\Sigma$ is not
necessarily small whereas $\Delta s$ is not necessarily large.
This anomalous mechanism thus provides a plausible and simple solution 
to the proton spin puzzle: It
explains the suppression of $\Gamma_1^p$ observed by EMC and brings the
improved parton model close to what expected from the quark model, 
provided that $\Delta G$ is
positive and large enough. But then we face a dilemma. According to the OPE
approach, which is a model-independent approach, $\Gamma_1^p$
does {\it not} receive contributions from hard gluons because only 
quark operators contribute to the first moment of $g_1^p$ at the 
twist-2 and spin-1 level. This conflict between the anomalous gluon 
interpretation and the sea-quark explanation of $\Gamma_1^p$ has been 
under hot debate over past many years.

   In spite of much controversy over the aforementioned issue, this
dispute was already resolved in 1990 by Bodwin and Qiu \cite{BQ} (see also
Manohar \cite{Man2}). They emphasized that the size of the hard-gluonic 
contribution to $\Gamma_1^p$ is purely a matter of the 
$k_\perp$-factorization convention chosen in defining the quark spin
density $\Delta q(x)$ and the hard cross section for photon-gluon
scattering. As a result, the above-mentioned two different popular
interpretations, corresponding to chiral-invariant and gauge-invariant
factorization schemes respectively, are on the same footing. Their equivalence
will be shown in details in Sec.~4 in the framework of perturbative QCD. 
The axial anomaly that breaks chiral symmetry can generate 
negative helicity even for massless sea quarks. Therefore,
a sizeable strange-sea polarization $\Delta s\approx -0.10$ is no 
longer a problem in the sea-quark interpretation. Despite this 
clarification, some of recent articles and reviews are still biased
towards or against one of the two popular explanations for $\Gamma_1^p$;
this is considerably unfortunate and annoying.

   One can imagine that after a certain point it is difficult to
make further theoretical or phenomenological progress without new
experimental inputs. Fortunately, this situation was dramatically changed
after 1993. Since then many new experiments using different targets 
have been carried out. In 1994 SMC and SLAC-E143 have reported independent
measurements of $g_1^p(x)$ and confirmed the validity of the EMC data.
The new world average $\Delta\Sigma\sim 0.30$ indicates that the
proton spin problem becomes less severe than before. The new measurements of 
polarized neutron and deuteron structure functions by SMC, SLAC-E142 and 
SLAC-E143
allowed one to perform a serious test on the Bjorken sum rule. This year
marks the 30th anniversary of this well-known sum rule. We learned that
it has been tested to an accuracy of 10\% level. Data on the transverse
spin structure function $g_2(x)$ have just become available. A probe of
$g_2$ might provide a first evidence of higher-twist effects. Finally,
the $x$-dependent spin distributions
for $u$ and $d$ valence quarks and for non-strange sea quarks have been
determined for the first time by measuring semi-inclusive spin asymmetry
for positively and negatively charged hadrons from polarized DIS. 
In short, the experimental progress in past few years is quite remarkable.

   On the theoretical side, there are also several fascinating progresses.
For example, two successful first-principles calculations of 
the quark spin contents based on lattice QCD were published last year. The
calculation revealed that sea-quark polarization arises from the 
disconnected insertion and is empirically SU(3)-flavor symmetric. This
implies that the axial anomaly is responsible for a substantial part of 
sea polarization. The lattice calculation also suggests that
the conventional practice of decomposing $\Delta q$ into valence and sea
components is not complete; the effect of ``cloud" quarks should be taken
into account. Other theoretical progresses will be surveyed in Sec.~6. 

  With the accumulated data of $g_1^p(x),~g_1^n(x)$ and $g_1^d(x)$ and with
the polarized two-loop splitting functions available very recently, it
became possible to carry out a full and consistent next-to-leading order
analysis of $g_1(x,Q^2)$ data. The goal is to determine the spin-dependent
parton distributions from DIS experiments as much as we can, especially
for sea quarks and gluons.

    There are several topics not discussed in this review. The transverse
spin structure function $g_2(x)$ is not touched upon except for a brief
discussion on the Wandzura-Wilczek relation in Sec.~4.2. The small or very 
small $x$
behavior of parton spin densities and polarized structure functions 
is skipped in this article. Perspective of polarized hadron colliders
and $ep$ colliders will not be discussed here. Some of the topics can
be found in a number of excellent reviews [12-24] on polarized structure 
functions and the proton spin problem.

%\newpage
\section{Polarized Deep Inelastic Scattering}
\setcounter{equation}{0}
\subsection{Experimental progress}
   Before 1993 it took averagely 5 years to carry out a new polarized DIS 
experiment (see Table I). This situation was dramatically changed after 1993. 
Many new experiments measuring the nucleon and deuteron spin-dependent 
structure functions became available. The experimental progress is certainly
quite remarkable in the past few years.

    In the laboratory frame the differential cross section for the
polarized lepton-nucleon scattering has the form
\be
{d^2\sigma\over d\Omega\,dE'}=\,{1\over 2M}\,{\alpha^2\over q^4}\,{E'\over E}
\,L_{\mu\nu}W^{\mu\nu},
\en
where $E~(E')$ is the energy of the incoming (outgoing) lepton, $L_{\mu\nu}$
and $W_{\mu\nu}$ are the leptonic and hadronic tensor, respectively. The
most general expression of $W_{\mu\nu}$ reads
\be
W_{\mu\nu} &=& W_{\mu\nu}^S+iW_{\mu\nu}^A   \non \\
 &=& F_1\left(-g_{\mu\nu}+{q_\mu q_\nu\over q^2}\right)+F_2\left(p_\mu-{
p\cdot qq_\mu\over q^2}\right)\left(p_\nu-{p\cdot qq_\nu\over q^2}\right)
/(p\cdot q)  \\
&&+\, i{M\over p\cdot q}\,\epsilon_{\mu\nu\rho\sigma}q^\rho\left\{s^\sigma g_1
+\left[s^\sigma-{s\cdot q p^\sigma\over p\cdot q}\right]g_2\right\},  \non
\en
that is, it is governed by two spin-averaged structure functions $F_1$ and
$F_2$ and two spin-dependent structure functions $g_1$ and $g_2$.

  Experimentally, the polarized structure functions $g_1$ and $g_2$ are 
determined by measuring two asymmetries:
\be
A_\Vert=\,{d\sigma^{\up\down}-d\sigma^{\up\up}\over d\sigma^{\up\down}+
d\sigma^{\up\up}},~~~~~A_\perp=\,{d\sigma^{\down\to}-d\sigma^{\up\to}\over 
d\sigma^{\down\to}+d\sigma^{\up\to}},
\en
where $d\sigma^{\up\up}$ ($d\sigma^{\up\down}$) is the differential cross
section for the longitudinal lepton spin parallel (antiparallel) to the
longitudinal nucleon spin, and $d\sigma^{\down\to}$ ($d\sigma^{\up\to}$)
is the differential cross section for the lepton spin antiparallel (parallel)
to the lepton momentum and nucleon spin direction transverse to the lepton
momentum and towards the direction of the scattered lepton. It is convenient
to recast the measured asymmetries $A_\Vert$ and $A_\perp$ in terms of the 
asymmetries $A_1$ and $A_2$ in the virtual photon-nucleon scattering:
\be
A_1=\,{\sigma_{1/2}-\sigma_{3/2}\over \sigma_{1/2}+\sigma_{3/2}},~~~~~A_2=\,
{2\sigma^{TL}\over \sigma_{1/2}+\sigma_{3/2}},
\en
where $\sigma_{1/2}$ and $\sigma_{3/2}$ are the virtual photon absorption
cross sections for $\gamma^*(1)+N(-{1\over 2})$ and $\gamma^*(1)+N({1\over 
2})$ scatterings, respectively, and
$\sigma^{TL}$ is the cross section for the interference between
transverse and longitudinal virtual photon-nucleon scatterings.
The asymmetries $A_1$ and $A_2$ satisfy the bounds
\be
|A_1|\leq 1,~~~~~|A_2|\leq\sqrt{R},
\en
where $R\equiv\sigma_L/\sigma_T$ and $\sigma_T\equiv(\sigma_{1/2}+
\sigma_{3/2})/2$.
The relations between the asymmetries $A_\Vert,~
A_\perp$ and $A_1,~A_2$ are given by
\be
A_\Vert=\,D(A_1+\eta A_2),~~~~~A_\perp=\,D(A_2-\xi A_1),
\en
where $D$ is a depolarization factor of the virtual photon, $\eta$ and $\xi$
depend only on kinematic variables. The asymmetries $A_1$ and $A_2$ in
the virtual photon-nucleon scattering are related to the polarized structure 
functions $g_1$ and $g_2$ via
\be
A_1=\,{g_1-\gamma^2g_2\over F_1},~~~~~A_2=\,{\gamma(g_1+g_2)\over F_1},
\en
where $\gamma\equiv Q/\nu=Q/(E-E')=2Mx/\sqrt{Q^2}$. Note that the 
more familiar relation $A_1=g_1/F_1$ is valid only when $\gamma\approx 0$
or $g_2\approx 0$. By solving (2.6) and (2.7), one obtains expressions
of $g_1$ and $g_2$ in terms of the measured asymmetries $A_\Vert$ and 
$A_\perp$. Since
$\gamma\to 0$ in the Bjorken limit, it is easily seen that to a good
approximation, $A_\Vert\simeq DA_1$ and
\be
g_1(x,Q^2)\simeq F_1(x,Q^2)\,{A_\Vert\over D}=\,{F_2(x,Q^2)\over 2x(1+R
(x,Q^2))}\,{A_\Vert\over D}.
\en

    Some experimental results on the polarized structure 
function $g_1^p(x)$ of the proton, $g_1^n(x)$ of the neutron, and 
$g_1^d(x)$ of the deuteron are summarized in Table I. The  
spin-dependent distributions for various targets are related by
\be
g_1^p(x)+g_1^n(x)=\,{2\over 1-1.5\omega_D}\,g_1^d(x),
\en
where $\omega_D=0.058$ is the probability that the deuteron is in a $D$ state.
Since experimental measurements only cover a limited kinematic range, an
extrapolation to unmeasured $x\to 0$ and $x\to 1$ regions is necessary.
At small $x$, a Regge behavior $g_1(x)\propto x^{\alpha(0)}$ with
the intercept value $0<\alpha(0)<0.5$ is conventionally assumed. In the EMC
experiment \cite{EMC}, $\alpha(0)=0$ is chosen so that $g_1^p(x)$
approximates a constant $\sim 0.2$ as $x<0.01$, and hence $\int^{0.01}_0
g_1^p(x)_{\rm EMC}dx=0.002\,$. However, the SMC data
\cite{SMC94} of $g_1^p$ show a tendency to rise at low $x$ ($x<0.02$),
and it will approach a constant $1.34\pm 0.62$ as $x<0.003$ if $\alpha(0)=0$
is chosen. Then we have $\int^{0.003}_0g_1^p(x)dx=0.004\pm 0.002\,$. Using 
the SMC
data at small $x$ and the above extrapolation yields $\int^{0.01}_0g_1^p(x)
_{\rm SMC}dx=0.017\pm 0.006\,$. This explains why $\Gamma_1^p$ obtained by SMC
is larger than that of EMC (see Table I).

\vskip 0.5cm
%\begin{table}
\centerline{{\small Table I. Experiments on the polarized structure functions
$g_1^p(x,Q^2),~g_1^n(x,Q^2)$ and $g_1^d(x,Q^2)$.}}
{\footnotesize
\begin{center}
\begin{tabular}{|c|c|c|c|c||c|c|} \hline
Experiment & Year & Target & $\la Q^2\ra$ & $x$ range & $\Gamma^{\rm
target}_1$ & Reference  \\  
  & & & (GeV$^2$) & & $=\int^1_0g_1^{\rm target}(x,\la Q^2\ra)dx$ & \\ \hline
E80/E130 & 1976/1983 & $p$ & $\sim 5$ & $0.1<x<0.7$ & $0.17\pm 0.05^*$ & 
\cite{E80,E130} \\
EMC & 1987 & $p$ & 10.7 & $0.01<x<0.7$ & $0.126\pm 0.010\pm 0.015^\dagger$ & 
\cite{EMC} \\
SMC & 1993 & $d$ & 4.6 & $0.006<x<0.6$ & $0.023\pm 0.020\pm 0.015$ & 
\cite{SMC93}  \\
SMC & 1994 & $p$ & 10 & $0.003<x<0.7$ & $0.136\pm 0.011\pm 0.011$ & 
\cite{SMC94} \\
SMC & 1995 & $d$ & 10 & $0.003<x<0.7$ & $0.034\pm 0.009\pm 0.006$ & 
\cite{SMC95} \\
E142 & 1993 & $n$ & 2 & $0.03<x<0.6$ & $-0.022\pm 0.011$ & \cite{E142}  \\
E143 & 1994 & $p$ & 3 & $0.03<x<0.8$ & $0.127\pm 0.004\pm 0.010$ &
\cite{E143a} \\
E143 & 1995 & $d$ & 3 & $0.03<x<0.8$ & $0.042\pm 0.003\pm 0.004$ & 
\cite{E143b} \\   \hline  
\end{tabular}
\end{center} }
{\footnotesize
\noindent $^*$ Obtained by assuming a Regge behavior $A_1\propto x^{1.14}$
for small $x$.   \\
\noindent $^\dagger$ Combined result of E80, E130 and EMC data. The EMC data
alone give $\Gamma_1^p=0.123\pm 0.013\pm 0.019\,$.}
%\end{table}
\vskip 0.45cm

   A serious test of the Bjorken sum rule for the difference $\Gamma_1^p-
\Gamma_1^n$ [\,$\Gamma_1$ being defined in (2.13)\,], which is a rigorous 
consequence of QCD,
became possible since 1993. The current experimental results are
\be
{\rm SMC}~\cite{SMC95}: && \Gamma_1^p-\Gamma_1^n=\,0.199\pm 0.038~~~{\rm at}~
Q^2=10\,{\rm GeV}^2,  \non \\
{\rm E143}~\cite{E143b}: && \Gamma_1^p-\Gamma_1^n=\,0.163\pm 0.010\pm 0.016
~~~{\rm at}~Q^2=3\,{\rm GeV}^2,  
\en
to be compared with the predictions\footnote{The theoretical value
$0.187\pm 0.003$ for $\Gamma_1^p-\Gamma_1^n$ quoted by the SMC paper
\cite{SMC95} seems to be obtained for three quark flavors rather than
for four flavors.}
\be
\Gamma_1^p-\Gamma_1^n &=& 0.187\pm 0.003~~~~{\rm with}~\alpha_s(10\,{\rm
GeV}^2)=0.24\pm 0.03\,,   \non \\
\Gamma_1^p-\Gamma_1^n &=& 0.171\pm 0.008~~~~{\rm with}~\alpha_s(3\,{\rm
GeV}^2)=0.35\pm 0.05\,,   
\en
obtained from the Bjorken sum rule evaluated up to $\alpha_s^3$ for 
three light flavors \cite{Lar91}
\be
\Gamma_1^p(Q^2)-\Gamma^n_1(Q^2)=\,{1\over 6}\,{g_A\over g_V}\left[1-
{\alpha_s(Q^2)\over \pi}-{43\over 12}\left({\alpha_s(Q^2)\over\pi}\right)^2-
20.22\left({\alpha_s(Q^2)\over \pi}\right)^3\right],
\en
where use of $g_A/g_V=F+D=1.2573\pm 0.0028$ \cite{PDG} has been made. 
Therefore, the Bjorken sum rule has been confirmed by data to an accuracy 
of 10\% level.

   Recently, data on the transverse spin structure function $g_2$ have
just become available \cite{SMCg2,E143g2}.
A probe of $g_2$ might provide a first evidence of
higher-twist effects. Finally, the $x$-dependent spin distributions
for $u$ and $d$ valence quarks and for non-strange sea quarks have been
determined for the first time by measuring semi-inclusive spin asymmetry
for positively and negatively charged hadrons from polarized DIS 
\cite{SMC96}. For some discussions, see Sec.~8.

\subsection{The proton spin crisis}
   From the parton-model analysis in Sec.~3 or from the OPE approach 
in Sec.~4, the first moment of the polarized proton structure function
\be
\Gamma_1^p(Q^2) \equiv\int^1_0 g_1^p(x,Q^2)dx,
\en
can be related to the combinations of the quark spin components via\footnote{
As will be discussed at length in Sec.~4, whether or not gluons contribute
to $\Gamma^p_1$ depends on the factorization convention chosen in defining
the quark spin density $\Delta q(x)$. Eq.(2.14) is valid in the
gauge-invariant factorization scheme. However, gluons are allowed to
contribute to $g_1^p(x)$ and to the proton spin, irrespective of the
prescription of $k_\perp$-factorization.}
\be
\Gamma_1^p=\,{1\over 2}\sum_q e^2_q\Delta q(Q^2)=-{1\over 2}\sum_q e^2_q
\la p|\bar{q}\gamma_\mu\gamma_5q|p\ra s^\mu,
\en
where $\Delta q$ represents the net helicity of the quark flavor $q$ along
the direction of the proton spin in the infinite momentum frame:
\be
\Delta q=\int^1_0\Delta q(x)dx\equiv\int^1_0\left[ q^\up(x)+\bar{q}^\up(x)-
q^\down(x)-\bar{q}^\down(x)\right]dx.
\en
For a definition of $\Delta q$ in the laboratory frame, see Sec.~4.5.
At the EMC energies $\la Q^2\ra=10.7\,{\rm GeV}^2$ or smaller, only 
three light flavors are relevant
\be
\Gamma_1^p(Q^2)=\,{1\over 2}\left({4\over 9}\Delta u(Q^2)+{1\over 9}\Delta
d(Q^2)+{1\over 9}\Delta s(Q^2)\right).
\en
Other information on the quark polarization is available from the
nucleon axial coupling constants $g_A^3$ and $g_A^8$:
\be
g_A^3(Q^2)=\Delta u(Q^2)-\Delta d(Q^2),~~~~~g_A^8(Q^2)=\Delta u(Q^2)+\Delta d
(Q^2)-2\Delta s(Q^2).
\en
Since there is no anomalous dimension associated with the axial-vector 
currents $A_\mu^3$ and $A_\mu^8$, the non-singlet couplings
$g_A^3$ and $g_A^8$ do not evolve with $Q^2$ and hence
can be determined at $q^2=0$ from low-energy neutron and
hyperon beta decays. Under SU(3)-flavor symmetry, the non-singlet 
couplings are related to the SU(3) parameters $F$ and $D$ by
\be
g_A^3=\,F+D,~~~~~~g_A^8=\,3F-D.
\en
We use the values \cite{CloseR}
\be
F=0.459\pm 0.008\,,~~~~D=\,0.798\pm 0.008\,,~~~~F/D=\,0.575\pm 0.016
\en
to obtain $g_A^3=0.579\pm 0.025$\,.

   Prior to the EMC measurement of polarized structure functions, a
prediction for $\Gamma_1^p$ was made based on the assumption that the
strange sea in the nucleon is unpolarized, i.e., $\Delta s=0$. It follows
from (2.16) and (2.17) that
\be
\Gamma_1^p(Q^2)=\,{1\over 12}g_A^3+{5\over 36}g_A^8.
\en
This is the Ellis-Jaffe sum rule \cite{EJ}. It is evident that the measured
results of EMC, SMC and E143
for $\Gamma_1^p$ (see Table I) are smaller than what expected from
the Ellis-Jaffe sum rule: $\Gamma_1^p=0.185\pm 0.003$ without QCD 
corrections ($=0.171\pm 0.006$ at $Q^2=10\,{\rm GeV}^2$ to leading-order
corrections).

   To discuss QCD corrections, it is convenient to recast (2.16) to
\be
\Gamma_1^{p(n)}(Q^2)=\,C_{NS}(Q^2)\left(\pm{1\over 12}g_A^3+{1\over 36}g_A^8
\right)+{1\over 9}\,C_S(Q^2)g_A^0(Q^2),
\en
where the isosinglet coupling is related to the quark spin sum: 
\be
g_A^0(Q^2)=\Delta\Sigma(Q^2)\equiv\Delta u(Q^2)+\Delta d(Q^2)+\Delta s(Q^2).
\en
Perturbative QCD corrections to $\Gamma_1$ have been calculated to
${\cal O}(\alpha_s^3)$ for the non-singlet coefficient $C_{NS}$ and
to ${\cal O}(\alpha_s^2)$ for the singlet coefficient $C_S$ 
\cite{Lar91,Lar94}:\footnote{The singlet coefficient is sometimes written as
\be
C_S(Q^2)=1-{1\over 3}\,{\alpha_s\over \pi}-0.56\left({\alpha_s\over\pi}\right)
^2   \non
\en
in the literature, but this is referred to the quark polarization in the
asymptotic limit, namely $\Delta q(Q^2)\to\Delta q(\infty)$. The above
singlet coefficient is obtained by substituting the relation
\be
g_A^0(Q^2)=\left[1+{2\over 3}{\alpha_s\over \pi}+1.21\left({\alpha_s\over\pi}
\right)^2+\cdots\right]g_A^0(\infty)   \non
\en
into (2.21). }
\be
C_{NS}(Q^2) &=& 1-{\alpha_s\over \pi}-{43\over 12}\left({\alpha_s\over\pi}
\right)^2-
20.22\left({\alpha_s\over \pi}\right)^3,  \non \\
C_{S}(Q^2) &=& 1-{\alpha_s\over \pi}-1.10\left({\alpha_s\over\pi}\right)^2
\en
for three quark flavors and for $\alpha_s=\alpha_s(Q^2)$. 

   From (2.17), (2.18) and the leading-order QCD correction to $\Gamma_1^p$
in (2.21) together with the EMC result $\Gamma_1^p(Q^2)=0.126
\pm 0.010\pm 0.015$ \cite{EMC}, we obtain
\be
\Delta u=\,0.77\pm 0.06\,,~~~\Delta d=-0.49\pm 0.06\,,~~~\Delta s=-0.15\pm 
0.06\,,
\en
and
\be
\Delta\Sigma=\,0.14\pm 0.17
\en
at $Q^2=10.7\,{\rm GeV}^2$.
The results (2.24) and (2.25) exhibit two surprising features: The
strange-sea polarization is sizeable and negative, and the total contribution 
of quark
helicities to the proton spin is small and consistent with zero.  This is
sometimes referred to as (though not pertinently) the ``proton spin crisis".

   The new data of SMC, E142 and E143 obtained from different targets are
consistent with each other and with the EMC data when higher-order 
corrections in (2.21) are taken into account \cite{EK95}. A global fit 
to all available data evaluated at a common $Q^2$ in a consistent treatment
of higher-order perturbative QCD effects yields \cite{EK95}
\be
\Delta u=\,0.83\pm 0.03\,,~~~\Delta d=-0.43\pm 0.03\,,~~~\Delta s=-0.10\pm 
0.03\,,
\en
and
\be
\Delta\Sigma=\,0.31\pm 0.07
\en
at $Q^2=10\,{\rm GeV}^2$. An updated analysis with most recent data (mainly
the E142 data) gives \cite{EK96}
\be
\Delta u=\,0.82\pm 0.03+\cdots\,,~~~\Delta d=-0.44\pm 0.03+\cdots\,,~~~\Delta 
s=-0.11\pm 0.03+\cdots\,,
\en
and
\be
\Delta\Sigma=\,0.27\pm 0.04+\cdots
\en
at $Q^2=3\,{\rm GeV}^2$, where dots in (2.28) and (2.29) represent further
theoretical and systematical errors remained to be assigned. Evidently,
the proton spin problem becomes less severe than before. Note that all
above results for $\Delta q$ and $\Delta\Sigma$ are extracted from data 
based on the assumption of SU(3)-flavor symmetry. It has been advocated
that SU(3) breaking will leave $\Delta\Sigma$ essentially intact but
reduce $\Delta s$ substantially \cite{Lip}. However, recent lattice
calculations indicate that not only sea polarization is of order 
$-0.10$ but also it is consistent with SU(3)-flavor symmetry within errors
(see Sec.~6.1). It is also worth remarking that elastic $\nu p$
scattering, which has been suggested to measure the strange-sea
polarization, actually measures the scale-independent combination
$(\Delta s-\Delta c)$ instead of the scale-dependent $\Delta s$ (see Sec.~8).

    The conclusions that only a small fraction of the proton spin is carried
by the quarks and that the polarization of sea quarks is negative and 
substantial lead
to some puzzles, for example, where does the proton get its spin ? why is that
the total quark spin component is small ? and why is the sea polarized ? 
The proton spin problem emerges in the sense that experimental results are
in contradiction to the naive quark-model's picture. The non-relativistic
SU(6) constituent quark model predicts that $\Delta u={4\over 3}$, $\Delta 
d=-{1\over 3}$ and hence $\Delta\Sigma=1$, but its prediction $g_A={5\over
3}$ is too large compared to the measured value $1.2573\pm 0.0028$ 
\cite{PDG}. In the relativistic quark model the proton is no longer 
a low-lying $S$-wave state since the quark orbital angular momentum
is nonvanishing due to the presence of quark transverse momentum in
the lower component of the Dirac spinor. The quark polarizations
$\Delta u$ and $\Delta d$ will be reduced by the same factor of 
${3\over 4}$ to 1 and $-{1\over 4}$, respectively, if $g_A^3$
is reduced from ${5\over 3}$ to ${5\over 4}$ (see also Sec.~6.1)
The reduction of the total
quark spin $\Delta\Sigma$ from unity to 0.75 by relativistic effects is
shifted to the orbital component of the proton spin so that the 
spin sum rule now reads \cite{Seh}
\be
{1\over 2}=\,{1\over 2}(\Delta u+\Delta d)+L_z^q.
\en
Hence, {\it it is expected in the relativistic constituent quark model that
3/4 of the proton spin arises from the quarks and the rest of the proton spin
is accounted for by the quark orbital angular momentum.}

   Let $\Delta q$ be decomposed into valence and sea components, $\Delta q
=\Delta q_v+\Delta q_s$. The experimental fact that $\Delta\Sigma\sim 0.30$,
much smaller than the quark-model expectation 0.75, can be understood
as a consequence of negative sea polarization:
\be
\Delta\Sigma=\,\Delta\Sigma_v+\Delta\Sigma_s=\,(\Delta u_v+\Delta d_v)+(
\Delta u_s+\Delta d_s+\Delta s).
\en
Nevertheless, we still encounter several problems. First, in the absence of 
sea polarization, we find from (2.17) and (2.18) that 
$\Delta u_v=0.92,~\Delta 
d_v=-0.34$ and $\Delta\Sigma_v=0.58\,$. As first noticed by Sehgal \cite{Seh},
even if sea polarization vanishes, a substantial part of the proton spin 
does not arise from the quark spin components. In fact, the Ellis-Jaffe
prediction $\Gamma_1^p(10\,{\rm GeV}^2)=0.171\pm 0.006$ is based on
the above ``canonical" values for $\Delta q_v$ and $\Delta\Sigma_v$. Our 
question is why the canonical $\Delta\Sigma_v$ still deviates significantly 
from the relativistic quark model expectation 0.75\,.
A solution to this puzzle will be discussed in Sec.~6.1. It turns out that
the canonical valence quark polarization is actually a combination
of ``cloud-quark" and truly valence-quark spin components.
Second, in the presence of sea-quark
polarization, the spin sum rule must be modified to include
all possible contributions to the proton spin:\footnote{It has been argued
that in the double limit, $m_q\to 0$ and $N_c\to\infty$, where $m_q$ and
$N_c$ are the light quark mass and the number of colors respectively,
one has $g_A^0(=
\Delta\Sigma)=0$ and $\Delta G=L_z^G=0$, so that the proton spin is orbit in 
origin \cite{Brod88,Ellis88}.}
\be
{1\over 2}=\,{1\over 2}(\Delta u+\Delta d+\Delta s)+\Delta G+L_z^q+L_z^G,
\en
where $\Delta G=\int^1_0\Delta G(x)dx\equiv\int^1_0 [G^\up(x)-G^\down(x)]dx$
is the gluon net helicity along the proton spin direction,
and $L_z^{q(G)}$ is the quark (gluon) orbital angular momentum.
It is a most great challenge, both experimentally
and theoretically, to probe and understand each proton spin content.

\begin{figure}[ht]
\vspace{-3.5cm}
\hskip 1cm
  \psfig{figure=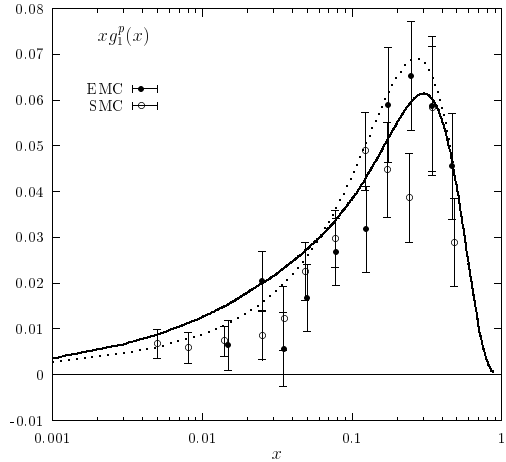,width=13cm}
\vspace{-7cm}
    \caption[]{\small Two theoretical curves for $xg_1^p(x)$. The solid curve 
is obtained by fitting to the EMC \cite{EMC} and SMC \cite{SMC94} data at 
$x\gsim 0.2$ with the polarized
valence quark distributions given by (7.11), and the dotted curve arises
from the valence spin distributions given by (2.33). At first sight, the 
latter curve appears to give a reasonable ``eye-fit" to the data, even though
its first moment is too large compared to the measured value.}
%    \label{fig1} 
\end{figure}

    Before closing this section, we wish to remark that experimentally it 
is important to evaluate the first moment of $g_1^p(x)$ in order to ensure
that the existence of sea polarization is inevitable. Suppose there is no
spin component from sea quarks, then
it is always possible to parametrize the valence quark spin
densities, for example\footnote{This parameterization is taken from 
\cite{GS95} except that we have made a different normalization in order
to satisfy the first-moment constraint: $\Delta u_v=0.92$ and $\Delta d_v
=-0.34\,$.}
\be
\Delta u_v(x) &=& 0.355\,x^{-0.54}(1-x)^{3.64}(1+18.36x),   \non \\
\Delta d_v(x) &=& -0.161\,x^{-0.54}(1-x)^{4.64}(1+18.36x),   
\en
in such a way that they make
a reasonable ``eye-fit" to the EMC \cite{EMC} and SMC \cite{SMC94}
data of $g_1^p(x)$ even at small $x$ (see the dotted curve in Fig.~1).
One cannot tell if there is truly
a discrepancy between theory and experiment unless $\Gamma_1^p$ is
calculated and compared with data [(2.33) leads to 0.171 for $\Gamma_1^p$\,]. 
This example gives a nice demonstration that an eye-fit to the data can
be quite misleading \cite{CLW}.
Since the unpolarized strange-sea distribution is
small at $x>0.2$, the positivity constraint $|\Delta s(x)|\leq s(x)$ 
implies that the data of $g_1^p(x)$ should be fully accounted for by
$\Delta u_v(x)$ and $\Delta d_v(x)$. In Sec.~7.1 we show that a best least
$\chi^2$ fit to $g_1^p(x)$ leads to a parametrization (7.11) for valence quark
spin densities. The theoretical curve of $g_1^p$ without sea and gluon 
contributions is depicted in Fig.~1. It is clear that a deviation of theory
from experiment for $g_1^p(x)$ manifests at small $x$ where sea 
polarization starts to play an essential role.

%\newpage
\section{Anomalous Gluon Effect in the Parton Model}
\setcounter{equation}{0}

\subsection{Anomalous gluon contributions from box diagrams}

   We see from Section II that the polarized DIS data indicate that
the fraction of the proton spin carried by the light quarks inside the proton
is $\Delta\Sigma\approx 0.30$ and the strange-quark polarization is
$\Delta s\approx -0.10$ at $Q^2=10\,{\rm GeV}^2$. The question is what kind of
mechanism can generate a sizeable and negative sea polarization. It is
difficult, if not impossible, to accommodate a large
$\Delta s$ in the naive parton model because massless quarks and
antiquarks created from gluons have opposite helicities owing to helicity
conservation. This implies that {\it sea polarization for massless quarks
cannot be induced perturbatively from hard gluons, irrespective of gluon
polarization.} (Recall that our definition of $\Delta q$ (2.15) 
includes both quark and antiquark contributions.) 
It is unlikely that the observed $\Delta s$ comes solely from 
nonperturbative effects or from chiral-symmetry breaking due to nonvanishing 
quark masses. (We will discuss in Sec.~4.4 the possible mechanisms for
producing sea polarization.)
It was advocated by Efremov and Teryaev \cite{Efr}, 
Altarelli and Ross \cite{AR}, Carlitz, Collins and Mueller \cite{CCM} 
(see also Leader and Anselmino \cite{Lead}) that 
the difficulty with the unexpected large sea polarization can be overcome by 
the anomalous gluon effect stemming from the axial anomaly, which we shall 
elaborate in this section.

    As an attempt to understand the polarized DIS data, we consider QCD
corrections to the polarized proton structure function $g_1^p(x)$. To the
next-to-leading order (NLO) of $\alpha_s$, the expression for
$g_1^p(x)$ is\footnote{Beyond NLO, it is necessary to decompose the quark
spin density into singlet and non-singlet components; see Eq.(7.6)
for a most general expression of $g_1^p(x,Q^2)$.}
\be
g_1^p(x,Q^2) &=& {1\over 2}\sum^{n_f}_i e^2_i\Big\{\ [\Delta q_i(x,Q^2)+{
\alpha_s(Q^2)\over 2\pi}\Delta f_q(x,Q^2)\otimes\Delta q_i(x,Q^2)] \non \\
&& +\gg_{\rm hard}(x,Q^2)
\otimes\Delta G(x,Q^2)\Big\},
\en
where $n_f$ is the number of active quark flavors, $\Delta q(x)\equiv 
q^\uparrow(x)+
\bar{q}^\uparrow(x)-q^\downarrow(x)-\bar{q}^\downarrow(x)$, $\Delta G(x)\equiv
G^\uparrow(x)-G^\downarrow(x)$, and $\otimes$ denotes the convolution
\be
f(x)\otimes g(x)=\int^1_x{dy\over y}f\left({x\over y}\right)g(y).
\en
There are two different types of QCD corrections in (3.1): the $\Delta f_q$ 
term arising from vertex and self-energy corrections (corrections due to 
real gluon emission account for the $\ln Q^2$ dependence
of quark spin densities) and the other from polarized photon-gluon scattering
[the last term in (3.1)]. As we shall see later, the QCD effect due to
photon-gluon scattering is very special: Unlike the usual QCD corrections, it
does not vanish in the asymptotic limit $Q^2\to\infty$. 
The $\Delta f_q(x)$ term in (3.1) depends on the regularization scheme 
chosen. Since the majority of unpolarized parton distributions is 
parametrized and fitted to data in the $\overline{\rm MS}$
scheme, it is natural to adopt the same regularization scheme for polarized
parton distributions in which \cite{Rat}\footnote{The expression of (3.3) for
$\Delta f_q(x)$ is valid for $\u^2=Q^2$, where $\u$ is a factorization
scale to be introduced below. When $\u^2\neq Q^2$, the contribution
\cite{Rat}
\be
{4\over 3}\left[2\left({1\over 1-x}\right)_+-1-x+{3\over 2}\delta(1-x)
\right]\ln{Q^2\over \u^2}   \non
\en
should be added to $\Delta f_q(x,Q^2)$.}
\be
\Delta f_q(x) &=& f_q(x)-{4\over 3}(1+x)  \non \\
 &=& {4\over 3}\Bigg[ (1+x^2)\left({\ln(1-x)\over 1-x}\right)_+-{3\over 
2}{1\over (1-x)_+}-\left({1+x^2\over 1-x}\right)\ln x  \non \\
&& +3+2x-\left({9\over 2}+{\pi^2\over 3}\right)\delta(1-x)\Bigg]-{4\over 3}
(1+x),
\en
where the ``$+$" distribution is given by
\be
\int^1_0 g(x)\left( {f(x)\over 1-x}\right)_+dx=\int^1_0 f(x)\,{g(x)-g(1)\over
1-x}dx.
\en
The first moment of $f_q(x)$ and $\Delta f_q(x)$ is 0 and $-2$, respectively.
Note that the first moment of $\Delta f_q(x)$ is scheme independent at least 
to NLO.

\begin{figure}[ht]
\vspace{-4cm}

    \centerline{\psfig{figure=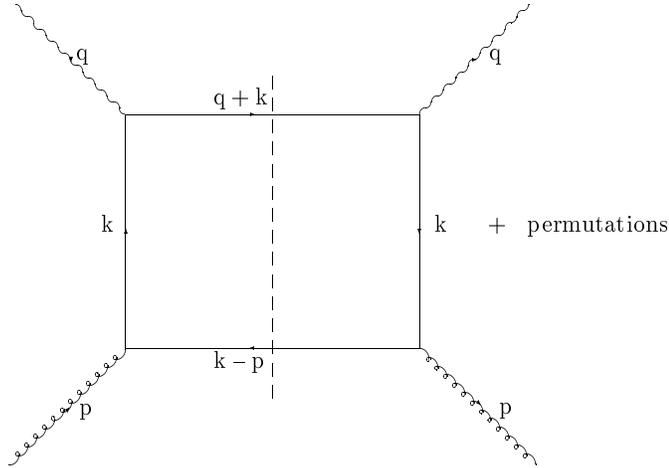,width=8cm}}
    \caption[]{\small The photon-gluon scattering box graph.}
    \label{fig1} 

\end{figure}

   To compute the polarized photon-gluon cross section $\gg$
amounts to evaluating the box diagram of photon-gluon scattering with a 
physical cutoff on the intermediate states (see Fig.~2). Using the relation
\be
\ep_i(+)\ep^*_j(+)-\ep_i(-)\ep^*_j(-)=-i\epsilon_{ij},
\en
where $\ep^\mu(\pm)=(0,0,1,\pm i)/\sqrt{2}$ is the transverse polarization of 
external gluons and $\epsilon_{ij}$ is an antisymmetric tensor with 
$\epsilon_{12}=1$, the contribution of Fig.~2 to $\gg$ for a single quark 
flavor is 
\cite{CCM}
\be
\gg &\equiv& \sigma(\gamma^{(+)}G^{(+)}\to q\bar{q})-\sigma(\gamma^{(+)}
G^{(-)}\to q\bar{q})   \non \\
&=& 2g^2T\int {d^2k_\perp dk^+dk^-\over (2\pi)^4}\,{ 2\pi\delta[(p-k)^2-m^2]
\delta[(q+k)^2-m^2]{\rm Tr}\{ \cdots \} \over (k^2-m^2+i\epsilon)^2 },
\en
with
\be
{\rm Tr}\{ \cdots \} &=& {1\over 4}\epsilon_{ij}\epsilon_{ln}{\rm Tr}\Big\{ 
\gamma^j(k\!\!\!/+m)\gamma^n (q\!\!\!/+k\!\!\!/+m)\gamma^l(k\!\!\!/+m)\gamma^i
(p\!\!\!/-k\!\!\!/-m)   \non \\
&& +\gamma^j(k\!\!\!/+q\!\!\!/-p\!\!\!/+m)\gamma^n (k\!\!\!/-p\!\!\!/+m)
\gamma^i(k\!\!\!/+m)\gamma^l(k\!\!\!/+q\!\!\!/+m)\Big\},
\en
where $T={1\over 2}$, $m$ is the mass of the quark, $p$ is the gluon momentum, 
and $k_\perp$ is the quark transverse momentum perpendicular to the virtual 
photon direction. It is convenient to evaluate the integral of (3.6) in the 
light-front coordinate $p^\mu=(p^-,p^+,p_\perp)$ with $p_\perp=0$ and
$p^\pm=(p^0\pm p^3)/\sqrt{2}$. A tedious but straightforward calculation
yields (for a derivation, see e.g.,
\cite{Dok} for the general case and \cite{AR} for on-shell gluons):
\be
\gg(x,Q^2)=-{\alpha_s\over 2\pi}\int^{K^2}_0{dk^2_\perp\over\sqrt{1-k^2_
\perp/K^2}}\left({(1-2x)
(k^2_\perp+m^2)-2m^2(1-x)\over [\,k^2_\perp+m^2-p^2x(1-x)]^2}-{1-2x\over 2K^2}
\right),
\en
where $K^2=[(1-x)/4x]Q^2$ and $x=k^+/p^+$. Note that all higher-twist
corrections of order $p^2/Q^2$ and $m^2/Q^2$ have been suppressed in (3.8).
It is evident that $\gg(x)$
has infrared and collinear singularities at $m^2=p^2=0$ and $k_\perp^2=0$.
Hence, we have to introduce a soft cutoff to make $\gg(x)$ finite.
For $Q^2>> m^2,-p^2$, Eq.(3.8) reduces to (for an exact expression of 
$\gg(x)$ after $k_\perp$ integration, see \cite{Vog1,Bass1})
\be
\gg(x,Q^2) &=& {\alpha_s\over 2\pi}\Bigg[ (2x-1)\left(\ln{Q^2\over m^2-p^2x
(1-x)}+\ln{1-x\over x}-1\right)   \non \\
&& +(1-x)\,{2m^2-p^2x(1-2x)\over m^2-p^2x(1-x)}\Bigg].
\en
Depending on the infrared regulators, we have
\be
\gg_{\rm CCM}(x,Q^2) &=& {\alpha_s\over 2\pi}\,(2x-1)\left(\ln{Q^2\over 
-p^2}+\ln{1\over x^2}-2\right),   \\
\gg_{\rm AR}(x,Q^2) &=& {\alpha_s\over 2\pi}\left[(2x-1)\left(\ln{Q^2\over 
m^2}+\ln{1-x\over x}-1\right)+2(1-x)\right],   \\
\gg_{\rm R}(x,Q^2) &=& {\alpha_s\over 2\pi}\left[(2x-1)\left(\ln{Q^2\over 
\ums^2}+\ln{1-x\over x}-1\right)+2(1-x)\right],   
\en
for the momentum regulator ($p^2\neq 0$) \cite{CCM}, the mass regulator 
($m^2\neq 0$) \cite{AR} and the dimensional regulator ($\ums\neq 0$) in
the modified miminal subtraction scheme \cite{Rat}, respectively. Note that 
the coefficient $(2x-1)$ in Eqs.(3.10-3.12) is nothing but the spin 
splitting function $2\Delta P_{qG}(x)$ [see (3.26)] and that 
the term proportional to $2(1-x)$ in (3.11) and (3.12) is an effect of chiral 
symmetry breaking:\footnote{The $2(1-x)$ term in (3.11) and (3.12) was 
neglected in the original work of Altarelli and Ross \cite{AR} and of 
Ratcliffe \cite{Rat}. One may argue that since this contribution is soft, 
it will not contribute to ``hard" $\gg$.  As shown below, the cross sections 
$\gg_{\rm AR}(x)$ and $\gg_{\rm R}(x)$ without the $2(1-x)$ term indeed give 
the correct result for the first moment of $\gg_{\rm hard}$ in
the chiral-invariant factorization scheme.}
It arises from the 
region where $k_\perp^2\sim m^2$ in the mass-regulator scheme, and from 
$k_\perp^2\sim\ums^2$ in the $\epsilon=n-4$ dimensions 
in the dimensional regularization scheme due to the violation of the identify
$\{\gamma_\mu,\,\gamma_5\}=0$. For the first moment of $\gg(x)$,
it is easily seen that
\be
\int^1_0\gg_{\rm CCM}(x)dx=-{\alpha_s\over 2\pi},~~~\int^1_0\gg_{\rm AR}(x)
dx=\int^1_0\gg_{\rm R}(x)dx=0.
\en
The result (3.13) can be understood as follows. The cutoff-dependent
logarithmic term, which is 
antisymmetric under $x\to 1-x$, makes no contribution to 
$\int^1_0\gg(x)dx$, a consequence of chiral symmetry or helicity 
conservation. As a result, $\int^1_0\gg(x)dx$ receives ``hard" contributions
from $k_\perp^2\sim Q^2$ in the momentum-regulator scheme, but it is
compensated by the soft part arising from $k_\perp^2\sim m^2$ in the 
mass-regulator scheme. 

   It is clear that the cross sections given by (3.10-3.12) are not 
perturbative QCD reliable since they are sensitive to the choice of the 
regulator. First, there are terms depending logarithmically on the soft 
cutoff. Second, the first moment of $\gg(x)$ is regulator dependent. It
is thus important to have a reliable perturbative QCD calculation for $\gg(x)$
since we are interested in QCD corrections to $g_1^p(x)$. To do this, we need
to introduce a factorization scale $\u$, so that
\be
\gg(x,Q^2)=\,\gg_{\rm hard}(x,Q^2,\u^2)+\gg_{\rm soft}(x,\u^2)
\en
and the polarized photon-proton cross section is decomposed into
\be
 \Delta\sigma^{\gamma p}(x,Q^2) &=& \sum_i^{n_f}\Big(\Delta\sigma^{\gamma 
q}(x,Q^2,\u^2)\otimes\Delta q_i(x,\u^2)   \non \\
&+& \gg_{\rm hard}(x,Q^2,\u^2)
\otimes\Delta G(x,\u^2)\Big).   
\en
That is, the hard piece of $\gg(x)$ which can be evaluated reliably by
perturbative QCD contributes to $g_1^p(x)$, while the 
soft part is factorized into the nonperturbative quark spin densities 
$\Delta q_i(x)$. Since $\Delta\sigma^{\gamma p}(x)$ is a physical quantity,
a different factorization scheme amounts to a different way of shifting the 
contributions between $\gg_{\rm hard}(x)$ and $\Delta q(x)$. An obvious 
partition of $\gg(x)$ is that the region where $k_\perp^2\gsim\u^2$ 
contributes to the hard cross section, whereas the soft part receives 
contributions from $k_\perp^2\lsim\u^2$ and hence can be interpreted
as the quark and antiquark spin densities in a gluon, i.e., 
$\gg_{\rm soft}(x,\u^2)=\Delta q^G(x,\u^2)$.
Physically, the quark and antiquark jets produced in deep inelastic
scattering with $k_\perp^2\lsim \u^2$ are not hard enough to satisfy the
jet criterion and thus should be considered as a part of one-jet cross
section \cite{CCM}. The choice of the ``ultraviolet" cutoff for soft 
contributions specifies the $k_\perp$ factorization
convention. There are two extremes of interest: the
chiral-invariant scheme in which the ultraviolet regulator respects chiral
symmetry, and the gauge-invariant scheme in which gauge symmetry is respected
but chiral symmetry is broken by the cutoff.

 The next task is to compute $\gg_{\rm soft}(x)$. It can be calculated from
the box diagram by making a direct cutoff $\sim \u$ on the $k_\perp$ 
integration. Note that for $k_\perp^2\lsim\u^2$, the box diagram for 
photon-gluon scattering is reduced under the collinear approximation for the 
quark-antiquark pair created by the gluon to a triangle diagram  
with the light-front cut vertex $\gamma^+\gamma^5$ combined with a trivial 
photon-quark scattering \cite {CCM,BQ}. As a result, $\Delta q^G(x)$  can
be also obtained by calculating the triangle diagram with the
constraint $k_\perp^2\lsim \u^2$. In either way, one obtains
\be
\gg_{\rm soft}(x,\u^2)_{\rm CI}=\Delta q^G_{\rm CI}(x,\u^2)=-{\alpha_s\over
2\pi}\int^{\u^2}_0 dk^2_\perp\,{(k^2_\perp+m^2)(1-2x)-2m^2(1-x)\over [\,k^2_
\perp+m^2-p^2x(1-x)]^2},
\en
where ${\cal O}(1/Q^2)$ corrections are negligible for $\u^2<< Q^2$ and the 
subscript CI indicates that we are working in the chiral-invariant 
factorization scheme. The result is \cite{Ste}
\be
\Delta q^G_{\rm CI}(x,\u^2) &=& {\alpha_s\over 2\pi}\Bigg\{ (2x-1)\ln{\u^2+
m^2-p^2x(1-x)\over m^2-p^2x(1-x)}  \non \\
&& +(1-x)\,{2m^2-p^2x(1-2x)\over m^2-p^2x(1-x)}\,{\u^2\over \u^2+m^2-p^2x
(1-x)}\Bigg\}.
\en
For $\u^2>>m^2,~-p^2$, it reduces to
\be
\Delta q^G_{\rm CI}(x,\u^2)=\cases{ 
{\alpha_s\over 2\pi}\left[(2x-1)\left(\ln(\u^2/ -p^2)+\ln{1\over x(1-x)}
\right)+1-2x\right];   \cr
{\alpha_s\over 2\pi}\left[(2x-1)\ln(\u^2/ m^2)+2(1-x)\right];   \cr
{\alpha_s\over 2\pi}\left[(2x-1)\ln(\u^2/\ums^2)+2(1-x)\right],   \cr}
\en
for various soft cutoffs. Note that, as stressed in 
\cite{Mank}, the soft cross sections or quark spin densities in a helicity 
$+$ gluon given by (3.17) or (3.18) do not make sense in QCD as they are 
derived using perturbation theory
in a region where it does not apply. Nevertheless, it is instructive to 
see that 
\be
\Delta q^G_{\rm CI}=\int^1_0\Delta q^G_{\rm CI}(x)dx=0,~~~{\rm for}~
m^2=0~~{\rm or}~ -p^2>>m^2,
\en 
as expected. Hence, a sea polarization for massless quarks, if any, 
must be produced nonperturbatively or via the anomaly
(see Sec.~4.4). Now it does make sense in 
QCD to subtract $\gg_{\rm soft}$ from $\gg$ to obtain a 
reliable perturbative QCD result for $\gg_{\rm hard}$:
\be
\gg_{\rm hard}(x,Q^2,\u^2)_{\rm CI}=\,{\alpha_s\over 2\pi}(2x-1)\left
(\ln{Q^2\over\u^2}+\ln{1-x\over x}-1\right).
\en
Evidently, $\gg_{\rm hard}(x)$ is independent of the infrared regulators 
as long as $\u^2>>\ums^2,m^2,-p^2$; terms depending on soft cutoffs are 
absorbed into the quark spin densities.
It is also clear that the soft $2(1-x)$ term in 
(3.11) and (3.12) drops out in $\gg_{\rm hard}(x)$. Therefore,
\be
\gg_{\rm hard}(Q^2,\u^2)_{\rm CI}=\int^1_0dx\gg_{\rm hard}(x,Q^2,\u^2)_{\rm 
CI}=-{\alpha_s\over 2\pi}.
\en
Since gauge invariance and helicity conservation in the quark-gluon vertex 
are not broken in the chiral-invariant factorization scheme, it is evident
that $\Delta q^G_{\rm CI}$ does not evolve, consistent with the naive 
intuition based on helicity conservation that the quark spin $\Delta q^G_{\rm 
CI}=\int^1_0\Delta q^G_{\rm GI}(x)dx$ for massless quarks is $Q^2$ 
independent.

   Substituting (3.3) and (3.20) into (3.1) leads to
\be
\Gamma_1^p(Q^2)\equiv\int^1_0 dxg_1^p(x,Q^2)=\,{1\over 2}\left(1-{
\alpha_s\over 
\pi}\right)\sum_i\left[\Delta q_i(Q^2)_{\rm CI}-{\alpha_s(Q^2)\over 
2\pi}\Delta G(Q^2)\right],
\en
where $\Delta q_{\rm CI}(Q^2)=\Delta q_{\rm CI}(Q^2_0)$ and use has been 
made of
\be
\int^1_0dx\int^1_x{dy\over y}f\left({x\over y}\right)g(y)=\int^1_0dxf(x)
\int^1_0dy\,g(y).
\en
The $(1-{\alpha_s\over \pi})$ term in (3.22) comes from the QCD loop 
correction,\footnote{Recall that perturbative QCD corrections to 
$\Gamma_1^p$ have been calculated up to the order of $\alpha_s^3$; see 
Eq.(2.21).}
while the $\alpha_s\Delta G$ term arises from the box diagram of photon-gluon 
scattering. If the gluon polarization inside the proton is positive, a 
partial cancellation between $\Delta q$ and
${\alpha_s\over 2\pi}\Delta G$ will explain why the observed $\Gamma_1^p$
is smaller than what naively expected from the Ellis-Jaffe sum rule. 
It is tempting to argue that the box-diagram QCD correction is negligible
at large $Q^2$ since $\alpha_s$ vanishes in the asymptotic limit 
$Q^2\to\infty$.
However, it is not the case. To see this, consider the Altarelli-Parisi (AP)
equation for flavor-singlet polarized parton distribution functions:
\be
{d\over dt}\left(\matrix{\Delta\Sigma_{\rm CI}(x,t)   \cr   \Delta G(x,t) \cr}
\right)=\,{\alpha_s(t)\over 2\pi}\left(\matrix{\Delta P_{qq}^{\rm S}(x) & 
2n_f\Delta P_{qG}(x)  \cr  \Delta P_{Gq}(x) & \Delta P_{GG}(x) \cr} 
\right)\otimes\left(\matrix
{\Delta \Sigma_{\rm CI}(x,t)   \cr   \Delta G(x,t) \cr} \right),
\en
where $t=\ln(Q^2/\Lambda^2_{\rm QCD})$. Although the leading-order splitting 
functions in
\be
\Delta P(x)=\Delta P^{(0)}(x)+{\alpha_s\over 2\pi}\Delta P^{(1)}(x)+\cdots,
\en
have been obtained long time ago \cite{Alt}, the NLO results are not 
available until very recently \cite{Mert}. To the leading order, the AP 
splitting functions read \cite{Alt}
\be
&&\Delta P_{{\rm S},qq}^{(0)}(x)={4\over 3}\left({1+x^2\over 1-x}\right)_+,~~~
\Delta P_{qG}^{(0)}(x)={1\over 2}(2x-1),~~~\Delta P_{Gq}^{(0)}(x)={4\over 3}
(2-x), \non \\
&& \Delta P_{GG}^{(0)}(x)=3\left[(1+x^4)\left({1\over x}+{1\over 
(1-x)_+}\right)-{(1-x)^3\over x}\right]+{\beta_0\over 2}\delta(1-x),
\en
with $\beta_0=(33-2n_f)/3$. Since $\alpha_s(Q^2)=4\pi/(\beta_0
\ln Q^2/\Lambda^2_{\rm QCD})$, it follows from (3.24) that
\be
{d\over d\ln Q^2}\left(\alpha_s(Q^2)\Delta G(Q^2)\right)={\alpha_s^2\over
\pi}\Delta\Sigma={\cal O}(\alpha_s^2).
\en
Consequently, $\alpha_s\Delta G$ is conserved to the leading-order QCD
evolution;\footnote{This constant behavior for $\alpha_s\Delta G$ also can be 
seen from the analysis of anomalous dimensions of the Chern-Simons current
(see Sec.~4.6).}
that is, $\Delta G$ grows with $\ln Q^2$, whereas $\alpha_s$ is
inversely proportional to $\ln Q^2$. Explicitly, a solution to (3.24) reads
\be
\Delta\Sigma(Q^2) &=& \Delta\Sigma(Q^2_0)\,,  \non \\
\Delta G(Q^2) &=& -{4\over \beta_0}\Delta\Sigma(Q^2_0)+{\ln Q^2/\Lambda^2_{_
{\rm QCD}}\over \ln Q^2_0/\Lambda^2_{_{\rm QCD}}}\left(\Delta G(Q^2_0)+{4
\over \beta_0}\Delta\Sigma(Q^2_0)\right).
\en
Hence, {\it hard gluons contribute to the
first moment of $g_1^p(x)$ even in the asymptotic limit}. As we shall see 
below, it is the axial anomaly that makes this QCD effect so special.

   Physically, the growth of the gluon spin with $Q^2$ can be visualized
in two different ways. From (3.26) we have 
$\int^1_0\Delta P_{Gq}^{(0)}(x)dx=\,2\,$.
This means that a polarized quark is preferred to radiate a gluon with
helicity parallel to the quark spin. Since the net quark spin component within
the proton is positive, it is clear that $\Delta G>0$ at least for the gluons 
perturbatively emitted from quarks. As $Q^2$ increases, the number of 
gluons with + helicity radiated from polarized quarks also increases
and this explains why $\Delta G$ grows with $Q^2$. Alternatively, this
growth also can be understood by considering the splitting of a helicity
+ gluon into a quark-antiquark pair or into two gluons.  Since
\be
\int^1_0\Delta P_{qG}^{(0)}(x)dx=0,~~~~\int^1_0\Delta P_{GG}^{(0)}(x)dx=
{1\over 2}\beta_0,
\en
the gluon helicity has a net gain with probability
$11/2-n_f/3>0$ in the splitting \cite{Ji1}. Thus the gluon spin component
increases with increasingly smaller distance scale. Now we see that
perturbative QCD provides all the necessary ingredients for understanding
the smallness of $\Delta\Sigma$. As a result of anomalous gluonic
contributions to $\Gamma_1^p$ in the chiral-invariant factorization scheme,
what measured in polarized DIS experiments is not
$\Delta q$, but rather a combination of $\Delta q$ and $\alpha_s\Delta G$
[cf. Eq.(3.22)]:
\be
\Delta q\to\Delta q-{\alpha_s\over 2\pi}\Delta G.
\en
Consequently, (2.26) and (2.27) are replaced by 
\be
\Delta u_{\rm CI}-{\alpha_s\over 2\pi}\Delta G &=& 0.83\pm 0.03\,,  \non \\
\Delta d_{\rm CI}-{\alpha_s\over 2\pi}\Delta G &=& -0.43\pm 0.03\,,   \\
\Delta s_{\rm CI}-{\alpha_s\over 2\pi}\Delta G &=& -0.10\pm 0.03\,,  \non
\en
and 
\be
(\Delta u+\Delta d+\Delta s)_{\rm CI}-{3\alpha_s\over 2\pi}\Delta G=\,0.31
\pm 0.07
\en
at $Q^2=10\,{\rm GeV}^2$.
(3.31) and (3.32) imply that in the presence of anomalous gluon 
contributions, $\Delta\Sigma_{\rm CI}$ is not necessarily small and $\Delta
s_{\rm CI}$ is not necessarily large. In the absence of sea polarization
and in the framework of perturbative QCD,
it is easily seen that $\Delta G\sim 0.10(2\pi/\alpha_s)\sim 2.5$ at $Q^2=
10\,{\rm GeV}^2$ and $\Delta\Sigma_{\rm CI}\sim 0.60\,.$ It thus provides
a nice and simple solution to the proton spin problem: This improved parton 
picture is reconciled, to a large degree,
with the constituent quark model and yet explains
the suppression of $\Gamma_1^p$, provided that $\Delta G$ is positive and
large enough. This anomalous gluon interpretation of the observed 
$\Gamma_1^p$, as first proposed in \cite{Efr,AR,CCM}, looks appealing and 
has become a popular explanation since 1988.
Note that $\Delta G\sim 2.5$ ought to be regarded as an upper 
limit for the magnitude of the gluon spin component within a proton, as it
is derived by assuming no intrinsic strange-sea polarization (see also
Sec.~4.4).

\subsection{Role of the axial anomaly}

    In order to understand the origin of the anomalous gluon contribution
to $\Gamma_1^p$, we consider an important consequence of the OPE which 
requires that \cite{CCM}
\be
\int^1_0\gg(x)dx=\,{1\over 2p^+}\Gamma^+_5,
\en
where $\Gamma^+_5$ is the contribution of the triangle diagram for the
axial-vector current $J_5^+$ between external gluons (see Fig.~3 in Sec.~4.2)
evaluated in the
light-front coordinate $p^\mu=(p^-,p^+,p_\perp)$. The relation (3.33)
ensures that the two different approaches, the OPE and the improved parton 
model, yield the same results. It has been shown in \cite{CCM} that
the integrands of both sides 
of (3.33) are equal in the low $k_\perp^2$ region. This in turn implies that
$\gg_{\rm soft}(x)=\Delta q^G(x)$, namely the soft part of the photon-gluon
scattering cross section equals to the soft part of the triangle diagram,
a relation which we have employed before for computing the quark spin 
densities inside a polarized gluon [see Eq.(3.16)].
Moreover, we have shown that $\int^1_0\gg_{\rm soft}(x)dx=\int^1_0\Delta q
^G_{\rm CI}(x)dx=0$ for $m^2=0$ or $-p^2>>m^2$ [cf. Eq.(3.19)]. This
means that only the integrands at large $k_\perp^2$ region contribute 
to (3.33).

    It is well known that the triangle diagram has an axial anomaly
manifested at $k_\perp^2\to\infty$ (see Sec.~4.3). Since only the $k^2_\perp
\sim Q^2$ region contributes to the nonvanishing first moment of $\gg(x)$
\cite{CCM},
it follows from (3.33) that {\it the anomalous gluon contribution $-{\alpha_s
\over 2\pi}\Delta G$ to $\Gamma_1^p$ is intimately related to the axial 
anomaly}. (Both sides of (3.33) have values $-\alpha_s/ 2\pi$.) That is, the 
gluonic anomaly occurs in the box diagram (Fig.~2) at $k_\perp^2=K^2
\equiv[(1-x)/4x]Q^2$ with $x\to 0$ and contributes to the first moment of
$\gg_{\rm hard}(x)$. This means that the upper quark line in the box 
diagram has 
shrunk to a point and this point-like behavior measures the gluonic
component of the quark Fock space \cite{CCM}, which is identified with 
the contribution $-{\alpha_s\over 2\pi}\Delta G$ to $\Gamma_1^p$. 

   At this point, it is instructive to compare unpolarized and polarized
structure functions. The unpolarized structure function $F_1(x,Q^2)$ has
a similar expression as Eq.(3.1) for the polarized one. However,
the first moments of unpolarized $f_q(x)$
[cf. Eq.(3.3)] and $\sigma^{\gamma G}(x)$ vanish so that QCD corrections to
$\int F_1(x)$ reside entirely in the $Q^2$ evolution of the first moment
of the unpolarized quark distributions:
\be
\int^1_0 F_1(x,Q^2)dx=\,{1\over 2}\sum_i^{n_f}e^2_iq_i(Q^2).
\en
It is mainly the anomalous gluon contribution that makes $\int 
g_1^p(x)dx$ behave so differently from $\int F_1(x)dx$.
We conclude that {\it it is the gluonic anomaly that accounts for the 
disparity between the first moments of $g_1^p(x)$ and $F_1(x)$.}

    We should remind the reader that thus far in this section we have only 
considered the chiral-invariant factorization scheme in which a brute-force
ultraviolet cutoff on the $k_\perp$ integration is introduced to the soft 
part of the box diagram. In this case, the axial anomaly manifests in the 
hard cross section for photon-gluon scattering. However, this is not the only 
$k_\perp$-factorization scheme we can have. In the next section, we shall see 
that it is
equally acceptable to choose a (gauge-invariant) factorization prescription
in which the anomaly is shifted from $\gg_{\rm hard}$ to the quark spin
density inside a gluon. Contrary to the aforementioned anomalous gluon 
effects, hard gluons in the gauge-invariant scheme do {\it not} make 
contributions to the first moment of $\gg_{\rm hard}(x)$.

   Before ending this section we would like to make two remarks. The first
one is a historical remark.

   1). The first consideration of the hard gluonic contribution to 
$\Gamma_1^p$ was put forward by Lam and Li \cite{Lam} long before the EMC 
experiment. The questions of the regulator dependence in the evaluation
of the photon-gluon scattering box diagram, the identification of $\Delta 
G$ with the
forward nucleon matrix element of the Chern-Simons current (see Sec.~4.5),
and the $Q^2$ behavior of $\alpha_s\Delta G,\cdots$ etc. were already 
addressed by them. A calculation of $\gg(x)$ using the dimensional
regularization was first made by Ratcliffe \cite{Rat} also before the EMC
paper.

    2). We see from (3.26) that $\int^1_0\Delta P_{qq}(x)=0$. This indicates
that $\Delta\Sigma_{\rm CI}$ is $Q^2$ independent. Physically, this is because
the quark helicity is conserved by the vector coupling of a gluon to a 
massless quark. However, $\Delta\Sigma_{\rm CI}$ and $\Delta q_{\rm CI}$ 
cannot be written as a nucleon matrix element of a local and gauge-invariant
operator (this will be discussed in Sec.~4.5). Since $\Delta\Sigma_{\rm CI}$ 
does not evolve and since $\Delta G$ induced by gluon emissions from
quarks increases with $Q^2$, conservation of angular momentum requires
that the growth of the gluon polarization with $Q^2$ be compensated by
the orbital angular momentum of the quark-gluon pair so that the spin 
sum rule (2.32) is $Q^2$ independent; that is, $L_z^q+L_z^G$ also 
increases with $Q^2$ with opposite sign. It is conjectured in Sec.~6.3 that
$L_z^q$ in the chiral-invariant factorization scheme could be negative.

%\newpage
\section{Sea Polarization Effect in the OPE Approach}
\setcounter{equation}{0}
\subsection{Preamble}

   We see from the last section that the anomalous gluon contribution to
$\Gamma_1^p$ furnishes a simple and plausible solution to the proton spin
problem. A positive and large gluon spin component will help explain the
observed suppression of $\Gamma_1^p$ relative to the Ellis-Jaffe conjecture
and in the meantime leave the constituent quark model as intact as possible, 
e.g., $\Delta\Sigma\sim 0.60$ and $\Delta 
s\sim 0.$ However, this is by no means the end of the proton-spin story.
According to the OPE analysis, only quark operators contribute to the first 
moment of $g_1^p$ at the twist-2, spin-1 level. As a consequence, hard gluons 
do not make contributions to $\Gamma_1^p$ in the OPE approach. This is in 
sharp conflict with the improved parton model discussed before. Presumably, 
the OPE is more trustworthy as it is a model-independent approach.
So we face a dilemma here: On the one hand, the
anomalous gluon interpretation sounds more attractive and is able to 
reconcile to a large degree with the conventional quark model; on the 
other hand, the sea-quark 
interpretation of $\Gamma_1^p$ relies on a solid theory of the OPE. In fact,
these two popular explanations for the $g_1^p$ data have been under hot
debate over the past years.

    Although the OPE is a first-principles approach, the sea-quark 
interpretation is nevertheless subject to two serious criticisms. First, how
do we accommodate a large and negative strange-quark polarization $\Delta s
\sim -0.10$ within a proton ? Recall that, as we have repeatedly emphasized,
no sea polarization for massless quarks is expected to be generated
perturbatively from hard gluons owing to helicity conservation. Second, the
total quark spin $\Delta\Sigma$ in the OPE has an anomalous dimension first
appearing at the two-loop level. This means that $\Delta\Sigma$ evolves with
$Q^2$, in contrast to the naive intuition that the quark helicity is not
affected by the gluon emission. In the last 7 years, there are over a
thousand papers triggered by the unexpected EMC observation. Because of
the above-mentioned criticisms and because of the deviation of the 
sea-polarization 
explanation from the quark-model expectation, the anomalous gluon
interpretation seems to be more favored in the past by many of the 
practitioners in the field.

    In this section we will point out that within the approach of the OPE it 
is precisely the axial anomaly that provides the necessary mechanism for 
producing a negative sea-quark polarization from gluons. Hence, the sea-quark 
interpretation of $\Gamma_1^p$ is as good as the anomalous gluon one. 
In fact, we will show in the next section that these two different popular
explanations are on the same footing; physics is
independent of how we define the photon-gluon cross section and the quark 
spin density.

\subsection{A mini review of the OPE}

   The approach of the operator product expansion provides relations between
the moments of structure functions and forward matrix elements of
local gauge-invariant operators (for a nice review, see \cite{Man1}).
For {\it inclusive} deep inelastic
scattering, the hadronic tensor $W_{\mu\nu}$ has the expression
\be
W_{\mu\nu}=\,{1\over 2\pi}\int d^4x\,e^{iq\cdot x}\la p,s|\,[J_\mu(x),~
J_\nu(0)]|p,s\ra
\en
for a nucleon state with momentum $p^\mu$ and spin $s^\mu$. Since $W_{\mu\nu}$ 
in the DIS limit is dominated by the light-cone region where $x^2\sim 0$
(but not necessarily $x^\mu\sim 0$), the structure of the current product
is probed near the light cone. In order to evaluate $W_{\mu\nu}$, it is 
convenient to consider the time-ordered product of two currents:
\be
t_{\mu\nu}=\,i\int d^4x\,e^{iq\cdot x}T(J_\mu(x)J_\nu(0))
\en
and the forward Compton scattering amplitude $T_{\mu\nu}=\la p,s|t_{\mu\nu}
|p,s\ra$, which is related to the hadronic tensor by the relation $W_{\mu\nu}
={1\over \pi}{\rm Im}T_{\mu\nu}$ via the optical theorem.

    In the limit $q\to\infty$, the operator product expansion allows us to
expand $t_{\mu\nu}$ in terms of local operators; schematically,
\be
\lim_{q\to\infty}t_{\mu\nu}=\sum_n C_{\mu\nu,n}(q)O_n(0).
\en
The Wilson coefficient functions $C_n$ can be obtained by computing the 
quark or gluon matrix elements of $J_\mu J_\nu$ and $O_n$. Consider $t_{\mu
\nu}$ in the complex $\omega~(=1/x=2p\cdot q/-q^2)$ plane. By analyticity, 
the Feynman 
amplitude $M_{\mu\nu}$ corresponding to the free quark (or gluon) matrix 
element of $J_\mu J_\nu$ can be calculated at $\omega$ near 0 (but not in 
the physical region $1<\omega<\infty)$ and expanded around $\omega=0$. 
Generically,
\be
M_{\mu\nu}=\la k,\lambda|t_{\mu\nu}|k,\lambda\ra \sim\sum_n C_{\mu\nu,n}
\omega^n\la k,\lambda|O_n|k,\lambda\ra
\en
for a quark state with momentum $k^\mu$ and spin $\lambda^\mu$. Since the
free quark matrix elements of the quark operators have the form
\be
\la k,\lambda|O_V^{\mu_1\cdots\mu_n}|k,\lambda\ra=k^{\mu_1}\cdots k^{\mu_n},
~~~\la k,\lambda|O_A^{\mu_1\cdots\mu_n}|k,\lambda\ra=hk^{\mu_1}\cdots
k^{\mu_n},
\en
for vector and axial-vector types of quark operators, where $h$ is a helicity
of the quark state, the coefficient functions $C_{\mu\nu,n}$ 
are thus determined.

    The leading-twist (=\,dimension$-$spin) contributions to the antisymmetric
(spin-dependent) part of $t_{\mu\nu}$ in terms of the operator product
expansion are
\be
t_{[\mu\nu]}=\sum_{n=1,3,\cdots}i\epsilon_{\mu\nu\rho\sigma}q^\rho\left({2
\over -q^2}\right)^n q_{\mu_1}\cdots q_{\mu_{n-1}}\sum_{i}2C_{i,n}O_{i,A}
^{\sigma\{\mu_1\cdots\mu_{n-1}\}},
\en
where the sum $\sum_i$ is over the leading-twist quark and gluon operators.
The twist-2 quark and gluon operators are given by
\be
O_{1,A}^{\sigma\{\mu_1\cdots \mu_{n-1}\}} &=& {1\over 2}\left({i\over 2}
\right)^{n-1}\overline{\psi}\gamma^\sigma D^{{\{\mu_1}\cdots} D^{\mu_{n-1}\}}
\gamma_5\psi,     \\
O_{2,A}^{\sigma\{\mu_1\cdots \mu_{n-1}\}} &=& {1\over 2}\left({i\over 2}
\right)^{n-1}{\rm Tr}\left(\epsilon^{\sigma\alpha\beta\gamma}G_{\beta\gamma} 
D^{{\{\mu_1}\cdots}D^{\mu_{n-2}}G_\alpha^{\mu_{n-1}\}}\right),
\en
with $G_{\mu\nu}$ a gluon field strength tensor, $D^\mu$ a covariant
derivative, and $\{\cdots\}$ a complete symmetrization of the enclosed 
indices. The corresponding Wilson
coefficients in (4.6) to the zeroth order of $\alpha_s$ are
\be
C_{1,n} &=& e_q^2,~~~~{\rm for~quark~operators~of~flavor}~q,   \non \\
C_{2,n} &=& 0,~~~~~{\rm for~gluon~operators}.
\en
It is useful to decompose the operator $O_A$ into a totally symmetric one 
and a one with mixed symmetry \cite{Jaffe}
\be
O_A^{\sigma\{\mu_1\cdots \mu_{n-1}\}}=\,O_A^{\{\sigma\mu_1\cdots \mu_{n-1}\}}
+O_A^{[\sigma\{\mu_1]\mu_2\cdots \mu_{n-1}\}},
\en
where $[\cdots]$ indicates antisymmetrization,
\be
O_A^{\{\sigma\mu_1\mu_2\cdots \mu_{n-1}\}}=\,{1\over n}\left[ O_A^{\sigma\{
\mu_1\cdots \mu_{n-1}\}}+O_A^{\mu_1\{\sigma\mu_2\cdots \mu_{n-1}\}}+O_A^{
\mu_2\{\mu_1\sigma\cdots \mu_{n-1}\}}+\cdots\,\right],
\en
for $n=1,3,5,\cdots$, is a twist-2 operator, and
\be
O_A^{[\sigma\{\mu_1]\mu_2\cdots \mu_{n-1}\}} &=& {1\over n}\Big[ O_A^{
\sigma\{\mu_1\cdots \mu_{n-1}\}}-O_A^{\mu_1\{\sigma\mu_2\cdots \mu_{n-1}\}}
\non \\
&+& O_A^{\sigma\{\mu_1\cdots \mu_{n-1}\}}-O_A^{\mu_2\{\mu_1\sigma\cdots 
\mu_{n-1}\}}+\cdots\,\Big],
\en
for $n=3,5,\cdots$, is a twist-3 operator. The proton matrix elements of
these two operators are
\be
\la p,s|O_{i,A}^{\{\sigma\mu_1\mu_2\cdots \mu_{n-1}\}}|p,s\ra &=& {a_{i,n}
\over n}(s^\sigma p^{\mu_1}\cdots p^{\mu_{n-1}}+s^{\mu_1}p^\sigma\cdots 
p^{\mu_{n-1}}+\cdots),  \non \\
\la p,s|O_{i,A}^{[\sigma\{\mu_1]\mu_2\cdots \mu_{n-1}\}}|p,s\ra &=& {d_{i,n}
\over n}[(s^\sigma p^{\mu_1}-s^{\mu_1}p^\sigma)p^{\mu_2}\cdots p^{\mu_{n-1}} 
 \\
&& +(s^\sigma p^{\mu_2}-s^{\mu_2}p^\sigma)p^{\mu_1}\cdots p^{\mu_{n-1}}+
\cdots\,],  \non 
\en
where $a_{i,n}$ and $d_{i,n}$ are unknown reduced matrix elements.

   Writing
\be
T_{[\mu\nu]}=\,i\tilde{g}_1\,{\epsilon_{\mu\nu\rho\sigma}q^\rho s^\sigma M
\over p\cdot q}+i\tilde{g}_2\,{\epsilon_{\mu\nu\rho\sigma}q^\rho(p\cdot q 
s^\sigma-s\cdot q p^\sigma)M\over (p\cdot q)^2}
\en
in analog to $W_{[\mu\nu]}$ [see Eq.(2.2)] and comparing with the proton 
matrix element of $t_{[\mu\nu]}$ [cf. Eq.(4.6)] gives
\be
\tilde{g}_1 &=& \sum_{n=1,3,5\cdots}\sum_i2C_{i,n}a_{i,n}\omega^n,   \non \\
\tilde{g}_2 &=& \sum_{n=1,3,5\cdots}\sum_i\left[\left({1-n\over n}\right)
2C_{i,n}a_{i,n}\omega^n+\left({n-1\over n}\right)2C_{i,n}d_{i,n}
\omega^n\right].
\en
It follows from (4.15) that
\be
\tilde{g}_2(\omega)=-\tilde{g}_1(\omega)+\int^\omega_0 {d\omega'\over \omega'}
\tilde{g}_1(\omega')+\sum_{n=1,3,5,\cdots}\left({n-1\over n}\right)
\sum_i2C_{i,n}d_{i,n}\omega^n.
\en
Using dispersion relations to relate $\tilde{g}_{1,2}$ in the unphysical
region ($\omega\sim 0$) to their values in the physical region 
$(1<\omega<\infty)$ finally yields the moment sum rules:
\be
\int^1_0dx\,x^{n-1}g_1(x) &=& {1\over 2}\sum_iC_{i,n}a_{i,n}\,,~~~
n=1,3,5,\cdots,    \\
\int^1_0dx\,x^{n-1}g_2(x) &=& -{1\over 2}\left({n-1\over n}\right)\sum_i
C_{i,n}(a_{i,n}-d_{i,n})\,,~~~n=3,5,\cdots,
\en
and the relation\footnote{It should be stressed that the relation
(4.19) is derived from (4.16) rather than from (4.17) and (4.18). It has
been strongly claimed in \cite{Ansel} that (4.19) is {\it a priori} not
reliable since its derivation is based on the dangerous assumption that
(4.17) and (4.18) are valid for {\it all} integer $n$. Obviously, this 
criticism is not applied to our case and (4.19) is valid as it stands.}
\be
g_2(x)=g_2^{\rm WW}(x)+\bar{g}_2(x),
\en
obtained from (4.16), where
\be
g_2^{\rm WW}(x)=-g_1(x)+\int^1_x{dy\over y}g_1(y)
\en
is a contribution to $g_2(x)$ fixed by $g_1(x)$, first
derived by Wandzura and Wilczek \cite{WW}, and $\bar{g}_2(x)$ is a truly
twist-3 contribution related to the twist-3 matrix elements $d_{i,n}$.
   
  For $n=1$, the moment sum rule (4.17) for $g_1$ is particularly simple: 
Gluons do not contribute to the first moment of $g_1^p$ as it is clear from 
(4.8) that there is no twist-2 gauge-invariant local gluonic operator for 
$n=1$, as stressed in \cite{JM}.
Since
\be
\la p,s|O_{1,A}^\mu|p,s\ra =\,a_{1,1}s^\mu
\en
from (4.13) and $C_{1,1}=e_q^2$ to the zeroth order of $\alpha_s$ [see 
(4.9)], it follows that
\be
\int^1_0 g_1(x)dx = {1\over 2}C_{1,1}a_{1,1}  
=-{1\over 2}\,\la p,s|\sum_q e_q^2\bar{q}
\gamma_\mu\gamma_5q|p,s\ra s^\mu.
\en
Denoting
\be
\la p,s|\bar{q}\gamma_\mu\gamma_5 q|p,s\ra =\,s_\mu\Delta q,
\en
(4.22) leads to the well-known naive parton-model result [cf. Eq.(2.16)]
\be
\int^1_0 g_1^p(x)dx=\,{1\over 2}\,({4\over 9}
\Delta u+{1\over 9}\Delta d+{1\over 9}\Delta s),
\en
which is rederived here from the OPE approach.

\subsection{Axial anomaly and sea-quark polarization}

     Contrary to the improved parton model discussed in Sec.~III, we see that
there is no any gluonic operator contributing to the
first moment of $g_1^p(x)$ according to the OPE analysis. The questions 
are then what is the deep 
reason for the absence of gluonic contributions to $\Gamma_1^p$ and how are
we going to understand a large and negative strange-quark polarization ?
The solution to these questions relies on the key observation that
the hard cross section $\gg_{\rm hard}(x)$ and hence the quark spin
density $\Delta q(x)$ are $k_\perp$-factorization scheme dependent. 
We have freedom to redefine $\gg_{\rm hard}(x)$ and $\Delta q(x)$ in 
accord with (3.15) but the physical cross section $\Delta\sigma^{\gamma p}(x)$
remains intact. Therefore, there must exist a factorization scheme that 
respects the OPE: Hard gluons make no contribution to $\Gamma_1^p$ and 
$\Delta q$ can be expressed as a nucleon matrix element of a local 
gauge-invariant operator. 
In this scheme, gluons can induce a sea polarization even for massless quarks. 
This can be implemented as follows. As discussed in the last section,
the quark spin density inside a gluon $\Delta q^G(x)$  can
be obtained by calculating the triangle diagram with an ultraviolet cutoff to
ensure that $k_\perp^2\lsim \u^2$. It is well known that in the presence of
the axial anomaly in the triangle diagram, gauge invariance and chiral 
symmetry
cannot coexist. So if the ultraviolet regulator respects gauge symmetry and
axial anomaly, chiral symmetry will be broken. As a consequence, 
quark-antiquark pairs created from the gluon via the gluonic anomaly can have 
the same helicities and give rise to a nonvanishing $\Delta q^G(x)$.
Since the axial anomaly resides at $k_\perp\to\infty$, evidently we
have to integrate over $k_\perp^2$ from 0 to $\infty$ to achieve the 
axial anomaly and hence chiral-symmetry breaking, and then
identify the ultraviolet cutoff with $\u$. We see that the desired
ultraviolet regulator must be gauge-invariant but chiral-variant owing
to the presence of the QCD anomaly in the triangle diagram. Obviously,
the dimensional and Pauli-Villars regularizations, which respect the axial 
anomaly, are suitable for our purposes.

\begin{figure}[ht]
\vspace{-5cm}

    \centerline{\psfig{figure=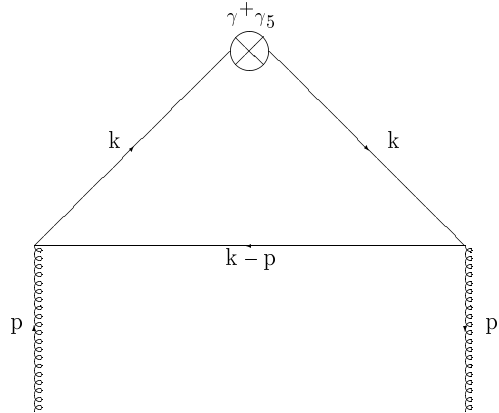,width=8cm}}
    \caption[]{\small The triangle graph for $j^\mu_5$ between 
               external gluons.}
    \label{fig3}

\end{figure}

The contribution of the triangle diagram Fig.~3 for a single quark flavor is 
\be
\Gamma^+_5 = 2ig^2T\int {d^{n-2}k_\perp dk^+dk^-
\over (2\pi)^4}\,
 { {\rm Tr}\{\ep\!\!\!/^* (k\!\!\!/+m)\gamma^+\gamma_5(k\!\!\!/+m)
\ep\!\!\!/[(p\!\!\!/ -k\!\!\!/)-m]\}\over (k^2-m^2+i\epsilon)^2[(p-k)^2-m^2
+i\epsilon]},
\en
where $T={1\over 2}$, $\ep^\mu=(0,0,1,\pm i)/\sqrt{2}$ is the transverse  
polarization of external gluons, the factor 2 comes from the fact the gluon 
in Fig.~3 can circulate in opposite direction, and the dimensional 
regularization is employed to regulate the ultraviolet divergence.
The quark spin density inside a gluon is then given by
\be
\Delta q^G_{\rm GI}(x)= {2ig^2T\over 2p^+}\int {d^{n-2}k_\perp dk^+dk^-
\over (2\pi)^4}\delta\left(x-{k^+\over p^+}\right)  
{ {\rm Tr}\{\cdots \}\over (k^2-m^2+i\epsilon)^2[(p-k)^2-m^2+i\epsilon]}.
\en
Note that $\int^1_0\Delta q^G_{\rm GI}(x)dx=\Gamma^+_5/(2 p^+)$ [cf. 
Eq.(3.33)]. We first perform the $k^-$ integral in (4.26) by 
noting that a pole of $k^-$ locating at
\be
k^-=\,p^--{k^2_\perp+m^2\over 2(p^+-k^+)}
\en
in the region $0<k^+<p^+$ contributes to contour integration.
The result for ``+" helicity external gluons is \cite{CCM}
\be
\Delta q^G_{\rm GI}(x) &=& -{\alpha_s\over 2\pi}\int{d^{n-2}k_\perp/\pi\over 
[\,k_\perp^2+m^2-p^2x(1-x)]^2} \non \\
&& \times\left[(k^2_\perp+m^2)(1-2x)-2m^2(1-x)-2\left(
{n-4\over n-2}\right)k_\perp^2(1-x)\right],
\en
where the subscript GI designates a gauge-invariant factorization scheme. The 
last term proportional to $(n-4)$ arises from the use of $\gamma_5$ 
in dimensional
regularization. The $\gamma_5$ matrix ($=i\gamma^0\gamma^1\gamma^2\gamma^3$)
anticommutes with the Dirac matrix in 4 dimensions but commutes with
the Dirac matrix in $n-4$ dimensions. This term originating from the axial 
anomaly thus survives at $k_\perp^2\to\infty$. By comparing (4.28) with 
(3.16),
it is clear that $\Delta q^G_{\rm GI}(x)$ has the same expression as that of
$\Delta q^G_{\rm CI}(x)$ except for the presence of an axial-anomaly term
in the former. It follows that \cite{Bass}
\be
\Delta q^G_{\rm GI}(x)-\Delta q^G_{\rm CI}(x) &=& -{\alpha_s\over 2\pi}\Bigg[
(2x-1)\ln\,{\u^2+m^2-p^2x(1-x)\over \u^2}  \non \\
&& +{2\u^2(1-x)\over \u^2+m^2-p^2x(1-x)}\Bigg]
\en
for mass and momentum cutoffs, and
\be
\Delta q^G_{\rm GI}(x)-\Delta q^G_{\rm CI}(x)=-{\alpha_s\over 2\pi}\left[
(2x-1)\ln\,{\u^2+\ums^2\over \u^2}+{2\u^2(1-x)\over \u^2+\ums^2}\right]
\en
for the dimensional infrared cutoff. Hence,
\be
\Delta q^G_{\rm GI}(x)-\Delta q^G_{\rm CI}(x)=-{\alpha_s\over \pi}(1-x)
\en
for $\u^2>>\ums^2,m^2,-p^2$. The difference between the quark spin densities
in gauge-invariant and chiral-invariant factorization schemes thus lies in
the gluonic anomaly arising at the region $k_\perp^2
\sim\u^2$. As noted in passing, the quark spin distribution in a gluon
cannot be reliably calculated by perturbative QCD; however, the difference
between $\Delta q^G_{\rm GI}(x)$ and $\Delta q^G_{\rm CI}(x)$ is trustworthy
in QCD. It is interesting to see from Eqs.~(4.31) and (3.19) that
\be
\Delta q^G_{\rm GI}(\u^2)=-{\alpha_s(\u^2)\over 2\pi}
\en
for massless quarks. Therefore, {\it the sea-quark polarization perturbatively 
generated by helicity $+$ hard gluons via the anomaly mechanism is 
negative} ! In other words, a polarized gluon is preferred to split into a 
quark-antiquark pair with helicities {\it antiparallel} to the gluon spin. 
As explained before, chiral-symmetry breaking induced by the gluonic anomaly
is responsible for the sea polarization produced perturbatively by
hard gluons.

    Since $\gg_{\rm hard}(x)=\gg(x)-\Delta q^G(x)$, it follows that the 
hard cross section has the form
\be
\gg_{\rm hard}(x,Q^2,\u^2)_{\rm GI} &=& \gg_{\rm hard}(x,Q^2,\u^2)_{\rm CI}+
{\alpha_s\over \pi}(1-x)   \non \\
&=& {\alpha_s\over 2\pi}\left[(2x-1)\left(\ln{Q^2
\over \u^2}+\ln{1-x\over x}-1\right)+2(1-x)\right].
\en
Hence,
\be
\int^1_0dx\gg_{\rm hard}(x,Q^2,\u^2)_{\rm GI}=0,
\en
and the gluonic contribution to $\Gamma_1^p$ vanishes. This is so
because {\it the axial anomaly characterized by the ${\alpha_s\over \pi}(
1-x)$ term is shifted from the hard cross section for photon-gluon
scattering in the chiral-invariant factorization scheme to the quark spin 
density in the gauge-invariant scheme.} It was first observed and 
strongly advocated by Bodwin and Qiu \cite{BQ} that the above conclusion is 
actually quite general: The hard gluonic contribution to the first moment 
of $g_1^p$ 
vanishes as long as the ultraviolet regulator for the spin-dependent 
quark distributions respects gauge invariance, Lorentz invariance, and 
the analytic structure of the unregulated distributions. Hence, the OPE 
result (4.24) for $\Gamma_1^p$ is a general consequence of the 
gauge-invariant factorization scheme.

  We wish to stress that the quark spin density $\Delta q^G(x)$ measures
the polarized sea-quark distribution in a helicity $+$ gluon rather than in a
polarized proton.
Consequently, $\Delta q^G(x)$ must convolute with $\Delta G(x)$ in order to be 
identified as the sea-quark spin distribution in a proton \cite{Cheng}:
\be
\Delta q^{\rm GI}_s(x,\u^2)-\Delta q^{\rm CI}_s(x,\u^2)=-{\alpha_s\over 
\pi}(1-x)\otimes\Delta G(x,\u^2).
\en
Since the valence quark spin distribution $\Delta q_v(x)=\Delta q(x)-
\Delta q_s(x)$ is $k_\perp$-factorization independent, it follows that 
\cite{ChengWai}
\be
\Delta q_{\rm GI}(x,\u^2)-\Delta q_{\rm CI}(x,\u^2)=-{\alpha_s\over 
\pi}(1-x)\otimes\Delta G(x,\u^2),
\en
which leads to
\be
\Delta q_{\rm GI}(Q^2)-\Delta q_{\rm CI}(Q^2)=-{1\over 2\pi}\,\alpha_s(Q^2)
\Delta G(Q^2),
\en
where we have set $\u^2=Q^2$.
Eqs.~(4.33) and (4.36) provide the necessary relations between the 
gauge-invariant and chiral-invariant factorization schemes. The reader may 
recognize that (4.37) is precisely the relation (3.30) 
obtained in the improved parton model.

\subsection{Sea-quark or anomalous gluon interpretation for $\Gamma_1$ ?}

    We have seen that there are two different popular explanations for
the data of $\Gamma_1$. In the sea-quark interpretation, the smallness
of the fraction of the proton spin carried by the quarks $\Delta\Sigma=
\Delta\Sigma_v+\Delta\Sigma_s\approx 0.30$ is ascribed to the negative sea 
polarization 
which partly compensates the valence-quark spin component. By contrast,
 a large and negative sea-quark polarization is not demanded in the
anomalous-gluon interpretation that the discrepancy between experiment
and the Ellis-Jaffe sum rule for $\Gamma_1^p$ is accounted for by 
anomalous gluon contributions. The issue of the contradicting
statements about the gluonic contributions to the first moment of $g_1(x)$
between the improved parton model and the OPE analysis has been under hot
debate over the past years. Naturally we would like to ask : Are these two
seemingly different explanations equivalent ? If not, then which scheme
is more justified and sounding ?

   In spite of much controversy on the aforementioned issue, this dispute
was actually resolved several years ago \cite {BQ}. The key point is that
a different interpretation for $\Gamma^p_1$ corresponds to a
different $k_\perp$-factorization definition for the quark spin density and
the hard photon-gluon cross section. The choice of the ``ultraviolet" cutoff 
for soft contributions specifies the factorization
convention. It is clear from (3.1), (4.33) and (4.36) that to NLO
\be
g_1(x,Q^2) &=& {1\over 2}\sum_q e_q^2\left[ \Delta q_{\rm GI}(x,Q^2)+{\alpha_s
\over 2\pi}\Delta f_q(x)\otimes\Delta q_{\rm GI}(x,Q^2)+\gg_{\rm hard}(x)_{\rm 
GI}\otimes\Delta G(x,Q^2)\right]   \non \\
&=& {1\over 2}\sum_q e_q^2\left[ \Delta q_{\rm CI}(x,Q^2)+{\alpha_s\over
2\pi}\Delta f_q(x)\otimes\Delta q_{\rm CI}(x,Q^2)+\gg_{\rm hard}(x)_{\rm CI}
\otimes\Delta G(x,Q^2)\right],  \non \\
&&
\en
where we have set $\u^2=Q^2$ so that the $\ln Q^2/\u^2$ terms in $\Delta 
f_q(x)$ and $\Delta\sigma^{\gamma G}_{\rm hard}(x)$ vanish. As will be
discussed in Sec.~7.1, the $Q^2$ evolution of $g_1(x,Q^2)$ in (4.38)
is governed by the parton spin distributions.
Therefore, the polarized structure function $g_1(x)$ is shown to be 
independent of the choice of the factorization convention up to the 
next-to-leading order of $\alpha_s$, as it should be. This is so because
a change of the factorization scheme merely shifts the axial-anomaly 
contribution between $\Delta q(x)$ and
$\gg(x)$ in such a way that the physical proton-gluon cross section remains
unchanged [cf. Eq.(3.15)]. It follows from (4.38) that
\be
\int^1_0 g_1(x,Q^2)dx &=& {1\over 2}\left(1-{\alpha_s\over \pi}\right)
\sum_q\Delta q_{\rm GI}(Q^2)   \non \\
&=& {1\over 2}\left(1-{\alpha_s\over \pi}\right)\sum_q\left(\Delta q_{\rm CI}
(Q^2)-{\alpha_s(Q^2)\over 2\pi}\Delta G(Q^2)\right).
\en
Hence, {\it the size of the hard-gluonic contribution to 
$\Gamma_1$ is purely a matter of the factorization convention chosen in 
defining $\Delta q(x)$ and $\gg(x)$}. This important observation
on the $k_\perp$-factorization dependence of the anomalous gluonic 
contribution to the first moment of $g_1(x)$ was first made by Bodwin and Qiu 
\cite{BQ} (see also Manohar \cite{Man2}, Carlitz and Manohar \cite{Man3},
Bass and Thomas \cite{BassT}, Steffens and Thomas \cite{Ste}).

 Thus far we have only considered two extremes of the $k_\perp$-factorization 
schemes: the
chiral-invariant scheme in which the ultraviolet regulator respects chiral
symmetry, and the gauge-invariant scheme in which gauge symmetry is respected
but chiral symmetry is broken by the cutoff.
Nevertheless, it is also possible to choose an intermediate
factorization scheme which is neither gauge nor chiral invariant, so
in general $\Delta q_{\rm GI}=\,\Delta q'-\lambda{\alpha_s\over 2\pi}\Delta G$ 
for an arbitrary $\lambda$ ($\lambda=0$ and $\lambda=1$ corresponding to 
gauge- and chiral-invariant schemes, respectively) \cite{Man4}. Experimentally
measured quantities do not depend on the value of $\lambda$.

   Although the issue of whether or not gluons contribute to $\Gamma_1$ was
resolved six years ago \cite{BQ,Man2},
the fact that the interpretation of $\Gamma_1$ is still under
dispute even today and that some recent articles and reviews are still biased
towards or against one of the two popular implications of the measured
$\Gamma_1$ is considerably unfortunate and annoying. As mentioned in 
Sec.~4.1, the anomalous gluon interpretation has been deemed to be 
plausible and more favored than the sea-quark one by many 
practitioners in the field over the past years. However, these two 
explanations are on the same footing and all the known criticisms to the 
gauge-invariant factorization scheme and the sea-quark interpretation 
of $\Gamma_1$ are in vain. Here we name a few:
\begin{itemize}
\item It has been often claimed \cite{Ball,Vog1,Mank}
that soft contributions are partly 
included in $\gg_{\rm hard}(x)_{\rm GI}$ rather than being factorized 
into parton spin densities because, apart from the soft-cutoff term, $\gg
_{\rm hard}(x)_{\rm GI}$ [see Eq.(4.33)] has exactly the same expression
as (3.11) or (3.12). Therefore, the last term proportional to $2(1-x)$
arises from the soft region $k_\perp^2\sim m^2<<\Lambda_{\rm QCD}$, and
hence it should be absorbed into the polarized quark distribution. 
This makes the gauge-invariant scheme pathological 
and inappropriate. However, this argument is fallacious.
It is true that the $2(1-x)$ term in (3.11) or (3.12) 
drops out in $\gg_{\rm hard}(x)_{\rm CI}$ because it stems from the
soft $k_\perp^2$ region, but it emerges again in the gauge-invariant 
scheme due to the axial anomaly being subtracted from $\gg_{\rm 
hard}(x)_{\rm CI}$ [see Eqs.(4.31-4.33)] and this time reappears in the hard 
region $k^2_\perp\sim \u^2$. As a result, the hard photon-gluon cross section
given by (4.33) is genuinely {\it hard} !
\item A sea-quark interpretation of $\Gamma_1^p$ with $\Delta s=-0.10$
at $Q^2=10\,{\rm GeV}^2$ has been criticized on the ground that a 
bound $|\Delta s|\leq 0.052^{+0.023}_{-0.052}$ \cite{Pre} can be 
derived based on the information of the behavior of $s(x)$ measured in 
deep inelastic neutrino experiments and on the positivity constraint that
$|\Delta s(x)|\leq s(x)$.
However, this claim is quite controversial \cite{Soffer} and not
trustworthy. Indeed, one can always find a polarized strange
quark distribution with $\Delta s\sim -0.10$ which satisfies positivity 
and experimental constraints \cite{CLW}. Moreover, a sea polarization of 
this order is also confirmed by lattice calculations \cite{Dong,Fuk}.
\end{itemize}

    By now, we wish to have convinced the reader that it does not make sense 
to keep disputing which factorization prescription is correct or 
which interpretation is superior as they are equivalent.
Once a set of $\Delta q_{\rm GI}(x),~\Delta G(x),~\gg
_{\rm hard}(x)_{\rm GI}$ or of $\Delta q_{\rm CI}(x),~\Delta G(x),~\gg_{\rm 
hard}(x)_{\rm CI}$ is chosen, one has to stick to the same scheme in all
processes. 

   It is worth emphasizing at this point that the equivalence of the 
sea-quark and anomalous-gluon interpretations is only applied to the
first moment of $g_1(x)$, but not to $g_1(x)$ itself. Suppose at a certain
$Q^2=Q_0^2$, the data of $g_1(x)$ are reproduced either by assuming 
$\Delta q_s(x)\neq 0$ but $\Delta G(x)=0$ in the sense of the sea-quark
interpretation, or by having $\Delta G(x)\neq 0$ but $\Delta q_s(x)=0$
in the sense of the anomalous gluon interpretation. It is clear that these two
explanations are no longer equivalent at $Q^2>Q^2_0$ as $\Delta
q_s(x,Q^2)$ and $\Delta G(x,Q^2)$ evolve differently. An equivalence
of the first moment of $g_1(x)$ does not imply the same results for the
higher moments of $g_1(x)$.
From (4.38) it is evident that {\it in spite of a vanishing gluonic
contribution to $\Gamma_1$ in the gauge-invariant scheme,
it by no means implies that $\Delta G$ vanishes in a polarized proton.}

   So far we have focused on the perturbative part of the axial anomaly.
The perturbative QCD results (4.35)-(4.37) indicate that the difference 
$\Delta q_s^{\rm GI}-\Delta q_s^{\rm CI}$ is induced perturbatively 
from hard gluons via the anomaly mechanism and its sign is predicted to 
be negative. By contrast,
$\Delta q_s^{\rm CI}(x)$ can be regarded as an intrinsic sea-quark spin 
density produced nonperturbatively.
As we have emphasized in passing (see Sec.~3.1), the sea-quark helicity
$\Delta q^{\rm CI}_s$ for massless quarks cannot be generated 
perturbatively from hard gluons due to helicity conservation. The question 
is what is the underlying mechanism
for producing an intrinsic negative helicity for sea quarks ? Does it
have something to do with the nonperturbative aspect of the axial anomaly ?
The well-known solution to the $U_A(1)$ problem in QCD involves two
important ingredients: the QCD anomaly and the QCD vacuum with a 
nontrivial topological structure, namely the $\theta$-vacuum constructed 
from instantons which are nonperturbative gluon configurations. Since
the instanton-induced interactions can flip quark helicity, in analog
to the baryon-number nonconservation induced by the 't Hooft mechanism,
the quark-antiquark pair created from the QCD vacuum via instantons 
can have a net helicity.
It has been suggested that this mechanism of quark helicity 
nonconservation provides a natural and nonperturbative way of generating
negative sea-quark polarization \cite{For,Dor,Dor93}.

 There are two extreme cases for the sea-quark spin component: In one case,
$\Delta q_s^{\rm CI}(Q^2)=0$ so that $\Delta q_s^{\rm GI}$ arises exclusively
from the perturbative anomaly mechanism. As a result, $\Delta G(Q^2)$
is equal to $-(2\pi/\alpha_s)
\Delta q_s^{\rm GI}(Q^2)$ [cf. Eq.(4.37)] and is of order 2.5 at 
$Q^2=10\,{\rm GeV}^2$. In the other extreme case,
$\Delta q_s^{\rm GI}(Q^2)=\Delta q_s^{\rm CI}(Q^2)$ so that
the sea-quark polarization is exclusively of 
nonperturbative nature and $\Delta G=0$, as advocated, for example, in 
the chiral soliton model \cite{Brod88,Ellis88}. The realistic case 
should be somewhere between these two extreme cases.

    In short, the sea-quark polarization $\Delta q_s^{\rm GI}$
consists of two components: the intrinsic nonperturbative part 
$\Delta q_s^{\rm CI}$ induced from the QCD vacuum
via instantons and the perturbative part i.e., $\Delta q_s^{\rm GI}-
\Delta q_s^{\rm CI}$ generated from the anomaly mechanism.
The lattice calculation (see Sec.~6.1) indicates that the sea
polarization is almost independent of light quark flavors and this 
suggests that it is indeed the perturbative and nonperturbative 
parts of the gluonic anomaly that account for the bulk of the negative
spin component of sea quarks.

\subsection{Operator definitions for $\Delta q$ and $\Delta G$}

    The quark spin component in the nucleon can be expressed as a matrix
element of a local and gauge-invariant operator in the gauge-invariant
$k_\perp$-factorization scheme. Since in the parton model $\Delta q$ 
given by (2.15) is defined
in the infinite momentum frame, we first consider such a
frame where the nucleon is moving in the $z$ direction with momentum
$p^3\equiv p_\infty\to\infty$ and helicity $+{1\over 2}$, so that
\be
\la p_\infty,\up|\bar{q}\gamma^3\gamma_5 q|p_\infty,\up\ra =\Delta q_{\rm GI}.
\en
This is equivalent to working in the light-front coordinate in the laboratory 
frame
\be
\la p,s|\bar{q}\gamma^+\gamma_5 q|p,s\ra =\,s^+\Delta q_{\rm GI},
\en
where ``$+$" is a good component in the light-front quantization formulation.
It should be stressed that $\Delta q$ is not equal to the net spin vector
sum $\int d^3p[q^\up(p)-q^\downarrow(p)]$ in the proton rest frame in the
equal-time quantization formulation, where $q^{\up(\downarrow)}(p)$ is the
probability of finding a quark flavor $q$ in the proton rest frame with 
momentum $p_\mu$ and spin parallel (antiparallel) to the proton spin 
\cite{Ma}. Technically, the helicity and spin components of the proton are
related to each other via the so-called Melosh transformation.
The quark spin $\Delta q_{\rm GI}$ is gauge invariant but 
it evolves with $Q^2$ since the flavor-singlet
axial-vector current $J^\mu_5=\sum_q\bar{q}\gamma^\mu\gamma_5q$ has an
anomalous dimension first appearing at the two-loop level \cite{Kod}. The
$Q^2$ dependence of $\Delta\Sigma_{\rm GI}(Q^2)$ will be discussed in 
Sec.~4.6. The evaluation of the nucleon matrix element of $J_\mu^5$ involves 
connected and disconnected insertions (see Fig.~4), which
are related to valence quark and vacuum (i.e.,
sea quark) polarizations, respectively, and are separately gauge invariant.
Thus we can make the identification:
\be
\la p,s|J_5^+ |p,s\ra=\la p,s|J_5^+|p,s\ra_{\rm con}+\la p,s|J_5^+|p,s
\ra_{\rm dis}=\sum_q(\Delta q^{\rm GI}_v+\Delta q_s^{\rm GI})s^+.
\en
Interestingly, lattice QCD calculations of $\Delta q^{\rm GI}_v$ and $\Delta 
q^{\rm GI}_s$ became available very recently \cite{Dong,Fuk}. It is found that
$\Delta u_s=\Delta d_s=\Delta s=-0.12\pm 0.01$ from the disconnected
insertion \cite{Dong}. This empirical SU(3)-flavor symmetry  
implies that the sea-quark polarization in the gauge-invariant scheme 
is indeed predominately generated by the axial anomaly. Recall that 
sea contributions in the unpolarized case are far from being SU(3) symmetric: 
$\bar{d}>\bar{u}>\bar{s}$.

\begin{figure}[ht]
\vspace{+1cm}

    \centerline{\psfig{figure=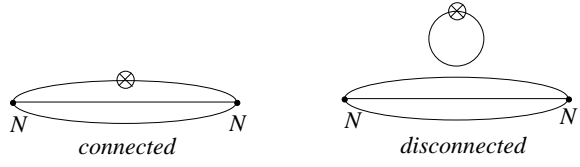,width=7.5cm}}
    \caption[]{\small Connected and disconnected insertions arising from the 
    flavor-singlet axial-vector current.}
    \label{fig4} 

\end{figure}

   In the chiral-invariant factorization scheme one is expected to have
\be
\la p,s|J_5^+|p,s\ra =\sum_q\left(\Delta q_{\rm CI}-{\alpha_s\over 2\pi}
\Delta G\right)s^+
\en
by virtue of (4.37). The question is that can one
define $\Delta q_{\rm CI}$ and $\Delta G$ separately in terms of a 
gauge-invariant local operator ? For this purpose we write
\be
J^\mu_5=\,J^\mu_5-n_fK^\mu+n_fK^\mu\equiv\,\tilde{J}^\mu_5+n_fK^\mu,
\en
with the Chern-Simons current 
\be
K^\mu={\alpha_s\over 2\pi}\,\epsilon^{\mu\nu\rho\sigma}A^a_\nu(\partial
_\rho A^a_\sigma-{1\over 3}gf_{abc}A_\rho^bA_\sigma^c)
\en
and $\epsilon_{0123}=1$. Since $K^\mu$ is made of gluon fields only and 
$\tilde{J}^\mu_5$ is conserved in the chiral limit so that $\la G|\tilde{J}
^\mu_5|G\ra=0$, it is tempted to make the identification:
\be
\la p,s|\tilde{J}^+_5|p,s\ra &=& \Delta\Sigma_{\rm CI}\,s^+, \non \\
\la p,s|K^+|p,s\ra &=& -{\alpha_s\over 2\pi}\Delta G\,s^+.
\en
It was originally claimed in \cite{AR} that although the topological operator
$K^\mu$ is gauge variant, its diagonal matrix element is nevertheless
gauge invariant. The argument goes as follows. Consider the matrix element
of $K^\mu$:
\be
\la N(p')|K^\mu|N(p)\ra &=& \bar{u}(p')[G_1(q^2)\gamma^\mu\gamma_5+G_2(q^2)
q^\mu\gamma_5]u(p).
\en
Since $\partial_\mu K^\mu=(\alpha_s /2\pi)G\widetilde{G}$ is gauge 
invariant, so is the expression $2m_NG_1(q^2)+q^2G_2(q^2)$. Consequently,
the absence of a Goldstone pole coupled to $J^\mu_5$ implies that $G_1(q^2
=0)$ and hence the matrix element of $K^\mu$ in the forward direction
becomes gauge invariant. Another argument relies on the observation
that under the gauge transformation, $K^\mu\to K^\mu+\epsilon^{\mu\nu\alpha
\beta}\partial_\nu(\cdots)$.
So the gauge-dependent term can be expressed as a four-derivative and thus
does not contribute to the diagonal matrix element of $K^\mu$. However,
both above-mentioned arguments are erroneous for the reason of the QCD U(1)
problem. In order to solve the U(1) problem, the SU(3)-singlet $\eta_0$
field must acquire a mass even in the chiral limit (see e.g., \cite{Chr}):
\be
m_{\eta_0}^2\sim -\lim_{q\to 0}{iq^\mu q^\nu\over f_{\eta_0}^2}\int d^4x\,
e^{iqx}\la 0|TK_\mu(x) K_\nu(0)|0\ra.
\en
This demands a ghost pole coupled to $K^\mu$. Hence, $G_2(q^2)q^2$ does not
vanish in the limit $q^2\to 0$. Also, under the ``large" gauge 
transformation,
\be
K^\mu\to K^\mu+\epsilon^{\mu\nu\alpha\beta}
\partial_\nu(\cdots)+{\alpha_s\over 2\pi}\epsilon^{\mu\nu
\alpha\beta}{\rm Tr}[(U^\dagger\partial_\nu U)(U^\dagger\partial_\alpha U)
(U^\dagger\partial_\beta U)].
\en
It is generally believed that a solution to the U(1) problem needs two
crucial ingredients: the axial anomaly and the instanton. The gauge
transformation $U(x)$ must be nontrivial if the instanton or the
topological structure of the vacuum exists. It follows from (4.49)
that the forward matrix element of $K^\mu$ is not gauge invariant under
the ``large" gauge transformation. (For an explicit example in the framework
of the Schwinger model, see \cite{Man5}.)

    Since the twist-2, spin-2 gluonic operator $G\widetilde{G}~(=G_{\mu\nu}^a
\widetilde{G}^{a\mu\nu})$ is gauge invariant, it has been proposed 
\cite{ChengLi} to utilize the divergence equation
\be
\partial_\mu J^\mu_5=\sum_q2m_q\bar{q}i\gamma_5 q+{\alpha_s n_f\over 4\pi}G
\widetilde{G}
\en
to define gauge-invariant quark and gluon spin components:
\be
\Delta\Sigma_{\rm CI}=\,{1\over 2m_N}\la p,s|\sum_q 2m_q\bar{q}i\gamma_5 
q|p,s\ra,~~~~\Delta G=-{1\over 2m_N}\la p,s|{1\over 2}G\widetilde{G}|p,s\ra.
\en
However, this local operator definition immediately encounters several 
insurmountable difficulties: for example, (i) the total light quark spin 
in a nucleon vanishes in the zero light quark mass limit, and (ii) 
$\Delta G$ and $\Delta s$ thus defined exhibit a large isospin violation, 
namely the gluon and sea-quark spin contents of the neutron are different
from that of the proton:
$\Delta s_n\neq \Delta s_p$ and $\Delta G_n\neq \Delta G_p$ (an explicit 
calculation shows $\Delta G_p<0$, while $\Delta G_n>0$) \cite{Hats}.
We conclude that there is no (spin-1 or spin-2) twist-2 gauge-invariant 
local operator definition for $\Delta G$ and $\Delta q_{\rm CI}$. This 
is consistent with the OPE statement that there is only one twist-2 
local gauge-invariant operator contributing to $\Gamma_1$.

    It turns out that although $K_\mu$ is not gauge invariant, its nucleon
matrix element can be related to $\Delta G$ defined below in (4.56) by 
choosing a specific gauge and coordinate. Spin and orbital angular momenta 
in QCD are governed by a rank-3 tensor $M^{\mu\nu\lambda}$ \cite{JM}:
\be
M^{\mu\nu\lambda} &=& {i\over 2}\overline{\psi}\gamma^\mu(x^\lambda\partial
^\nu-x^\nu\partial^\lambda)\psi+{1\over 2}\epsilon^{\mu\nu\lambda\rho}
\overline{\psi}\gamma_\rho\gamma_5\psi -G^{\mu\alpha}(x^\nu\partial^\lambda
-x^\lambda\partial^\nu)A_\alpha  \non \\
&+& (G^{\mu\lambda}A^\nu+G^{\nu\mu}A^\lambda)-{1\over 4} G^2(x^\nu g^{\mu
\lambda}-x^\lambda g^{\mu\nu}),
\en
where the color indices are implicit. The fourth term in (4.52) 
is relevant to the gluon spin and the generator of gluon spin rotations 
has the form\footnote{Note that the generators of gluon spin and orbital
rotations corresponding to the respective fourth and third terms in (4.52), 
were originally incorrectly identified in \cite{JM} with $\vec{A}\times
\vec{E}$ and $E^l(\vec{x}\times \vec{\nabla})A^l$, respectively. However,
the gluon's total angular momentum operator $\vec{J}_G=\vec{x}\times(\vec{E}
\times\vec{B})$ given in \cite{JM} is correct.}
\be
M^{0ij}_G({\rm spin})=\left(\vec{E}\times\vec{A}\right)^k.
\en
However, the gluon spin and orbital terms in $M^{\mu\nu\lambda}_G$ are
separately gauge variant and hence a choice of gauge fixing is necessary.
It appears that in the infinite momentum frame and in the temporal axial gauge
$A^0=0$, the operator $\vec{E}\times\vec{A}$ measures the gluon spin,
that is \cite{JM}
\be
\la p_\infty,\up|\left(\vec{E}\times\vec{A}\right)^3|p_\infty,\up\ra_{A^0=0}
=\Delta G.
\en
It is easy to check that the Chern-Simons current $K^3$ in temporal axial 
gauge is proportional to $\left(\vec{E}\times\vec{A}\right)^3$. We could also 
define the same $\Delta G$ in the laboratory frame using the 
light-front coordinate to obtain \cite{Jaffe96}
\be
\la p,s|M^{+12}_G({\rm spin})|p,s\ra_{A^+=0}=\la p,s|\left(\vec{E}\times
\vec{A}\right)^3+\vec{A}_\perp\cdot\vec{B}_\perp|p,s\ra_{A^+=0}=s^+\Delta G,
\en
with $B_i={1\over 2}\epsilon_{ijk}G^{jk}$, by noting that the gauge 
condition $A^0=0$ in the infinite momentum frame is modified to the 
light-front gauge $A^+=0$ in the light-front coordinate. Therefore, 
in light-front gauge \cite{CCM,Man6,Jaffe96}
\be
\la p,s|K^+|p,s\ra_{A^+=0}=-{\alpha_s\over 2\pi}\Delta G\,s^+.
\en
This is the local operator definition for the gluon spin component. 
Consequently, we also have
\be
\la p,s|\tilde{J}^+_5|p,s\ra_{A^+=0}=\,\Delta \Sigma_{\rm CI}\,s^+.
\en
We see that (4.46) is valid in the light-front coordinate and in light-front
gauge.

   The gluon spin $\Delta G$ (and likewise for $\Delta \Sigma_{\rm CI}$) also
can be recast as a nucleon matrix element of a string-like {\it nonlocal
gauge-invariant} operator \cite{Bali}. Of course, this nonlocal operator
will be reduced to the local operator $K^+$ or $M^{+12}_G({\rm spin})$ in 
light-front gauge. Moreover, it is also possible to have operator
representations for $\Delta G(x)$ and $\Delta q(x)$. The interested reader
is referred to \cite{Collins,Man6,BQ}.

    From (4.57) it is clear that $\Delta\Sigma_{\rm CI}$ does not evolve as 
the current $\tilde{J}^\mu_5$ is conserved in the chiral limit. In the
improved parton-model picture discussed in Sec.~III, this is so because 
the ultraviolet cutoff for
$\Delta q_{\rm CI}(x)$ is chiral invariant. Hence it is consistent with
the naive intuition that the quark spin is not affected by gluon emission.
Applying (4.56) and (4.57) to the axial-current matrix element leads to
\be
\la p,s|J_5^+ |p,s\ra &=& \la p,s|J_5^+|p,s\ra_{\rm con}+\la p,s|\tilde{J}_5
^+|p,s\ra_{\rm dis}+\la p,s|n_fK^+|p,s\ra_{\rm dis}   \non \\
& \buildrel A^+=0 \over\longrightarrow & \sum_q(\Delta q^{\rm CI}_v+
\Delta q_s^{\rm CI}-{\alpha_s\over 2\pi}\Delta G)s^+,
\en
where use of $\Delta q^{\rm GI}_v=\Delta q^{\rm CI}_v$ has been made. This
is in agreement with (4.43), as it should be.

\subsection{Anomalous dimensions of $\Delta\Sigma$ and $\Delta\Gamma$}
   It is pointed out in Sec.~3 that in the improved parton model there
is an anomalous gluonic contribution to the first moment of $g_1(x)$ even 
in the asymptotic limit. This can be seen by solving the spin-dependent
Altarelli-Parisi equation (3.24). However, it can be also understood in 
the OPE by considering the anomalous dimension of the Chern-Simons
current $K^\mu$. The QCD evolution equation for $J^\mu_5$ and $K^\mu$ is
given by
\be
{d\over dt}\left(\matrix{ J^\mu_5  \cr K^\mu}\right)=\,{\alpha_s(t)\over 2\pi}
\left( \matrix{ \gamma_
{11} & \gamma_{12}   \cr   \gamma_{21} & \gamma_{22} }\right)\left(\matrix{ 
J^\mu_5  \cr K^\mu}\right),
\en
where $t=\ln Q^2/\Lambda^2_{\rm QCD}$, and $\gamma_{ij}$ are anomalous 
dimensions:
\be
\gamma=\,\gamma^{(0)}+{\alpha_s\over 2\pi}\gamma^{(1)}+\cdots.
\en
Obviously, $\gamma_{12}=0$ due to the absence of $J^\mu_5$ and $K^\mu$ mixing
(the latter being gauge variant). Also, $\gamma_{22}=0$ because $\partial_\mu
K^\mu\sim G\widetilde{G}$ and $G\widetilde{G}$ does not get renormalized. 
Moreover, 
the fact that the Adler-Bardeen relation $\partial_\mu J^\mu_5=n_f\partial_\mu
K^\mu$ must be true at any renormalization scale $\mu$ implies that $\gamma
_{11}=n_f\gamma_{21}$. Next consider the evolution equation ${d\over dt}K^\mu
={\alpha_s\over 2\pi}\gamma_{21} J^\mu_5$ and take quark matrix elements. 
Since $K^\mu$ is 
of order $\alpha_s$, it is evident that $\gamma_{21}$ is also of order 
$\alpha_s$. As a result, (4.59) reduces to
\be
{d\over dt}\left(\matrix{ J^\mu_5  \cr K^\mu}\right)=\left({\alpha_s(t)\over 
2\pi}\right)^2\left( \matrix{ n_f\gamma^{(1)}
& 0 \cr  \gamma^{(1)} & 0 }\right)\left(\matrix{ 
J^\mu_5  \cr K^\mu}\right).
\en
Therefore, the anomalous dimension of $J^\mu_5$ starts at the 2-loop level.
The observation in Sec.~3.2 that $\alpha_s\Delta G$ is conserved to the
leading-order QCD evolution is now ascribed to the fact that the anomalous
dimension of $K^\mu$ starts at the order of $\alpha_s^2$ and that
$\alpha_s\Delta G$ has the same anomalous dimension as that of $K^\mu$ since
it is related to the nucleon matrix element of $K^+$ via (4.56).

   Now $\gamma^{(1)}$ can be calculated at the 2-loop level (i.e., 
$\gamma_{11}$) with $J^\mu_5$ or at the 1-loop level (i.e., $\gamma_{21}$)
with $K^\mu$. A direct calculation of $\gamma_{11}$ by Kodaira {\it et al.}
\cite{Kod}
gives $\gamma_{11}^{(1)}=-2n_f$, while $\gamma_{21}^{(1)}$ is computed
in \cite{Kap} to be $-2$. Hence the relation $\gamma_{11}=n_f\gamma_{21}$
is indeed obeyed. A solution of the renormalization group equation
\be
\left(-{\partial\over \partial t}+\beta{\partial\over \partial g}
+\gamma\right)J^\mu_5(t,g)=0
\en
yields 
\be
J^\mu_5(t)=\exp\left(\int{\gamma(\bar{g})\over\beta(\bar{g})} 
d\bar{g}\right)J^\mu_5(0)=\left(1+{\alpha_s(0)-\alpha_s(t)\over 2\pi\beta_0}\,
\gamma^{(1)}\right)J^\mu_5(0)
\en
with $\beta_0=(33-2n_f)/3$ and $\alpha_s(Q^2)=4\pi/(\beta_0
\ln{Q^2/\Lambda^2_{\rm QCD}})$. Hence, the total quark spin $\Delta\Sigma_{\rm 
GI}$ defined in the gauge-invariant $k_\perp$-factorization scheme begins
evolution with $Q^2$ at order $\alpha_s^2$. Since the anomalous dimension
$\gamma$ is negative, $\Delta\Sigma_{\rm GI}(Q^2)$ decreases with $Q^2$.

   From various operator definitions for $\Delta\Sigma_{\rm GI},~\Delta
\Sigma_{\rm CI}$ and $\Delta\Gamma\equiv(\alpha_s/2\pi)\Delta G$ given in 
Sec.~4.4, it is easily shown from (4.61) that (see also \cite{AL})
\be
{d\over dt}\left(\matrix{ \Delta\Sigma_{\rm GI} \cr \Delta\Gamma}\right)=
\left({\alpha_s\over 2\pi}\right)^2\left( \matrix{ -2n_f & 0 \cr  2 & 0 }
\right)\left(\matrix{ \Delta\Sigma_{\rm GI}  \cr \Delta\Gamma }\right)
\en
in the gauge-invariant scheme, and
\be
{d\over dt}\left(\matrix{ \Delta\Sigma_{\rm CI} \cr \Delta\Gamma}\right)=
\left({\alpha_s\over 2\pi}\right)^2\left( \matrix{ 0 & 0 \cr  2 & -2n_f }
\right)\left(\matrix{ \Delta\Sigma_{\rm CI}  \cr \Delta\Gamma }\right)
\en
in the chiral-invariant scheme. It is evident that $\Delta\Sigma_{\rm CI}$
is conserved. For parton spin densities $\Delta q(x,Q^2)$ and 
$\Delta G(x,Q^2)$, the anomalous dimensions are related to 
spin-dependent splitting functions, which we will discuss in Sec.~6.2.

\subsection{A brief summary}
   It is useful to summarize what we have learned from Secs.~3 and 4.
Depending on how we factorize the photon-gluon cross section into hard and
soft parts and how we specify the ultraviolet cutoff on the spin-dependent
quark distributions, we have considered two extremes of
$k_\perp$-factorization schemes.

   In the chiral-invariant factorization scheme, the ultraviolet regulator
respects chiral symmetry and gauge invariance but not the axial anomaly.
Consequently, $\Delta q_{\rm CI}$ does not evolve with $Q^2$ and is close
to the conventional parton-model intuition. There is an anomalous gluonic
contribution to the first moment of $g_1(x)$ due to the gluonic 
anomaly resided in the box diagram of photon-gluon scattering at 
$k_\perp^2=[(1-x)/4x]Q^2$ with $x\to 0$. Although $\Delta q_{\rm CI}$ 
cannot be written as a nucleon matrix element of a local gauge-invariant
operator, a gauge-variant local operator definition for $\Delta q_{\rm CI}$
does exist [cf. (4.57)] in the light-front coordinate and in the light-front
gauge $A^+=0$ (or in the infinite momentum frame and in temporal axial gauge). 
Since sea polarization cannot be perturbatively produced from hard gluons
due to helicity conservation, it is expected to be small. In the
extreme case that $\Delta s_{\rm CI}=0$, $\Delta G$ is of order 2.5 at 
$Q^2=10\,{\rm GeV}^2$, and it leads to the so-called anomalous gluon 
interpretation of $\Gamma_1$.

   Contrary to the above scheme, the ultraviolet cutoff in the 
gauge-invariant scheme satisfies gauge symmetry and the axial anomaly but
breaks chiral symmetry. As a result, $\Delta q_{\rm GI}$ is gauge 
invariant but $Q^2$ dependent. Hard gluons do not contribute to 
$\Gamma_1$ because the axial anomaly is shifted from the hard photon-gluon 
cross section to the spin-dependent quark distribution. Of course, this does
not imply a vanishing $\Delta G$. By contrast, the gluon spin component 
could be large enough to perturbatively generate a sizeable negative sea
polarization via the anomaly mechanism. Indeed, $\Delta G$ is 
$k_\perp$-factorization independent, and it does not make sense to say that 
$\Delta G$ is small in one scheme and large in the other scheme. For
a given $\Delta G(x)$, $\Delta q_{\rm GI}$ and $\Delta q_{\rm CI}$ are
related via (4.36), which is a rigorous consequence of perturbative QCD.
We have explicitly shown that $g_1(x)$ (not just $\Gamma_1$) is
independent of the factorization prescription up to NLO.

  In order to produce sea-quark polarization for massless quarks,
there are two mechanisms allowing for chiral-symmetry breaking and 
quark helicity flip: the nonperturbative way via 
instanton-induced interactions, and
the perturbative way through the anomaly mechanism. The empirical
lattice observation of SU(3)-flavor symmetry for spin components of 
sea quarks (Sec.~6.1) suggests that it is indeed the perturbative and 
nonperturbative parts of the axial anomaly, which are 
independent of light quark masses, that account for the bulk of 
sea polarization.

   Although the choice of $\Delta q_{\rm GI}(x),~\gg
_{\rm hard}(x)_{\rm GI}$ or $\Delta q_{\rm CI}(x),~\gg_{\rm 
hard}(x)_{\rm CI}$ is on the same footing, in practice it appears that 
the use of $\Delta q_{\rm GI}(x)$ is more
convenient than $\Delta q_{\rm CI}(x)$. First of all, $\Delta q_{\rm GI}$ 
corresponds to a nucleon matrix element of a local and gauge-invariant 
operator, and its
calculation in lattice QCD became available recently. For $\Delta q_{\rm CI}$,
one has to compute the matrix element of $\tilde{J}^+_5$ in light-front
gauge, which will require sizeable lattice gauge configurations. Second, NLO
polarized splitting functions have been determined very recently in 
the gauge-invariant scheme, and it is straightforward to study the evolution 
of $\Delta q_{\rm GI}(x,Q^2)$ through AP evolution equations.

\section{U(1) Goldberger-Treiman Relation and Its Connection to the 
Proton Spin}
\setcounter{equation}{0}
\subsection{Two-component U(1) Goldberger-Treiman relation}
  In the gauge-invariant and chiral-invariant factorization schemes 
the flavor-singlet axial coupling $g_A^0$ has the expression
\be
g_A^0 &=& \Delta\Sigma_{\rm GI}   \\
&=& \Delta\Sigma_{\rm CI}-{n_f\alpha_s\over 2\pi}\Delta G.  
\en
The smallness of the observed $g_A^0$ is attributed either to the negative
sea polarization or to the anomalous gluonic contribution. However,
the question of what is its magnitude still remains unanswered. The
well-known isotriplet Goldberger-Treiman (GT) relation
\be
g_A^3(0)=\,{\sqrt{2}f_\pi\over 2m_N}g_{_{\pi_3NN}},
\en
with $f_\pi=132$ MeV, indicates that the coupling $g_A^3$ is fixed in 
terms of the strong coupling constant $g_{_{\pi_3NN}}$. It is natural
to generalize this relation to the $U_A(1)$ case to see if we can 
learn something about the magnitude of $g_A^0$. 

Many discussions on the isosinglet GT relation around the period of 1989-1992 
\cite{Sch90,Bart,Shore92,Vene89,Shore90,Efr90,Hat91}
were mainly motivated by the desire of trying to understand why the 
axial charge $g_A^0$ inferred from the EMC experiment \cite{EMC} is so 
small, $g_A^0(0)=0.14\pm 0.17$ at $Q^2=10.7\,{\rm GeV}^2$ 
(pre-1993). (The $q^2$ of the form factor should not be confused 
with the momentum transfer $Q^2$ occurred in deep inelastic scattering.)
At first sight, the U(1) GT relation seems not to be in
the right ballpark as the naive SU(6) quark-model's prediction $g^{(0)}
_{_{\eta_0NN}}=(\sqrt{6}/5)g_{_{\pi NN}}$ yields a too large value of $g_A^0
(0)=0.80\,$. Fortunately, in QCD the ghost field $G\equiv \partial^\mu K_\mu$, 
which is necessary for solving the $U_A(1)$ problem, is 
allowed to have a direct $U_A(1)$-invariant interaction with the nucleon. 
This together with the mixing of $\partial^\mu K_\mu$ with the
$\eta_0$ implies that the net ``physical" $\eta_0-N$ coupling $\getao$ is 
composed of the bare coupling $\getao^{(0)}$ and the ghost
coupling $\gghost$. As a consequence, a possible cancellation between
$\getao$ and $\gghost$ terms will render $g_A^0$ smaller. However, this 
two-component expression for the axial charge is not free of ambiguity.
For example, $\gghost$ is sometimes assumed to be the 
coupling between the glueball and the nucleon in the literature. Moreover,
unlike the couplings $g_A^3$ and $g_A^8$, a prediction for $g_A^0$ is lost.

    Since the earlier parton-model analysis of polarized deep inelastic
scattering seems to indicate a decomposition of $g_A^0$ in terms of the 
quark and gluon spin components \cite{Efr,AR,CCM}, this has motivated many 
authors to identify the term $(\sqrt{3}f_\pi/2m_N)\getao$ with the total quark 
spin $\Delta\Sigma$ in a proton, and the other term with the anomalous
gluon contribution. However, this identification holds only in the 
chiral-invariant scheme. We will address this problem below.

One important thing we have learned from the derivation of the 
isotriplet Goldberger-Treiman (GT) relation (5.3)
is that this relation holds irrespective of the 
light quark masses. For $m_\pi^2\neq 0$, it is derived through the 
use of PCAC; while in the chiral limit, $g_A^3(q^2)$ is related to the 
form factor $f_A^3(q^2)q^2$, which receives a nonvanishing pion-pole 
contribution even in the $q^2\to 0$ limit. By the same token, it
is tempting to contemplate that the U(1) GT relation
should be also valid irrespective of the meson masses and the axial 
anomaly. This is indeed the case: {\it the U(1) GT relation (5.6) given below
remains totally unchanged no matter how one varies the anomaly and the 
quark masses}. This salient feature was first explicitly shown in 
\cite{Sch90,Bart}. It was 
also pointed out in \cite{Vene89} that this U(1) relation is independent 
of the interaction of the ghost field $\partial^\mu K_\mu$ with the nucleon.
The easist way of deriving the U(1) GT relation is thus to first work 
in the chiral limit. Defining the form factors
\be
\la N(p')|J^5_\mu|N(p)\ra= \,\bar{u}(p')[g_A^0(q^2)\gamma_\mu\gamma_5
+f_A^0(q^2)q_\mu\gamma_5]u(p),
\en
we obtain
\be
2m_Ng_A^0(0)=\la N|\partial^\mu J^5_\mu|N\ra=3\la N|\dk|N\ra.
\en
Assuming the $\eta_0$ pole dominance for $\dk$, namely
$\dk={1\over\sqrt{3}}m_{\eta_0}^2f_\pi\eta_0$, where the $\eta_0$
mass $m_{\eta_0}$ arises entirely from the axial anomaly, we are led to the 
isosinglet GT relation\footnote{It is argued in \cite{Shore90,Shore92}
that the U(1) GT relation (5.6) holds only when the $\eta_0$ is a massless 
Goldstone boson obtained in the large-$N_c$ or OZI limit. In general
the decay constant $f_{\eta_0}$ can be related to the topological
susceptibility $\chi'(0)$ of the QCD vacuum so that the U(1) GT relation
reads
\be
2m_Ng_A^0(0)=6\sqrt{\chi'(0)}\,g_{_{\eta_0NN}}^{(0)}.   \non
\en
In the OZI limit, $\sqrt{\chi'(0)}=f_\pi/(2\sqrt{3})$. The smallness of
the observed $g_A^0$ can be attributed either to the anomalously small
value of the first moment of QCD topological susceptibility 
\cite{Shore90,Shore92} (for an estimate of $1/N_c$ corrections to $\chi'(0)$,
see \cite{Grun}) or to the 
suppression of the coupling $g_{_{\eta_0NN}}^{(0)}$. The smallness of
$g_A^0$ in the former case is a generic QCD effect related to the 
anomaly and is independent of the target \cite{Narison}, whereas it can 
be quite target dependent in the latter case.}
\be
g_A^0(0)=\,{\sqrt{3}f_\pi\over 2m_N}g_{_{\eta_0NN}}^{(0)},
\en
with $\getao^{(0)}$ a bare direct coupling between $\eta_0$ and the nucleon.

When the quark masses are turned on, chiral 
symmetry is explicitly broken but the GT relation in terms of the $\eta_0$ 
remains intact, as shown in \cite{Sch90,Bart}. Nevertheless, the $\eta_0$ is 
no longer a physical meson, and it is related to the mass eigenstates via
\be
\left(\matrix{\pi_3 \cr \eta_8 \cr \eta_0 \cr}\right)=\left(\matrix{ 1 &
\theta_1\cos\theta_3+\theta_2\sin\theta_3 & \theta_1\sin\theta_3-\theta_2
\cos\theta_3 \cr  -\theta_1 & \cos\theta_3 & \sin\theta_3  \cr  \theta_2 &
-\sin\theta_3 & \cos\theta_3 \cr}\right)\left(\matrix{\pi^0 \cr  \eta  \cr
\eta'}\right),
\en
where $\theta_1,~\theta_2$ and $\theta_3$ are the mixing angles of 
$\pi^0-\eta$, are given in \cite{Efr90} with the numerical values
\be
\theta_1=-0.016\,,~~~\theta_2=0.0085\,,~~~\theta_3=-18.5^\circ\,.
\en
In Eq.(5.7) only terms linear in small angles $\theta_1$ and
$\theta_2$ are retained. Consequently, the complete GT relations in terms
of physical coupling constants read \cite{Bart}
\footnote{For the axial charge $g_A^0$, the authors of \cite{Efr90} 
obtained a result of the form (see Eq.(24) of the second reference of 
\cite{Efr90})
\be
{\sqrt{3}f_\pi\over 2m_N}\left({\getap\over \cos\theta_3}-\Delta m_{\eta'}
g_{_{QNN}}\right)-{1\over\sqrt{2}}
g_A^8\tan\theta_3\pm\sqrt{3\over 2}g_A^3(\theta_2-\theta_1\tan\theta_3)
\en
and claimed that in the limit of $\theta_1,~\theta_2\to 0$ but $\theta_3
\neq 0$, it will reproduce the result of Veneziano \cite{Vene89} only if 
the first-order correction from $\theta_3$ (i.e., the $g_A^8\theta_3$ term) 
is neglected. However, using Eqs.(5.11-12) and (5.16) one can show that 
(5.9) is nothing but 
$(\sqrt{3}f_\pi/2m_N)\getao^{(0)}$, as it should be.}
\be
g_A^3(0) ={\sqrt{2}f_\pi\over 2m_N}g_{_{\pi_3NN}} &=& {\sqrt{2}f_\pi\over
2m_N}[\gpi\pm\getap(\theta_1\sin\theta_3-\theta_2\cos\theta_3)  \non \\
&& \pm\,\geta(\theta_1\cos\theta_3+\theta_2\sin\theta_3)],   \\
g_A^8(0)={\sqrt{6}f_\pi\over 2m_N}g_{_{\eta_8NN}} &=& {\sqrt{6}f_\pi\over
2m_N}(\geta\cos\theta_3+\getap\sin\theta_3\mp\gpi\theta_1),   \\
g_A^0(0)={\sqrt{3}f_\pi\over 2m_N}g_{_{\eta_0NN}}^{(0)} &=& {\sqrt{3}f_\pi
\over 2m_N}(\getap\cos\theta_3-\geta\sin\theta_3\pm\gpi\theta_2)+\cdots, 
\en
where the first sign of $\pm$ or $\mp$ is for the proton and the second
sign for the neutron, and the ellipsis in the GT relation for $g_A^0$ is
related to the ghost coupling, as shown below.
Since the mixing angles $\theta_1$ and $\theta_2$ are 
very small, it is evident that
isospin violation in (5.10-5.12) is unobservably small.

   As we have accentuated before, the isosinglet GT relation in terms of the
$\eta_0$ remains unchanged no matter how one varies the quark masses and 
the axial anomaly. (A smooth extrapolation of the strong coupling constant 
from on-shell $q^2$ to $q^2=0$ is understood.)
However, the $\eta_0$ field is subject to a different interpretation
in each different case. For example, when the anomaly 
is turned off, the mass of $\eta_0$ is the same as the pion (for 
$f_{\eta_0}=f_\pi$). When both quark masses and anomaly are switched off,
the $\eta_0$ becomes a Goldstone boson, and the axial charge at $q^2=0$ 
receives its contribution from the $\eta_0$ pole.

    When the SU(6) quark model is applied to the coupling $\getao^{(0)}$, it
is evident that the predicted $g_A^0=0.80$ via the GT relation is too large.
This difficulty could be resolved by the observation that {\it a priori}
the ghost field $G\equiv\dk$ is allowed in QCD to have a direct
coupling with the nucleon
\be
{\cal L}=\cdots+{\gghost\over 2m_N}\partial^\mu G\,{\rm Tr}(\bar{N}\gamma_\mu
\gamma_5N)+{\sqrt{3}\over f_\pi}(\dk)\eta_0+\cdots,
\en
so that
\be
\dk=\,{1\over\sqrt{3}}m_{\eta_0}^2f_\pi\eta_0+{1\over 6}\gghost m^2_{\eta_0}
f_\pi\partial^\mu{\rm Tr}(\bar{N}\gamma_\mu\gamma_5N).
\en
However, the matrix element $\la N|\dk|N\ra$ remains unchanged
because of the presence of the $\dk-\eta_0$ mixing, as
schematically shown in Fig.~5 :
\be
\la N|\dk|N\ra &=& {1\over\sqrt{3}}f_\pi\getao^{(0)}-{1\over 3}m^2_{\eta_0}
f_\pi\gghost+{1\over 3}m^2_{\eta_0}f_\pi\gghost   \non \\
&=& {1\over\sqrt{3}}f_\pi\getao^{(0)}.
\en
We see that although it is still the bare coupling $\getao^{(0)}$ that 
relates to the axial charge $g_A^0$, the ``physical" $\eta_0-N$ coupling 
is modified to (see Fig.~5)
\be
\getao=\,\getao^{(0)}+{1\over\sqrt{3}}m_{\eta_0}^2f_\pi\gghost,
\en
where the second term arises from the $\eta_0-\dk$ mixing. As a consequence,
the quark model should be applied to $\getao$ rather than to $\getao^{(0)}$,
and we are led to
\be
g_A^0(0)=\,{\sqrt{3}f_\pi\over 2m_N}(\getao-{1\over \sqrt{3}}m^2_{\eta_0}
f_\pi\gghost).
\en
This two-component expression for the U(1) GT relation was first 
put forward by Shore and Veneziano \cite{Shore90}.

\begin{figure}[ht]
%\vspace{-4cm}

    \centerline{\psfig{figure=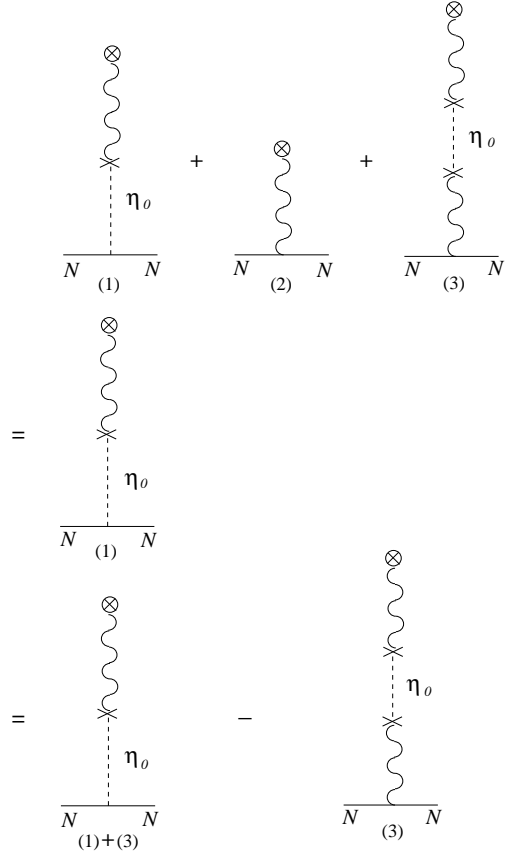,width=6.5cm}}
    \caption[]{\small Contributions to the matrix element $\la N|\dk|N\ra$ 
from (1) the $\eta_0$ pole dominance, (2) a direct coupling of the ghost field
with the nucleon, and (3) the $\dk-\eta_0$ mixing.}
    \label{fig5} 

\end{figure}

    It has been proposed that the smallness of $g_A^0$ may be explained
by considering the pole contributions to $\dk$ from higher single particle
states $X$ above the $\eta_0$, so that the isosinglet GT relation has the
form (see e.g., Chao {\it et al.} \cite{Vene89}, Ji \cite{Vene89}, 
Bartelski and Tatur \cite{Bart93})
\be
g_A^0(0)=\,{\sqrt{3}\over 2m_N}(f_{\eta_0}\getao+\sum_X f_Xg_{_{XNN}}).
\en
The state $X$ could be the radial excitation state of $\eta_0$ or a $0^{-+}$
glueball. (Note that the ghost field $\dk$ is {\it not} a physical glueball
as it can be eliminated via the equation of motion.)
However, we will not pursue this possibility further for two reasons: (i)
It is entirely unknown whether or not the $X$ states contribute destructively
to $g_A^0$. (ii) As we shall see later, the contribution from a direct
interaction of the ghost field with the nucleon corresponds to a 
disconnected insertion, which is shown to be negative according to 
recent lattice QCD calculations \cite{Dong,Fuk}. Therefore, the 
ghost-field effect is realistic, and if the contributions due
to the states $X$ are taken into account, one should make the following
replacement
\be
\getao\to\getao-{1\over\sqrt{3}}m^2_{\eta_0}f_\pi\gghost,~~~~g_{_{XNN}}\to
g_{_{XNN}}-{1\over \lambda}m_X^2g_{_{XNN}}
\en
in Eq.(5.18), where $\lambda$ is the $\dk\!-\!X$ mixing.

\subsection{Interpretation of the U(1) Goldberger-Treiman relation}
By comparing (5.17) and (5.2), it is tempting to identify the two 
components of the U(1) GT relation as
\be
\Delta\Sigma_{\rm CI}={\sqrt{3}f_\pi\over 2m_N}\getao,~~~~\Delta\Gamma=
{m_{\eta_0}^2f_\pi^2\over 2n_fm_N}\gghost.
\en
However, this identification is not unique and sensible because it does
not hold in the gauge-invariant factorization definition for $\Delta q
_{\rm GI}$. One may ask can one have a physical interpretation valid for
both $k_\perp$-factorization schemes for the $\getao$ and $\gghost$ 
terms in the two-component isosinglet 
GT relation (5.17) ? As noted in Sec.~4.5, the
evaluation of the hadronic flavor-singlet current involves connected and 
disconnected insertions (see Fig.~4) which are related to valence-quark 
and sea-quark contributions respectively and are separately 
gauge invariant. A recent lattice calculation \cite{Dong} indicates
an empirical SU(3)-flavor symmetric sea polarization; this
implies that the disconnected insertion is dominated by the
axial anomaly of the triangle diagram. Since the triangle contribution is
proportional to $\dk$, the ghost field, it is thus quite natural to make
the gauge-invariant identification:
\be
{\sqrt{3}f_\pi\over 2m_N}\getao=\,{\rm connected~insertion},~~~~-{m_{\eta_0}
^2f_\pi^2\over 2m_N}\gghost=\,{\rm disconnected~insertion},
\en
which is valid in both factorization schemes. Note that this identification
is basically an assumption since it is possible to add and substract
some part of the disconnected contribtuion in (5.21) and the resultant
identification is still gauge invariant. In the gauge-invariant 
factorization scheme, the disconnected insertion, which is responsible for
the smallness of $g_A^0$, should be interpreted as a screening effect for
the axial charge owing to the negative sea polarization rather than an 
anomalous gluonic effect.

  Having identified the two-component U(1) GT relation (5.17) with
connected and disconnected insertions, we are now able to extract the
physical coupling constants $\getap$ and $\geta$. This is because the
connected insertion (CI) corresponds to the total ``valence" quark 
contribution (strictly speaking, the valence-quark plus cloud-quark
contributions; see Sec.~6.1)
to the proton spin, so it is related to the quark model expectation; 
that is,
\be
{\sqrt{3}f_\pi\over 2m_N}\getao=g_A^0({\rm CI})=\,\Delta u_v+\Delta d_v=\,
3F-D,
\en
where the last identity follows from the fact that $g_A^8=3F-D=
\Delta u+\Delta d-2\Delta s\to\Delta u_v+\Delta d_v$ due to the aforementioned
SU(3) symmetry for sea polarization. Unlike the previous identification 
(5.20), $g_A^0({\rm CI})$ here is {\it not} identified with the total 
quark spin $\Delta\Sigma$. In the non-relativistic quark limit, 
$F={2\over 3},~D=1$, and hence $\Delta u_v+
\Delta d_v=1$. With the inclusion of the relativistic effects and cloud-quark
polarization (see Sec.~6.1),
$F$ and $D$ are reduced to $0.459$ and $0.798$, respectively,
and $g_A^0({\rm CI})$ is reduced to a value of $0.579\,$. 

   From Eqs.(5.11), (5.12) and (5.22), the GT relations for $g_A^8$ 
and $g_A^0$ are recast to
\be
&& 3F-D={\sqrt{6}f_\pi\over 2m_N}g_{_{\eta_8NN}}= {\sqrt{6}f_\pi\over
2m_N}(\geta\cos\theta_3+\getap\sin\theta_3),   \non \\
&& 3F-D={\sqrt{3}f_\pi\over 2m_N}g_{_{\eta_0NN}} = {\sqrt{3}f_\pi
\over 2m_N}(\getap\cos\theta_3-\geta\sin\theta_3),   
\en
where the tiny isospin-violating effect has been neglected.
Note that we have $\getao$ instead of $\getao^{(0)}$ on the second line
of the above equation. Using $\theta_3=-18.5^\circ$ [see Eq.(5.8)], it 
follows from (5.23) that \cite{Cheng96}
\be
\getap=3.4\,,~~~~~\geta=4.7\,,
\en
while
\be
\getao=4.8\,,~~~~~g_{_{\eta_8NN}}=3.4\,.
\en 
It is interesting to note that we have $\getap<\geta$, whereas $\getao>
g_{_{\eta_8NN}}$.
Phenomenologically, the determination of $\getap$ and $\geta$ is rather
difficult and subject to large uncertainties. The analysis of the $NN$ 
potential yields $\getap=7.3$ and $\geta=6.8\,$ \cite{Dumb}, while the forward 
$NN$ scattering analyzed using dispersion relations gives $\getap,~\geta
<3.5$\, \cite{Brein}. But these analyses did not take into account the ghost
pole contribution. An estimate of the $\eta'\to 2\gamma$ decay rate through
the baryon triangle contributions yields $\getap=6.3\pm 0.4$ \cite{Bag}.

   Finally, the ghost coupling is determined from the disconnected insertion
(DI)
\be
-{m_{\eta_0}^2f_\pi^2\over 2m_N}\gghost=\,g_A^0({\rm DI})
=\,\Delta u_s+\Delta d_s+\Delta s\to 3\Delta s.
\en
Using $g_A^0(0)=\Delta\Sigma=\,0.31\pm 0.07$ [see (2.27)] and (5.26) we obtain
\be
\gghost\approx 55\,{\rm GeV}^{-3}.
\en
In principle, this coupling constant can be inferred from the low-energy
baryon-baryon scattering in which an additional SU(3)-singlet contact
interaction arises from the ghost interaction \cite{Sch90}.

 To summarize, the U(1) GT relation (5.6) in 
terms of the $\eta_0$ remains totally unchanged no matter how one varies
the quark masses and the axial anomaly, while its
two-component expression (5.17) can be identified
with the connected and disconnected insertions.
Since $(\sqrt{3}f_\pi/2m_N)\getao$ is related to the
total valence quark contribution to the proton spin, we have determined the
physical coupling constants $\getap$ and $\geta$ from the GT relations for
$g_A^0$ and $g_A^8$ and found that $\getap=3.4$ and $\geta=4.7\,$.

%\newpage

\section{Other Theoretical Progresses}
\setcounter{equation}{0}
\subsection{Lattice calculation of proton spin content}
   The spin-dependent DIS experiments indicate that $\Delta u\sim 0.83,~
\Delta d\sim -0.43$ and $\Delta s\sim -0.10$ at $Q^2=10\,{\rm GeV}^2$ [cf.
(2.26)]. We learn from Secs.~3 and 4
that the axial anomaly plays an essential role for the smallness of $g_A^0$ or
the suppression of $\Gamma_1$ relative to the Ellis-Jaffe conjecture. 
However, many questions still remain unanswered,
for example: (i) what is the sea polarization of the non-strange light 
quarks (i.e., $\Delta u_s,~\Delta d_s$) ? (ii) what are spin components
of valence quarks $\Delta u_v,~\Delta d_v$ ? Are they consistent with
the expectation of quark models ? (iii) what is the magnitude and
sign of the gluon spin component in a proton ? (iv) what is the
orbital angular momentum content of quarks and gluons ? and (v) what are
spin-dependent parton distributions $\Delta q(x),~\Delta G(x)$ ?
A truly theoretical or experimental progress should address some of the
above-mentioned questions.  Obviously, a first-principles calculation
based on lattice QCD will, in principle, be able to provide some answers.
Indeed, the present lattice calculation is starting to shed light on 
the proton spin contents.

   After the 1987 EMC experiment, there existed several attempts of computing
$\Delta G$ and $g_A^0$ using lattice QCD (for a nice review, see Liu 
\cite{Liu}, Okawa \cite{Okawa} and references therein). A first
direct calculation of the quark spin content $\Delta q$ was made in
\cite{Mandu} but without final results. Fortunately, two successful
lattice calculations in the quenched approximation just became available
very recently \cite{Dong,Fuk}. A more ambitious program of computing
the polarized structure functions $g_1(x),~g_2(x)$ and their moments is
also feasible and encouraging early results were reported in \cite{Gock}.

   What computed in \cite{Dong,Fuk} is the gauge-invariant quark spin
component $\Delta q_{\rm GI}$ defined by $s^\mu\Delta q_{\rm GI}=\la
p,s|\bar{q}\gamma^\mu\gamma_5 q|p,s\ra$ (recall that $\Delta q$ has the
conventional partonic interpretation only in the ``+" component in the
light-front coordinate). An evaluation of $\Delta q_{\rm GI}$ involves
a disconnected insertion in addition to the connected insertion (see
Fig.~4; the infinitely many possible gluon lines and additional quark 
loops are implicit). The sea-quark spin contribution comes from the
disconnected insertion. It is found that
\be
\cite{Dong}: && ~~~\Delta u_{\rm dis}=\Delta d_{\rm dis}=-0.12\pm 0.01\,,~~~~
\Delta s=-0.12\pm 0.01\,,   \non \\
\cite{Fuk}: && ~~~\Delta u_{\rm dis}=\Delta d_{\rm dis}=-0.119\pm 0.044\,,~~~~
\Delta s=-0.109\pm 0.030\,.   
\en
Note that the results of \cite{Fuk} are gauge independent although the gauge
configurations on the $t=0$ time slice are being fixed to the Coulomb gauge
(see a discussion in \cite{Okawa}). It is evident that the disconnected
contribution is independent of the sea-quark mass in the loop within errors.
Therefore, this empirical SU(3)-flavor invariance for sea polarization
implies that {\it the disconnected insertion is dominated by the axial 
anomaly of the triangle diagram; that is, it is the gluonic anomaly
which accounts for the bulk of the negative sea polarization.} This is
consistent with the picture described in Sec.~4.3, namely a substantial
polarization of sea quarks is produced from gluons via the perturbative 
anomaly mechanism and from nonperturbative effects via instantons.

   It has been emphasized in \cite{Liu94} that the connected 
insertion involves not only valence quarks but also cloud quarks. In
the time-ordered diagrams, one class of the connected insertion involves
an antiquark propagating backward in time between the currents and
is defined as the ``cloud" antiquark as depicted in Fig.~6.
Another class involves a quark 
propagating forward in time between the currents and is defined to be 
the sum of valence and cloud quarks. Hence the quark spin 
distribution can be written as
\be
\Delta q(x)=\,\Delta q_V(x)+\Delta q_c(x)+\Delta q_s(x)=\,\Delta q_v(x)+
\Delta q_s(x),
\en
where $\Delta q_v(x)$ as conventionally referred to as the ``valence" quark 
spin density 
is actually a combination of cloud and truly valence contributions, i.e.,
$\Delta q_v(x)=\Delta q_V(x)+\Delta q_c(x)$. The concept of cloud quarks,
which is familiar to the nuclear-physics community, appears to be
foreign to the particle-physics community. As shown in \cite{Liu94}, the
presence of cloud quarks and antiquarks is the key for understanding the 
origin of deviation of the Gottfried sum rule from experiment, namely the
difference of $\bar{u}$ and $\bar{d}$ distributions in the nucleon.

\begin{figure}[ht]
\vspace{1cm}

    \centerline{\psfig{figure=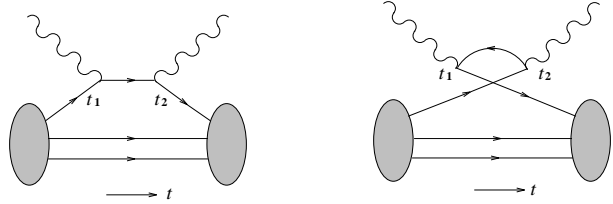,width=8cm}}
    \caption[]{\small Time-ordered diagrams of the connected insertion
involving quark and antiquark propagators between the currents.}
    \label{fig6} 

\end{figure}

  A very important lattice observation is made in \cite{Dong} that the SU(6) 
relation in the quark model is recovered in the valence approximation under
which cloud quarks in the connected insertion are turned off. For example,
the ratio $R_A=g^0_{\rm A,con}/g_A^3$ is found to be ${3\over 5}$ in the
lattice calculation when the presence of cloud quarks and antiquarks is 
eliminated by disallowing quarks from propagating backward in time,
while theoretically it is reduced under the valence approximation to
$(\Delta u_{\rm con}+\Delta d_{\rm con})/(\Delta u_{\rm con}-\Delta d_{
\rm con})$, which is equal to ${3\over 5}$ in the relativistic 
or non-relativistic quark model. Of course, the prediction $(g_A^3)^{\rm 
NR}={5\over 3}$ in the non-relativistic quark model is 
too large compared to the experimental value $(g_A^3)_{\rm expt}=1.2573\pm
0.0028$ \cite{PDG}. Presumably, $(g_A^3)^{\rm NR}$ is reduced by a factor 
of 3/4 due to relativistic effects. In other words, the above 
lattice observation
implies that {\it relativistic quark model results should be recovered
in the valence approximation.} Based on this observation, a picture
for the smallness of $\Delta\Sigma_{\rm GI}$ or $g_A^0$ emerges.
In the relativistic quark model, the non-relativistic SU(6) predictions 
$\Delta u^{\rm NR}={4\over 3}$ and $\Delta d^{\rm NR}=-{1\over 3}$ are
reduced by the same factor to $\Delta u_V=1$ and $\Delta d_V=-{1\over 4}$,
where the subscript ``V" denotes a genuine valence spin component. Since
the quark orbital angular momentum is nonvanishing in the presence of
quark transverse momentum in the lower component of the Dirac spinor,
the reduction of the spin component from $\Delta u^{\rm NR}
+\Delta d^{\rm NR}=1$ to $\Delta u_V+\Delta d_V=0.75$ is shifted to the
orbital component of the proton spin.
Assuming SU(3)-symmetric sea polarization, as suggested by lattice
calculations, one obtains from (2.26) that
\be
\Delta u_v=\Delta u_V+\Delta u_c\simeq 0.93\,,~~~~\Delta d_v=\Delta d_V+\Delta 
d_c\simeq -0.33\,,
\en
and hence 
\be
\Delta u_c\simeq -0.07\,,~~~~~\Delta d_c\simeq -0.08\,.
\en
{\it The cloud-quark polarization $\Delta q_c$ is thus negative in sign and 
comparable in magnitude to the sea polarization $\Delta q_s$}. Now we have
\be
\underbrace{\Delta u_{\rm GI}+\Delta d_{\rm GI}+\Delta s_{\rm GI}}_{0.30}
=\underbrace{\Delta u_V+\Delta d_V}
_{0.75}+\underbrace{\Delta u_c+\Delta d_c}_{-0.15}+\underbrace{\Delta 
u_s+\Delta d_s+\Delta s}_{-0.30}.
\en
We conclude that {\it the deviation of $\Delta\Sigma_{\rm GI}$ or $g_A^0$
from the relativistic quark model's value 0.75 is ascribed to the negative
cloud-quark and sea-quark polarizations}. In the future, it will be of
great importance to calculate $\Delta q_V$ and $\Delta q_c$ directly
by lattice QCD.

   The lattice results of \cite{Dong} and \cite{Fuk} for $\Delta q_{\rm GI},~
g_A,~F$ and $D$ are presented in Table II; in general they agree with 
experiments within errors. Moments
of polarized structure functions $g_1(x)$ and $g_2(x)$ from the
connected insertion are reported in \cite{Gock}. It is found that
twist-3 operators characterized by the matrix element $d_3$ 
[cf. Eq.(4.18)] provide the dominant contribution to $\int^1_0 x^2g_2(x)
dx$.

\begin{table}
{\small Table II. Axial couplings and quark spin contents of the proton
from lattice calculations and from experiments [see (2.19) and (2.26)].}
\begin{center}
\begin{tabular}{|c||c|c||c|} \hline
 & \cite{Dong} & \cite{Fuk} & Experiment   \\  \hline
$g_A^0$ & 0.25(12) & 0.18(10) & 0.31(7)  \\
$g_A^3$ & 1.20(10) & 0.985(25) & 1.2573(28)  \\
$g_A^8$ & 0.61(13) & --- & 0.579(25) \\
$\Delta u$ & 0.79(11) & 0.638(54) & 0.83(3) \\
$\Delta d$ & -0.42(11) & -0.347(46) & -0.43(3) \\
$\Delta s$ & -0.12(1) & -0.109(30) & -0.10(3) \\
$F$ & 0.45(6) & 0.382(18) & 0.459(8) \\
$D$ & 0.75(11) & 0.607(14) & 0.798(8) \\
\hline
\end{tabular}
\end{center}
\end{table}

   As for the chiral-invariant quantity $\Delta\Sigma_{\rm CI}$, it involves
the matrix element of $\tilde{J}^+_5$ in light-front gauge [see Eq.(4.57)]
and hence sizeable gauge configurations are needed in lattice calculations
for $\Delta\Sigma_{\rm CI}$.
Nevertheless, it is conceivable to have lattice results for $\Delta G$ and
$\Delta q_{\rm CI}$ soon in the near future.

\subsection{Two-loop spin-dependent splitting functions}
   The experimental data of $g_1(x,Q^2)$ taken at different $x$-bin 
correspond to different ranges of $Q^2$, that is, $Q^2$ of the data is
$x$-bin dependent. To the zeroth order in QCD, $g_1$ simply reads
$g_1(x)={1\over 2}\sum_i e^2_i\Delta q_i(x)$ without scaling violation.
To the leading order (LO), it becomes $g_1(x,Q^2)={1\over 2}\sum_i 
e^2_i\Delta q_i(x,Q^2)$ with scaling violation arising from
gluon bremsstrahlung and quark-antiquark pair creation from gluons.
In other words, $\Delta G(x)$ enters into $g_1$ at LO only via the $Q^2$
evolution governed by the LO polarized AP equation.
To the next-to-leading order (NLO), $g_1(x,Q^2)$ is given by (3.1). 
At this order, gluons contribute directly to the polarized structure 
function $g_1$. A full NLO QCD analysis of the $g_1$ data is thus not 
possible until the two-loop splitting functions $\Delta P_{ij}^{(1)}(x)$ 
in the NLO $Q^2$ evolution equation are known. Since the
complete results for $\Delta P_{ij}^{(1)}(x)$ are not available until 
very recently \cite{Mert}, all pre-1995 analyses based on the 
NLO expression (3.1) for $g_1(x,Q^2)$ are not complete and fully consistent.

The $Q^2$ dependence of parton spin densities is determined
by the spin-dependent Altarelli-Parisi equations:
\be
&& {d\over dt}\Delta q_{\rm NS}(x,t)=\,{\alpha_s(t)\over 2\pi}\Delta P_{qq}^
{\rm NS}(x)\otimes\Delta q_{\rm NS}(x,t),   \non \\
&& {d\over dt}\left(\matrix{\Delta q_{\rm S}(x,t)   \cr   \Delta G(x,t) \cr}
\right)=\,{\alpha_s(t)\over 2\pi}\left(\matrix{\Delta P_{qq}^{\rm S}(x) & 
2n_f\Delta P_{qG}(x)  \cr  \Delta P_{Gq}(x) & \Delta P_{GG}(x) \cr} 
\right)\otimes\left(\matrix
{\Delta q_{\rm S}(x,t)   \cr   \Delta G(x,t) \cr} \right),
\en
with $t=\ln(Q^2/\Lambda^2_{_{\rm QCD}})$, 
\be
\Delta q_{\rm NS}(x)=\Delta q_i(x)-\Delta q_j(x),~~~~~\Delta q_{\rm S}(x)=
\sum_i\Delta q_i(x),
\en
and  
\be
\Delta P_{ij}(x)=\Delta P_{ij}^{(0)}(x)+{\alpha_s\over 2\pi}\Delta P_{ij}^{
(1)}(x)+\cdots.
\en
The spin-dependent anomalous dimensions are defined as
\be
\Delta\gamma_{ij}^n=\int^1_0\Delta P_{ij}(x)x^{n-1}dx=\Delta\gamma_{ij}^{(0),
n}+{\alpha_s\over 2\pi}\Delta\gamma_{ij}^{(1),n}+\cdots.
\en
The leading-order polarized splitting functions $\Delta P_{ij}^{(0)}$ 
are given by (3.26) and the corresponding anomalous dimensions for $n=1$ are
\be
\Delta\gamma^{(0),1}_{qq}=\Delta\gamma^{(0),1}_{qG}=0,~~~\Delta\gamma^{(0),1}
_{Gq}=2,~~~\Delta\gamma^{(0),1}_{GG}={1\over 2}{\beta_0},
\en
where $\beta_0=11-2n_f/3$.
To the NLO, $\Delta P_{qq}^{(1)}$ and $\Delta P_{qG}^{(1)}$ were
calculated in the $\overline{\rm MS}$ scheme by Zijlstra and van Neerven
\cite{Zijl}. However, the other two polarized splitting functions
$\Delta P_{Gq}^{(1)}$ and $\Delta P_{GG}^{(1)}$ were not available until last
year. The detailed results of $\Delta P_{ij}^{(1)}(x)$ are given in 
\cite{Mert}. Here we just list the anomalous dimensions for $n=1$:
\be
&& \Delta\gamma^{(1),1}_{{\rm NS},qq}=0,~~~~\Delta\gamma^{(1),1}_{{\rm S},qq}=
-3C_FT_f=-2n_f,  \non \\
&& \Delta\gamma^{(1),1}_{qG}=0,~~~~\Delta\gamma^{(1),1}_{Gq}=
-{1\over 8}\left(-6C_F^2-{142\over 3}C_AC_F+{8\over 3}C_FT_f\right)=25-{2\over 
9}n_f,   \\
&& \Delta\gamma^{(1),1}_{GG}={\beta_1\over 4}=-{1\over 8}\left(-{68\over 3}
C_A^2+8C_FT_f+{40\over 3}C_AT_f\right)={1\over 4}\left(102-{38\over 3}
n_f\right),   \non
\en
where $C_F={4\over 3},~C_A=3,~T_f=n_f/2$. To this order,
\be
{\alpha_s\over 4\pi}=\,{1\over \beta_0\ln Q^2/\Lambda^2_{_{\overline{\rm 
MS}}}}-{\beta_1\ln\ln Q^2/\Lambda^2_{_{\overline{\rm MS}}}\over \beta_0^3(\ln 
Q^2/\Lambda^2_{_{\overline{\rm MS}}})^2}\,.
\en

   Note that the $\overline{\rm MS}$ regularization scheme is
a gauge-invariant $k_\perp$-factorization scheme as it respects the axial
anomaly in the triangle diagram. Therefore, the NLO evolution of parton
spin distributions in the gauge-invariant factorization scheme is
completely determined. Explicitly, the AP equation for the first 
moment of flavor-singlet parton spin densities reads 
\be
{d\over dt}\left(\matrix{\Delta\Sigma_{\rm GI}(t) \cr \Delta G(t)  \cr}\right)
=\,{\alpha_s(t)\over 2\pi}\left(\matrix{ {\alpha_s\over 2\pi}(-2n_f) & 0  \cr  
2+{\alpha_s
\over 2\pi}(25-{2\over 9}n_f)  &  {\beta_0\over 2}+{\alpha_s\over 2\pi}\,{
\beta_1\over 4}   \cr}\right)\left(\matrix{ \Delta\Sigma_{\rm GI}(t)  \cr  
\Delta G(t)\cr}\right).
\en
Defining $\Delta\Gamma\equiv(\alpha_s/2\pi)\Delta G$, it is easily seen that
\be
{d\over dt}\left(\matrix{\Delta\Sigma_{\rm GI}(t) \cr \Delta\Gamma(t)}\right)=
\left({\alpha_s\over 2\pi}\right)^2\left( \matrix{ -2n_f & 0 \cr  2 & 0 }
\right)\left(\matrix{ \Delta\Sigma_{\rm GI}(t)  \cr \Delta\Gamma(t) }\right)
+{\cal O}(\alpha_s^3),
\en
which is in agreement with (4.64). A derivation of (6.14) does not need
the information of $\Delta\gamma^{(1),1}_{Gq}$ and $\Delta\gamma^{(1),1}_
{GG}$, however. The NLO $Q^2$ evolution of parton spin densities has been
studied in \cite{Ball,Gluck96,Gehr96,Weig}. It is found that the difference
between LO and NLO evolution for $\Delta\Sigma(x,Q^2)$ and $\Delta G(x,Q^2)$
is sizeable at small $x$, $x\lsim 5\times 10^{-3}$ (see Figs.~4 and 5 of
\cite{Weig}). This feature can be understood from the $x\to 0$ behavior
of the splitting functions $\Delta P_{ij}(x)$. As $x\to 0$, we find from
(3.26) that
\be
\Delta P_{qq}^{(0)}\to {4\over 3},~~~\Delta P_{qG}^{(0)}\to -{1\over 2},~~~
\Delta P_{Gq}^{(0)}\to {8\over 3},~~~~\Delta P_{GG}^{(0)}\to 12,
\en
and from \cite{Mert} that
\be
&& \Delta P_{qq}^{(1)}\to \left(4C_FC_A-8C_FT_f-6C_F^2\right)\ln^2 x=\,
{16\over 3}(1-n_f)\ln^2x,  \non \\
&& \Delta P_{qG}^{(1)}\to -\left(2C_A+C_F\right)\ln^2x=-{22\over 3}\ln^2x, 
\non \\
&& \Delta P_{Gq}^{(1)}\to \left(8C_AC_F+4C_F^2\right)\ln^2x=\,{352\over 9}
\ln^2x,   \\
&& \Delta P_{GG}^{(1)}\to \left(16C_A^2-8C_FT_f\right)\ln^2x=\,{16\over 3}
(27-n_f)\ln^2x. \non
\en
It is evident that for small enough $x$, the NLO $\Delta P_{ij}^{(1)}(x)$
can overcome the suppression factor $(\alpha_s/2\pi)$ and become comparable
to the LO splitting functions.

\subsection{Orbital angular momentum}
   We have discussed the operator definitions for $\Delta q$ and $\Delta G$
that are accessible with experiment in Sec.~4.5 and their $Q^2$ evolution
in Secs.~3.3 and 4.6. It is natural to see if the similar analysis can be
generalized to the orbital angular momenta of quarks and gluons. So far
we have noticed two places where the orbital angular momentum plays a role.
One is the compensation of the growth of $\Delta G$ with $Q^2$ by 
the angular momentum of the quark-gluon pair (see Sec.~4.3). The other
is the reduction of the total spin component $\Delta\Sigma$ due to
the presence of the quark transverse momentum in the lower component of the
Dirac spinor is traded with the quark orbital angular momentum 
(see Sec.~6.1).

   The generators associated with rotation invariance are
\be
J^{\mu\nu}=\int d^3x\,M^{0\mu\nu},
\en
where $M^{\alpha\mu\nu}$ is the angular momentum density given by (4.52).
The first and third terms in (4.52) contribute to the quark and
gluon orbital angular momentum, respectively. The angular momentum operator
in QCD is related to the generators by
\be
J^i=\,{1\over 2}\epsilon^{ijk}J^{jk}.
\en
Explicitly \cite{JM},
\be
J_z^q &=& S_z^q+L_z^q=\int d^3x\,[\,\bar{\psi}\gamma^3\gamma_5\psi
+\psi^\dagger(\vec{x}\times i\vec{\partial})^3\psi],   \non \\
J_z^G &=& S_z^G+L_z^G=\int d^3x\,[\,(\vec{E}\times\vec{A})^3-E_i(\vec{x}
\times\vec{\partial})^3A_i].
\en
Except for the quark helicity operator $S_z^q$, the other three operators
$L_z^q,~S_z^G,~L_z^G$ 
are not separately gauge and Lorentz invariant. Very recently, Ji \cite{Ji2}
has obtained gauge-invariant expressions:
\be
L_z^q=\int d^3x\,\psi^\dagger(\vec{x}\times i\vec{D})^3\psi,~~~~J_z^G=\int 
d^3x\,\left[\vec{x}\times(\vec{E}\times\vec{B})\right]^3.
\en
The gluon total angular momentum $J_z^G$ does not permit further
gauge-invariant decomposition into spin and orbital pieces. However, in
the infinite momentum frame and in the temporal axial gauge $A^0=0$,
$S_z^G$ measures the gluon spin component $\Delta G$ which is accessible
experimentally [cf. Eq.(4.54)].
As a result, the nucleon matrix element of $L_z^G$
in the infinite momentum frame and in $A^0=0$ gauge (or in the light-front
coordinate and in light-front gauge) can be deduced from the matrix elements
of $J_z^G$ and $S_z^G$ \cite{Ji2}. However, whether this definition of the
gluon orbital angular momentum (and likewise $L_z^q$) contacts with
experiment is still unknown.

   The evolution of the quark and gluon orbital angular momenta was
first discussed by Ratcliffe \cite{Rat87}. Using the operators given in
(6.19), Ji, Tang and Hoodbhoy \cite{Ji1} recently have derived a complete
leading-log evolution equation:
\be
{d\over dt}\left(\matrix{ L_z^q  \cr L_z^G \cr}\right)={\alpha_s(t)\over 2\pi}
\left(\matrix{ -{4\over 3}C_F & {n_f\over 3} \cr {4\over 3}C_F & -{n_f\over 3} 
\cr}\right)\left(\matrix{ L_z^q  \cr L_z^G \cr}\right)+{\alpha_s(t)\over 2\pi}
\left(\matrix{ -{2\over 3}C_F & {n_f\over 3} \cr -{5\over 6}C_F & -{11\over 2}
\cr}\right)\left(\matrix{ \Delta\Sigma \cr \Delta G \cr}\right),
\en
with the solutions
\be
&& L_z^q(Q^2) = -{1\over 2}\Delta\Sigma+{1\over 2}\,{3n_f\over 16+3n_f}+
f(Q^2)\left(L_z^q(Q^2_0)+{1\over 2}\Delta\Sigma
-{1\over 2}\,{3n_f\over 16+3n_f}\right),   \non \\
&& L_z^G(Q^2) = -\Delta G(Q^2)+{1\over 2}\,{16\over 16+3n_f}+
f(Q^2)\left(L_z^G(Q^2_0)+\Delta G(Q^2_0)-{1\over 2}\,
{16\over 16+3n_f}\right), \non \\
&&
\en
where
\be
f(Q^2)=\left({\ln Q^2_0/\Lambda^2_{_{\rm QCD}}\over \ln Q^2/\Lambda^2_{_{\rm 
QCD}} }\right)^{{32+6n_f\over 33-2n_f}}
\en
and $\Delta\Sigma$ is $Q^2$ independent to the leading-log approximation. 
We see that the growth of $\Delta G$ with $Q^2$ is compensated by the
gluon orbital angular momentum, which also increases like $\ln Q^2$ but with 
opposite sign. The solution (6.22) has an interesting implication
in the asymptotic limit $Q^2\to \infty$, namely
\be
&& J_z^q(Q^2) = {1\over 2}\Delta\Sigma+L_z^q(Q^2)\to~~~ {1\over 2}\,{3n_f\over 
16+3n_f}\,,    \non \\
&& J_z^G(Q^2)=\Delta G(Q^2)+L_z^G(Q^2)\to ~~~{1\over 2}\,{16\over 16+3n_f}\,.
\en
Thus, history repeats herself: The partition of the nucleon spin between
quarks and gluons follows the well-known partition of the nucleon
momentum. Taking $n_f=6$, we see that $J_z^q:J_z^G=0.53\,:\,0.47\,$. If
the evolution of $J_z^q$ and $J_z^G$ is very slow, which is empirically 
known to be true for the momentum sum rule that half of the proton's momentum
is carried by gluons even at a moderate $Q^2$,
then $\Delta\Sigma\sim 0.30$
at $Q^2=10\,{\rm GeV}^2$ implies that $L_z^q\sim 0.10$ at the same
$Q^2$, recalling that the quark orbital angular momentum is expected to
be of order 0.125 in the relativistic quark model.

   Finally, it is worthy remarking that the spin sum rule
\be
{1\over 2}=\,{1\over 2}\Delta\Sigma_{\rm GI}+(L_z^q)_{\rm GI}+\Delta G+
L_z^G
\en
so far is defined in the gauge-invariant $k_\perp$-factorization scheme.
In the chiral-invariant scheme we have $\Delta\Sigma_{\rm CI}=
\Delta\Sigma_{\rm GI}+(n_f\alpha_s/2\pi)\Delta G$, but $g_1(x)$ and 
$\Gamma_1$ remain unchanged [cf. Eq.(4.38)]. Since $\Delta G$ and $L_z^G$
are independent of the $k_\perp$-factorization, a replacement of 
$\Delta\Sigma_{\rm GI}$ by $\Delta\Sigma_{\rm CI}$ in the spin sum rule 
(6.25) requires that the difference $\Delta\Sigma_{\rm CI}-\Delta
\Sigma_{\rm GI}=(n_f\alpha_s/2\pi)\Delta G$ be compensated by a counterpart
in the gluon orbital angular momentum; that is,
\be
(L_z^q)_{\rm CI}=(L_z^q)_{\rm GI}-{n_f\alpha_s\over 4\pi}\Delta G.
\en
This relation also can be visualized as follows (see \cite{Ji1}). Suppose
we first work in the chiral-invariant scheme and consider a gluon
with +1 helicity splitting into a massless quark-antiquark pair. The total 
helicity of the gluon is entirely transferred to the orbital angular 
momentum of the pair due to helicity conservation or chiral symmetry.
Now, shifting the axial anomaly from the hard part of the
photon-gluon box diagram
to the triangle diagram so that a negative sea-quark
polarization is produced via the anomaly mechanism [see Eqs.(4.29)-(4.32)].
In order to preserve the total angular momentum, this sea-quark
polarization must be balanced by the same amount of the quark orbital angular
momentum induced from the anomaly.
It is interesting to note that when $\Delta G$ is of order 2.5\,,
one will have $\Delta\Sigma_{\rm CI}\sim 0.60$ [cf. Eq.(3.31)] but
$(L_z^q)_{\rm CI}\sim -0.05$\,. In other words, while $\Delta\Sigma_{\rm
CI}$ is close to the quark-model value, $(L_z^q)_{\rm CI}$ deviates
more from the quark model and even becomes negative !

%\newpage
\section{Polarized Parton Distribution Functions}
\subsection{Prelude}
\setcounter{equation}{0}
    One of the main goals in the study of polarized hadron structure
functions measured in DIS is to determine the spin-dependent valence-quark,
cloud-quark, sea-quark and gluon distributions and to understand the
spin structure of the nucleon. In spite of the recent remarkable progress 
in polarized DIS experiments, the extraction of 
spin-dependent parton distribution functions, especially for sea quarks and
gluons, from the measured polarized hadron structure functions remains largely
ambiguous and controversial. We shall see that a full NLO analysis of the
$g_1(x,Q^2)$ data just became possible recently and it indicates that
the sea-quark and gluon spin distributions are, to a large degree, still
unconstrained by current experimental data. Nevertheless, we are entering
the phase of having the parton spin densities parametrized and determined
to the NLO.

  In general the polarized proton structure function
$g_1(x,Q^2)$ has the form \cite{Zijl}
\be
g_1(x,Q^2) &=& {1\over 2}\sum_q^{n_f}e^2_q\Bigg\{ \Delta q(x,\u^2)+{\alpha_s
(\u^2)\over 2\pi}\Big[\Delta C^{\rm S}_q(x,Q^2,\u^2)\otimes\Delta q_{\rm S}
(x,\u^2)  \\
&& +~\Delta C^{\rm NS}_q(x,Q^2,\u^2)\otimes\Delta q_{\rm NS}(x,\u^2)+
\Delta C_G(x,Q^2,\u^2)\otimes\Delta G(x,\u^2)\Big]\Bigg\},   \non
\en
with
\be
\Delta C_{q,G}(x)=\,\Delta C_{q,G}^{(0)}(x)+{\alpha_s\over 2\pi}\Delta 
C_{q,G}^{(1)}(x)+\cdots.
\en
Note that $(\alpha_s/2\pi)\Delta C_G(x)$ is equal to the hard 
photon-gluon cross section $\gg_{\rm hard}(x)$ in (3.1) and $\otimes$ denotes
convolution. The gluon coefficient function $\Delta C_G(x)$ and the quark spin 
density $\Delta q(x)$ depend on the $k_\perp$-factorization scheme,
while the quark coefficient function $\Delta C_q(x)$ depends on
the regularization scheme chosen. In the $\overline{\rm MS}$ scheme, which is 
also a gauge-invariant factorization scheme, $\Delta C_{{\rm S},q}^{(0)}(
x,Q^2)=\Delta C_{{\rm NS},q}^{(0)}(x,Q^2)=\Delta f_q(x,Q^2)$ [cf. Eq.(3.3)] 
and
\be
\Delta C_G^{(0)}(x,Q^2,\u^2)_{\rm GI}=\,(2x-1)\left(\ln{Q^2\over \u^2}
+\ln{1-x\over x}-1\right)+2(1-x)
\en
and the NLO $\Delta C_{q,G}^{(1)}$ are given in \cite{Zijl}. In the
chiral-invariant scheme, the leading order $\Delta C_G(x)$ is calculated to be
[cf. Eq.(3.20)]
\be
\Delta C_G^{(0)}(x,Q^2,\u^2)_{\rm CI}=\,(2x-1)\left(\ln{Q^2\over \u^2}
+\ln{1-x\over x}-1\right).
\en
The first moments of the coefficient functions are
\be
\int^1_0\Delta C_q^{(0)}(x)_{\rm GI}=-2,~~~\int^1_0\Delta C_G^{(0)}(x)_{
\rm GI}=0,~~~\int^1_0\Delta C_G^{(0)}(x)_{\rm CI}=-1.
\en
The $Q^2$ dependence of the parton spin densities is determined by the AP
equation (6.6). As mentioned in Sec.~6.2, at the zeroth order of 
$\alpha_s$, $\Delta C_{q,G}(x)=0$ and $\Delta P_{ij}(x)=0$. At NLO we still 
have $\Delta C_{q,G}(x)=0$ but $\Delta P_{ij}^{(0)}(x)\neq 0$; that is, there 
is a scaling violation in $g_1(x,Q^2)$ but $\Delta G(x)$ enters indirectly
via the $Q^2$ evolution. A complete NLO analysis of $g_1(x,Q^2)$ requires
the information of $\Delta P_{ij}^{(1)}(x)$ in addition to $\Delta 
C_{q,G}^{(0)}(x)$. At this order, gluons start to contribute directly to
the polarized structure function.
For a next-to-next-to-leading order description we have to 
await three-loop results for $\Delta P_{ij}^{(2)}(x)$, although $\Delta
C_{q,G}^{(1)}(x)$ have been calculated. \\

   Several remarks are in order.

\begin{itemize}
\item It is clear from (7.1) that with the input of parton spin distributions
at $Q^2=\u^2$, the $Q^2$ evolution is governed by the logarithmic term 
$\ln(Q^2/\u^2)$ in the coefficient functions, as long as $\alpha_s(\u^2)\ln(
Q^2/\u^2)<<1$. For $Q^2>>\u^2$, the logarithmic terms have to be resummed 
using renormalization group methods \cite{Zijl}. For a fixed $\u^2$ and for
$Q^2$ not deviating too much from $\u^2$, the $\ln(Q^2/\u^2)$ terms in 
$\Delta C_{q,G}^{(0)}$ give rise to the leading-log (LL) $Q^2$ evolution
to $g_1(x,Q^2)$, and the $\ln^2(Q^2/\u^2)$ terms in $\Delta C_{q,G}^{(1)}$
determine the $Q^2$ dependence to the next-to-leading log (NLL) 
approximation. When $\u^2$ is set to be $Q^2$, the $\ln(Q^2/\u^2)$ terms
appearing in coefficient functions are equal to zero and (7.1) becomes
\be
g_1(x,Q^2) &=& {1\over 2}\sum_q^{n_f}e^2_q\Bigg\{ \Delta q(x,Q^2)+{\alpha_s
(Q^2)\over 2\pi}\Big[\Delta C^{\rm S}_q(x,\alpha_s)\otimes\Delta q_{\rm S}
(x,Q^2) \non \\
&+& \Delta C^{\rm NS}_q(x,\alpha_s)\otimes\Delta q_{\rm NS}(x,Q^2)+
\Delta C_G(x,\alpha_s)\otimes\Delta G(x,Q^2)\Big]\Bigg\}.
\en
In this case, the $Q^2$ evolution of $g_1(x,Q^2)$ is taken over by the parton
spin distributions. Using $F_2(x,Q^2)$ as a testing example, it is shown
explicitly in \cite{Zijl} that the $Q^2$ dependence determined by the LL
(NLL) parametrization of parton densities in which LL (NLL) logs 
are resummed to all orders of perturbation theory is indeed consistent
with the leading (next-to-leading) $Q^2$ evolution obtained from
fixed order perturbation theory (i.e., $\u$ being kept fixed). In short,
generally we have to solve the spin-dependent AP equation (6.6) to
determine the $Q^2$ dependence of spin-dependent parton distributions and 
hence the $Q^2$ evolution of $g_1$ via (7.6). However, for
$Q^2$ not deviating too much from $\u^2$, (7.1) provides a
good approximation to the $Q^2$ evolution of $g_1(x,Q^2)$ through the
$\ln(Q^2/\u^2)$ terms in coefficient functions. 

\item Although the contribution $\Delta C_G\otimes\Delta G$ in (7.1) 
or (7.6) is formally of order $\alpha_s$, it actually does not vanish in the
asymptotic limit due to the axial anomaly. It is thus expected that
the NLO corrections to sea-quark and gluon spin distributions are important.

\item Before the availability of the two-loop splitting functions $\Delta
P_{ij}^{(1)}(x)$, some analyses of $g_1(x,Q^2)$ were strictly done at 
the leading order, namely $g_1(x,Q^2)={1\over 2}\sum e_q^2\Delta 
q(x,Q^2)$ with the gluon-spin effects entering via the $Q^2$ evolution
(see e.g., \cite{GRV95,GS95}).
As $\alpha_s(Q^2)\Delta G(Q^2)$ is of order $\alpha_s^0$, several analyses 
have been performed using a hybrid expression for $g_1$
\be
g_1(x,Q^2)=\,{1\over 2}\sum e^2_q\left[\Delta q(x,Q^2)+{\alpha_s\over 2\pi}
\,\Delta C_G(x)\otimes\Delta G(x,Q^2)\right]
\en
in the chiral-invariant factorization scheme. However, the
gluon coefficient function employed in many earlier studies is often 
incorrect. For example, $\Delta C_G^{\rm CI}(x)=-\delta(1-x)$ was used in
\cite{Altar89} and $\Delta C_G^{\rm CI}(x)=(2x-1)\ln[(1-x)/x]$ in  
\cite{Ellis89,RR90}.

\item In spite of the fact that the combination $\Delta q(x,\u)+(\alpha_s
(\u)/2\pi)\Delta C_G\otimes\Delta G(x,\u^2)$ in (7.1) is 
$k_\perp$-factorization independent [see Eqs.(4.33) and (4.37)], {\it the
lack of knowledge on the splitting functions $\Delta P_{ij}^{(1)}(x)$
in the chiral-invariant scheme indicates that, in practice, we should work 
entirely in the gauge-invariant factorization scheme in which hard gluons do
not make contributions to $\Gamma_1$}. This is further reinforced by the
observation that in the literature most of NLO parametrizations of 
unpolarized parton
distributions, which are needed to satisfy the positivity constraints
$|\Delta q(x,Q^2)|\leq q(x,Q^2)$ and $|\Delta G(x,Q^2)|\leq G(x,Q^2)$, are
performed in the $\overline{\rm MS}$ scheme.

\end{itemize}

\subsection{Constraints on polarized parton distributions}
    As stressed in Sec.~6.1, the quark spin density
$\Delta q(x)$ consists of valence-quark, cloud-quark and sea-quark 
components: $\Delta q_V(x),~\Delta q_c(x)$ and $\Delta q_s(x)$. 
Unfortunately, there is no any experimental and theoretical guidelines on
the shape of the spin-dependent cloud-quark distribution, though it is
argued in Sec.~6.1 that the cloud-quark polarization is comparable to
sea-quark polarization in sign and magnitude. Since the spin component of
cloud quarks
originates from valence quarks (there is no cloud strange quark), we will
proceed by considering the combination $\Delta q_v(x)=\Delta q_V(x)+
\Delta q_c(x)$, which is commonly (but not appropriately) referred to as the
``valence" quark contribution. Since the sea polarization is found to 
be SU(3)-flavor symmetric empirically in lattice calculations 
\cite{Dong,Fuk}, we will make the plausible assumption of SU(3)-symmetric
sea-quark spin components. This assumption is justified since
the disconnected insertions from which the sea-quark
spin component originates are dominated by the triangle diagram
and hence are independent of the light quark masses in the loop.\footnote{
It was first noticed in \cite{Gluck96} that even if sea-quark polarization is
SU(3) symmetric at, say $Q^2=Q^2_0$, the SU(3)-flavor symmetry
will be broken at $Q^2>Q^2_0$ due to a nonvanishing NLO $\Delta\gamma_{qq}^
{(1)}$ in (6.11). However, the degree of SU(3) breaking is so small that 
we can neglect it.}
Therefore, for SU(3)-symmetric sea polarization, we obtain from (2.26) that
$\Delta u_v=\,0.93\,,~\Delta d_v=-0.33$ at $Q^2=10\,{\rm GeV}^2$ 
[cf. Eq.(6.3)]. As explained
in Sec.~6.1, the deviation of the result $\Delta u_v+\Delta d_v=0.60$ from
the relativistic quark model's prediction $\Delta u_V+\Delta d_V=0.75$
stems from the negative cloud-quark polarization.

   The valence quark spin density at $x\to 1$ is subject to a 
model-independent constraint. According to the perturbative QCD argument
\cite{Farrar}, the valence quarks at $x=1$ remember the spin of the 
parent proton, i.e., $\Delta u_v(x)/u_v(x),~\Delta d_v(x)/d_v(x)\to 1$ 
as $x\to 1$, as originally conjectured by Feynman \cite{Feynman}.
Since $\Delta d_v$ is negative while $\Delta d_v(x)$ is positive 
as $x\to 1$, it means that the sign of $\Delta d_v(x)$ flips somewhere between 
$0<x<1$ \cite{Callaway}. A model for valence-quark spin distributions has
been proposed sometime ago by Carlitz and Kaur \cite{CK}. According to this 
model, $\Delta d_v(x)/ d_v(x)\to -{1\over 3}$ as $x\to 1$, which disagrees
with what expected from perturbative QCD. Experimentally, it is possible
to carry out a straightforward measurement of the ratio $\Delta d_v(x)/
d_v(x)$ to test various predictions by measuring the longitudinal spin
asymmetry in the inclusive $W^-$ production in proton-proton collisions
\cite{Nad}. This spin asymmetry is proportional to $\Delta d_v(x)/ d_v(x)$
in the appropriate kinematic range (see Sec.~8).

   In terms of valence and sea spin distributions, $g_1^p$ can be recast 
to the form
\be
g_1^p(x,Q^2) &=& {1\over 2}\int^1_x{dy\over y}\Bigg\{
\left[{4\over 9}\Delta u_v(y,Q^2)+{1\over 9}\Delta d_v(y,Q^2)
+{2\over 3}\Delta s(y,Q^2)\right]   \\
&\times& \left[\delta\left(1-{x\over y}\right)+{\alpha_s(Q^2)\over 2\pi}
\Delta C_q^{(0)}(x/ y)\right]
+{\alpha_s(Q^2)\over 6\pi}\,\Delta C_G^{(0)}(x/y)
\Delta G(y,Q^2)\Bigg\}.   \non
\en
In general, both sea quarks and gluons contribute to $g_1^p(x)$.
Since the unpolarized sea distribution and the unpolarized gluon 
distribution multiplied by $\alpha_s/(6\pi)$ are small at $x\gsim 0.2$, the 
positivity constraints $|\Delta s(x)|\leq s(x)$ and $|\Delta G(x)|\leq
G(x)$ imply that the data of $g_1^p(x)$
at $x\gsim 0.2$ should be almost accounted for by $\Delta u_v(x)$ and 
$\Delta d_v(x)$. Therefore, the shape of the spin-dependent valence quark 
densities is nicely restricted by the measured $g_1^p(x)$ at $x\gsim 0.2\,$
together with the first-moment constraint (6.3) and the perturbative QCD 
requirement\footnote{It was assumed in \cite{ChengWai,Cheng84} that 
$\Delta u_v(x)=\alpha(x) u_v(x)$, 
$\Delta d_v(x)=\beta(x)d_v(x)$ with $\alpha(x),~\beta(x)\to 1$ as $x\to 1$ and
$\alpha(x),~\beta(x)\to 0$ as $x\to 0$. However, the constraint at $x=0$ is
not a consequence of QCD. In the present work we find that $\Delta u_v(x)/
u_v(x)=0.41$ and $\Delta d_v(x)/d_v(x)=-0.136$ at $x=0$. As a result, $|\Delta
q_v(x)|$ is usually larger than $|\Delta s(x)|$ even at very small $x$.}
that $\Delta q_v(x)/ q_v(x)\to 1$ at $x=1$. In order to ensure the
validity of the positivity condition $|\Delta q_v(x)|\leq q_v(x)$, we choose 
the NLO Martin-Roberts-Stirling MRS(A$'$) set \cite{MRS94} parametrized 
in the $\overline{\rm MS}$ scheme 
at $Q^2=4\,{\rm GeV}^2$ as unpolarized valence quark distributions:
\be
u_v(x,Q^2=4\,{\rm GeV}^2) &=& 2.26\,x^{-0.441}(1-x)^{3.96}(1-0.54\sqrt{x}
+4.65x),   \non \\
d_v(x,Q^2=4\,{\rm GeV}^2) &=& 0.279\,x^{-0.665}(1-x)^{4.46}(1+6.80\sqrt{x}
+1.93x).   
\en
Accordingly, we must employ the same $\overline{\rm MS}$ scheme for polarized
parton distributions in order to apply the positivity constraint.
For the spin-dependent valence distributions we assume that they have the form
\cite{CLW}
\be
\Delta q_v(x)=\,x^\alpha(1-x)^\beta(a+b\sqrt{x}+cx+dx^{1.5}),
\en
with $\alpha$ and $\beta$ given by Eq.(7.9). We find that an additional 
term proportional to $x^{1.5}$ is needed in (7.10) in order to satisfy the 
above three constraints.

   For the data of $g_1^p(x)$, we will use the SMC \cite{SMC94} and EMC 
\cite{EMC} results, 
both being measured at the mean value of $Q^2_0=10\,{\rm GeV}^2$. Following 
the SMC analysis we have used the new $F_2(x)$ structure function measured by 
NMC \cite{NMC}, which has a better
accuracy at low $x$, to update the EMC data. Assuming that $Q^2=\la Q^2_0\ra
=10\,{\rm GeV}^2$ for each $x$ bin of the $g_1(x,Q^2)$ data, a best least 
$\chi^2$ fit to $g_1^p(x)$ at $x\gsim 0.2$ by (7.10) is found to be 
\cite{CLW}
\be
\Delta u_v(x,Q^2_0) &=& x^{-0.441}(1-x)^{3.96}(0.928+0.149\sqrt{x}-1.141x
+11.612x^{1.5}),   \non \\
\Delta d_v(x,Q^2_0) &=& x^{-0.665}(1-x)^{4.46}(-0.038-0.43\sqrt{x}-5.260x+
8.443x^{1.5}), 
\en
which satisfies all aforementioned constraints. 
Note that we have evoluted $q_v(x,Q^2)$ from $Q^2=4\,{\rm GeV}^2$ to $10\,{\rm
GeV}^2$ in order to compare with $\Delta q_v(x,Q^2_0)$ and that the sign
of $\Delta d_v(x)$ in our parametrization flips at $x_0=0.496\,$ (see Fig.~7).

    The NLO parametrization (7.11) for valence quark spin densities
is obtained by assuming $Q^2=\la Q^2_0\ra$ for each $x$ bin of the
$g_1(x,Q^2)$ data. However, a full NLO analysis should take into account
the measured $x$ dependence of $Q^2$ at each $x$ bin by considering the NLO 
evolution of parton spin distributions. At present, there already exist
several such analyses \cite{Ball,Gluck96,Gehr96,Weig}. For example, by fitting
some parametrizations for spin-dependent parton distributions to all
available world data on $g_1(x,Q^2)$, Gehrmann and Stirling \cite{Gehr96}
obtained (set A)
\be
\Delta u_v(x,Q^2) &=& 0.918A_u\,x^{-0.488}(1-x)^{3.96}(1-4.60\sqrt{x}+11.65x),
\non \\
\Delta d_v(x,Q^2) &=& -0.339A_d\,x^{-0.220}(1-x)^{4.96}(1-3.48\sqrt{x}+7.81x),
\en
to NLO at $Q^2=4\,{\rm GeV}^2$, where $A_u=1.3655$ and $A_d=3.8492$ are
normalization factors ensuring that the first moments of $\Delta u_v(x)$
and $\Delta d_v(x)$ are 0.918 and $-0.339$, respectively. 
However, the $x\to 1$ behavior of the valence quark spin distributions (7.12)
: $\Delta u_v(x)/ u_v(x)\to 0.87$ and $\Delta d_v(x)/d_v(x)\to -2.56$
is not consistent with above-mentioned QCD constraint  (the 
latter also seems to violate the positivity constraint). Three different
NLO parametrizations of valence quark spin distributions are shown in Fig.~7.

\begin{figure}[ht]
\vspace{-4cm}
\hskip 1cm
  \psfig{figure=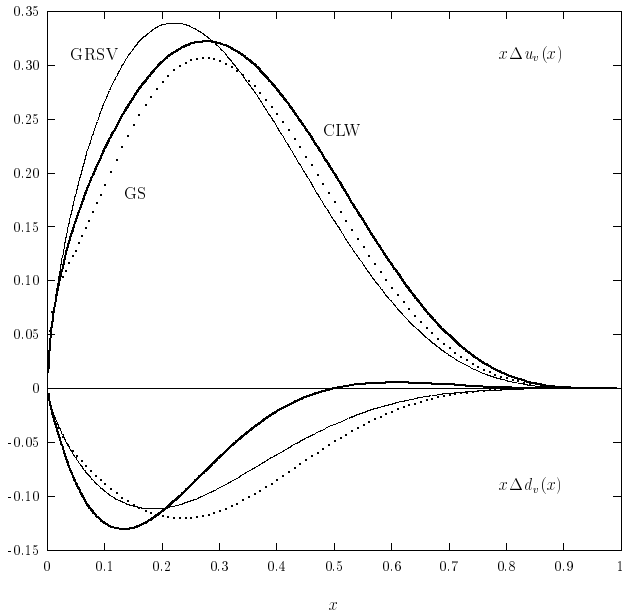,width=12cm}
\vspace{-3cm}
    \caption[]{\small NLO valence-quark spin distributions at $Q^2=10\,{\rm
GeV}^2$ for three different parametrizations: (7.11) (thick solid
curve denoted by CLW \cite{CLW}), the standard set of GRSV \cite{Gluck96}
(solid curve) and set A of GS \cite{Gehr96} (dotted curve).}
    \label{fig7} 

\end{figure}

    A comparison between the theoretical curve of $xg_1^p(x)$ 
fitted to the EMC and SMC data at $x\gsim 0.2$ with the polarized valence 
quark distribution given by (7.11) is shown in Fig.~1 (see Sec.~2.2).
The discrepancy between theory and experiment for $g_1^p(x)$ at small $x$ 
is presumably accounted for by
sea quarks and gluons. For sea-quark polarization, we know that its size is of
order $-0.10$, but there is no any information on the size of gluon
polarization. The sea-quark and gluon spin distributions cannot be
separately determined by current experimental data \cite{Gehr96,Gluck96};
they are correlatively constrained by (7.8). In other words, {\it while the 
shapes of the spin-dependent valence quark distributions are fairly 
constrained by the data, the sea-quark and gluon spin densities are almost 
completely undetermined}. In principle, measurements of scaling violation 
in $g_1(x,Q^2)$ via, for example, the derivative of $g_1(x,Q^2)$ with
respect to $Q^2$, in next-generation experiments will allow an estimate of the
gluon spin density and the overall size of gluon polarization. Of course,
the data should be sufficiently accurate in order to study the gluon spin 
density. Meanwhile, it is even more important to probe $\Delta G(x)$ 
independently in those hadron-hadron collision processes where gluons play a 
dominant role (see Sec.~8).

   As stressed in passing, the fact that gluons make no contributions to
$\Gamma_1$ in the gauge-invariant factorization scheme
does not imply a vanishing gluon contribution to $g_1(x)$.
Quite opposite to the naive sea-quark interpretation for $g_1(x)$, if 
there is no sea polarization
in the chiral-invariant scheme, then the size of the gluon spin component in a
proton must numerically obey the relation
$\Delta G(Q^2)=-(2\pi/ \alpha_s(Q^2))\Delta q_s^{\rm GI}(Q^2)$ [cf. Eq.(4.37)]
in order to perturbatively generate a negative sea-quark polarization 
$\Delta q_s^{\rm CI}(Q^2)$ via the anomaly mechanism. In other words, even 
gluons do not contribute to $\Gamma_1^p$, the gluon spin can be as large as 
2.5 for $\Delta q_s^{\rm GI}=-0.10\,$ at $Q^2=10\,{\rm GeV}^2$ provided
that $\Delta q_s^{\rm CI}=0$. Recall that the gluon polarization induced
from quark's bremsstrahlung is positive [cf. Eq.(3.29)].
Recently, a full NLO analysis was 
performed in the chiral-invariant factorization scheme \cite{Ball} and it 
was claimed that the present data of $g_1$ are sufficient to determine
the first moment of the gluon spin distribution, namely $\Delta G(Q^2=1
\,{\rm GeV}^2)=1.5\pm 0.8$ and is roughly twice as large at $Q^2=10\,{\rm
GeV}^2$. Since the shape and size of the spin-dependent gluon distribution is
$k_\perp$-factorization independent, evidently there is a contradiction 
between the conclusions of {\cite{Ball} and of \cite{Gluck96,Gehr96}. 
Because $\Delta P_{ij}^{(1)}(x)$ and $\Delta C_q^{(1)}(x)$ are available 
only in the $\overline {\rm MS}$ scheme, it has been attempted to 
introduce a modification on NLO anomalous dimensions and 
hard coefficient functions to transfer from 
the GI scheme to the CI prescription \cite{Zijl}. In the CI scheme, 
$\Delta\Sigma_{\rm CI}$ does not evolve with $Q^2$ and this 
requires that
\be
\int^1_0\Delta C_G^{(1)}(x)_{\rm CI}dx=-1,~~~~\Delta\gamma_{{\rm S},
qq}^{(1),1}=0.
\en
However, this transformation cannot be unique since it is only subject to 
the constraints (7.13). Indeed, three different scheme changes have been 
constructed in 
\cite{Ball}. As a consequence, the NLO evolution of polarized 
parton distributions in the CI scheme obtained in this manner \cite{Ball} 
is ambiguous as it depends on the scheme of transformation. \\

    Finally, the interested reader is referred to \cite{Ladi} for a 
collection of polarized parton distributions up to 1995.

\section{Experimental Signatures of Parton Polarizations}
\setcounter{equation}{0}
   It is concluded in Sec.~7.2 that while the present experimental data
put a useful constraint on the shape of the valence-quark spin 
distributions, the sea-quark and gluon spin densities are only loosely
constrained. The question of what is the magnitude and even the sign of
the gluon spin remains unanswered. In view of this, we shall survey the
processes which can be used to probe the parton spin densities, especially
for sea quarks and gluons. For the purposes of this section, we will
use $\Delta q$ and $\Delta\bar{q}$ to denote the spin components of 
quarks and antiquarks, respectively.\\

\noindent {\bf valence-quark spin distribution $\Delta q_v(x)$}
\vskip 0.3cm
$\bullet$~{\it semi-inclusive DIS} \cite{Frank,Strik,Flor96}~~Consider the 
semi-inclusive decays $\vec{e}+\vec{p}\to e+(\pi^\pm,K^\pm,K^0,$$\bar{K}^0)+X$ 
with the longitudinally polarized lepton beam and proton target. The 
differential cross section for the production of a hadron in DIS of 
charged lepton has the form
\be
{d^3\sigma\over dxdydz}\sim {dN^h\over dz}\,\propto\,\sum_{i=q,\bar{q}} e^2_i 
q_i(x,Q^2)D_i^h(z,Q^2),
\en
where $y=\nu/E=(E-E')/E$, $z$ is the fraction of the parent parton's momentum 
carried by the final hadron $h$, $N^h$ is the number of hadrons produced
with a value of $z$, and $D_i^h$ is the fragmentation function of 
a quark $i$ into the hadron $h$. Therefore,
\be
{dN^h_{\up\down}\over dz}-{dN^h_{\up\up}\over dz}\,\propto\, \sum e^2_i\Delta 
q_i(x,Q^2)D^h_i(z,Q^2).
\en
Based on (8.2), two semi-inclusive asymmetries of interest are
\be
{\cal A}^h(x,Q^2) =\, {N^h_{\up\down}-N^h_{\up\up}\over N^h_{\up\down}+
N^h_{\up\up}}=\,{\sum e^2_i\Delta q_i(x,Q^2)D_i^h(Q^2)\over \sum e^2_i
q_i(x,Q^2)D^h_i(Q^2)},
\en
where $D^h_i(Q^2)=\int^1_0dz\,D^h_i(z,Q^2)$, and
\be
{\cal A}^{h^+-h^-}(x,z,Q^2) =\,{dN^{h^+-h^-}_{\up\down}/dz-dN^{h^+-h^-}_{\up
\up}/dz\over dN^{h^+-h^-}_{\up\down}/dz+dN^{h^+-h^-}_{\up\up}/dz}.
\en
It follows that the asymmetries in difference of deep inelastic
$\pi^+$ and $\pi^-$ productions given by
\be
{\cal A}_p^{\pi^+-\pi^-}(x,Q^2)=\,{4\Delta u_v(x,Q^2)-\Delta d_v(x,Q^2)\over 
4u_v(x,Q^2)-d_v(x,Q^2)}
\en
for the proton target, and
\be
{\cal A}_d^{\pi^+-\pi^-}(x,Q^2)=\,{\Delta u_v(x,Q^2)+\Delta d_v(x,Q^2)\over 
u_v(x,Q^2)+d_v(x,Q^2)}
\en
for the deuteron target,
are completely independent of the fragmentation function and can be used to
extract the polarized valence quark densities. Likewise, for kaon production
\be
{\cal A}^{K^+-K^-}_p(x,Q^2)=\,{\Delta u_v(x,Q^2)\over u_v(x,Q^2)},~~~~~{\cal 
A}^{K^0-\bar{K}^0}_p(x,Q^2)=\,{\Delta d_v(x,Q^2)\over d_v(x,Q^2)}.
\en
A measurement of the asymmetry in difference of $K^0$ and $\bar{K}^0$
productions is thus very useful to test the large $x$ behavior of $\Delta 
d_v(x)/d_v(x)$, which is expected to approach unity in perturbative QCD, but
to $-{1\over 3}$ according to the Carlitz-Kaur model \cite{CK}. 
Theoretically, NLO corrections to semi-inclusive asymmetries were
recently studied in \cite{Flor96}. Experimentally, asymmetries ${\cal A}_{
p,d}^\pi$ and ${\cal A}_{p,d}^{\pi^+-\pi^-}$ have been measured by 
SMC recently \cite{SMC96},
from which valence quark and non-strange quark spin distributions are
extracted with the results: $\Delta u_v=1.01\pm 0.19\pm 0.14$ and
$\Delta d_v=-0.57\pm 0.22\pm 0.11\,$. 

\vskip 0.3cm
$\bullet$~{\it $W^\pm$ and $Z$ production} \cite{Bourr89,Bourr}~~In 
high-energy hadron-hadron collisions, the single-spin asymmetry 
${\cal A}_L$ defined by
\be
{\cal A}_L=\,{d\sigma^\up-d\sigma^\down\over d\sigma^\up+d\sigma^\down},
\en
with $d\sigma^{\up(\down)}$ denoting the inclusive cross section where 
one of the initial hadron beams is longitudinally polarized and has $+(-)$
helicity, is expected to vanish to all orders in strong interactions unless
some of the parton-parton scatterings involve parity-violating 
weak interactions. Therefore, a nonzero ${\cal A}_L$ arises from the
interference between strong and weak amplitudes and usually is small, of
order $10^{-4}$ (for a recent analysis of parity-violating asymmetries, see
\cite{CHW,Taxil}). The only exception is the direct $W^\pm$ and $Z$ 
productions in proton-proton collisions where a large ${\cal A}_L$ of 
order 10\% is
expected to be seen at RHIC energies \cite{Bourr}. In the parton model,
$p\vec{p}\to W^\pm+X$ proceeds dominantly via $u\bar{d}\to W^+~(\bar{u}d\to
W^-)$
\be
{\cal A}_L^{W^+} &=& {\Delta u(x_a,M_W^2)\bar{d}(x_b,M_W^2)-\Delta \bar{d}
(x_a,M_W^2)u(x_b,M_W^2)\over u(x_a,M_W^2)\bar{d}(x_b,M_W^2)+
\bar{d}(x_a,M_W^2)u(x_b,M_W^2)}, \non \\
{\cal A}_L^{W^-} &=& {\Delta d(x_a,M_W^2)\bar{u}(x_b,M_W^2)-\Delta \bar{u}
(x_a,M_W^2)d(x_b,M_W^2)\over  d(x_a,M_W^2)\bar{u}(x_b,M_W^2)+
\bar{u}(x_a,M_W^2)d(x_b,M_W^2)}, 
\en
where $x_a=\sqrt{\tau}\exp(y),~x_b=\sqrt{\tau}\exp(-y),~\tau=M^2_W/s$. As
$y$ is near 1, we have $x_a>>x_b$ and
\be
{\cal A}_L^{W^+}\sim\,{\Delta u(x_a,M^2_W)\over u(x_a,M_W^2)},~~~~{\cal A}_L^{
W^-}\sim\,{\Delta d(x_a,M^2_W)\over d(x_a,M_W^2)},
\en
where $x_a=e\sqrt{\tau}$. The (valence) quark spin distributions at 
large $x$ and at $Q^2=M^2_W$ thus can be determined at the kinematic limit
$y\sim 1$. \\

\noindent {\bf antiquark and sea-quark distributions $\Delta \bar{q}(x),~
\Delta q_s(x)$} \footnote{As mentioned in the beginning of this section,
we employ a different definition for quark spin densities here: 
$\Delta q_s=q_s^\up(x)-q
_s^\down(x)$ and $\Delta\bar{q}(x)=\bar{q}^\up(x)-\bar{q}^\down(x)$. {\it
A priori} $\Delta\bar{q}$ can be different from $\Delta q_s$ if they are
not produced from gluons. Based on the measurements of octet
baryon magnetic moments in conjunction with the quark polarization 
deduced from DIS, it has been claimed in \cite{ChengLi96} that $\Delta
\bar{q}\approx 0$. In principle, a measurement of the correlations between
the target polarization and the $\Lambda$ and $\bar{\Lambda}$ polarizations
in DIS will provide a way of discriminating between $\Delta\bar{s}$ and
$\Delta s$ (see e.g., \cite{Brod96}). Another nice method is to measure
the single asymmetry in charmed meson production in the semi-inclusive
DIS process as discussed below.}
\vskip 0.3cm
$\bullet$ {\it semi-inclusive DIS} \cite{Frank,Close}~~Assuming $D^{\pi^+}
_u(z)>D^{\pi^+}_d(z)>D^{\pi^+}_s(z)$, we can neglect the
strange-quark contribution in (8.2) and obtain
\be
{\cal A}^{\pi^+}(x,z)={N^{\pi^+}_{\up\down-\up\up}\over N^{\pi^+}_{\up\down
+\up\up}}=\,{4\Delta u(x)+\Delta\bar{d}(x)+[4\bar{u}(x)+\Delta d(x)]D_d^{
\pi^+}(z)/D_u^{\pi^+}(z)\over 4u(x)+\bar{d}(x)+[4\bar u(x)+d(x)]D_d^
{\pi^+}(z)/D_u^{\pi^+}(z)},
\en
and ${\cal A}^{\pi^-}$ from ${\cal A}^{\pi^+}$ with the replacement
$u\leftrightarrow d$ and $\bar{u}\leftrightarrow \bar{d}$.
Hence, the polarized non-strange antiquark distribution is determined
provided that valence quark spin densities and the ratio $D_d^{\pi^+}(z)/
D_u^{\pi^+}(z)$ are known. For other strategies, see \cite{Frank,Strik}.
Another possibility is to tag fast-moving $K^-$
produced in semi-inclusive DIS to probe the strange-quark polarization
\cite{Close,Strik}. 
\vskip 0.3cm
$\bullet$ {\it $W^\pm$ and $Z$ production} \cite{Bourr89,Bourr}~~ In the 
other extreme kinematic limit $y\to -1$, we have $x_a<< x_b$ and the
parity-violating asymmetries (8.9) become
\be
{\cal A}_L^{W^+}\sim -{\Delta \bar{d}(x_a,M^2_W)\over \bar{d}(x_a,M_W^2)},
~~~~{\cal A}_L^{W^-}\sim -{\Delta \bar{u}(x_a,M^2_W)\over \bar{u}(x_a,M_W^2)},
\en
with $x_a=\sqrt{\tau}/e$. The parity-violating asymmetry in the kinematic
region $y\sim -1$ provides information on $\Delta \bar{u}(x)$ and 
$\Delta\bar{d}(x)$ at small $x$ and large $Q^2$, $Q^2=M^2_W$. 

\vskip 0.3cm
$\bullet$ {\it Drell-Yan process} \cite{CL,Chia,Gupta,Kamal}~~The 
double-spin asymmetry defined by
\be
{\cal A}_{LL}^{\rm DY}=\,{d\sigma^{\up\up}/dQ^2-d\sigma^{\up\down}/dQ^2\over
d\sigma^{\up\up}/dQ^2+d\sigma^{\up\down}/dQ^2}
\en
measured in the Drell-Yan process $\vec{p}\vec{p}\to\ell^+\ell^-+X$, where
$d\sigma^{\up\up(\up\down)}$ designates the Drell-Yan cross section for the
configuration where the incoming proton helicities are parallel
(antiparallel), is sensitive to the sea spin densities. In the parton 
model, the asymmetry reads
\be
{\cal A}_{LL}^{\rm DY}=\,{4\pi\alpha^2\over 9Q^2s}\int^1_0{dx_1\over x_1}
{dx_2\over x_2}\,\delta\left(1-{Q^2\over sx_1x_2}\right)\sum_q e^2_q\Delta
q(x_1,Q^2)\Delta\bar{q}(x_2,Q^2)+(1\leftrightarrow 2).
\en
The sign of ${\cal A}_{LL}^{\rm DY}$ is expected to be negative as $\Delta
u(x)>0$ and $\Delta\bar{u}(x)<0$. A recent analysis of NLO effects
in \cite{Kamal} indicates that the $q\bar{q}$ subprocess exhibits great
perturbative stability, whereas the $qG$ subprocess is important and 
contributes destructively. For a discussion of single-spin asymmetry in the
Drell-Yan process, see \cite{Bourr89}. 
\vskip 0.3cm
$\bullet$ {\it inclusive DIS with charged current} \cite{Morii,Ansel96,Maul}~~
High energy lepton-proton scattering at large $Q^2$ allows to probe spin 
effects in charged current interactions in the DIS process 
$\ell^\mp p\to\nu(\bar{\nu})X$. Consider the 
parity-violating DIS of unpolarized charged lepton on longitudinally
polarized proton: $\ell^++\vec{p}\to\bar{\nu}+X$. It is shown in
\cite{Morii} that the single-spin asymmetry in this process is sensitive
to $\Delta d_v(x)$ and to antiquark/sea-quark spin densities: $\Delta 
\bar{u}(x),~\Delta d_s(x)$ and $\Delta s(x)$. With longitudinally polarized 
lepton and proton beams, the double asymmetries defined by
\be
{\cal A}_{LL}^{W^\mp}=\,{d\sigma^{\ell^\mp p}_{\up\down}-d\sigma^{\ell^\mp
p}_{\up\up}\over d\sigma^{\ell^\mp p}_{\up\down}+d\sigma^{\ell^\mp
p}_{\up\up}}
\en
have the expressions \cite{Ansel96,Maul}
\be
{\cal A}_{LL}^{W^-} &=& {\Delta u(x)+\Delta c(x)-(1-y)^2[\Delta \bar{d}(x)+
\Delta \bar{s}(x)]\over u(x)+c(x)+(1-y)^2[\bar{d}(x)+\bar{s}(x)]},  \non\\
{\cal A}_{LL}^{W^+} &=& {(1-y)^2[\Delta {d}(x)+
\Delta {s}(x)]-\Delta\bar{u}(x)-\Delta\bar{c}(x)\over (1-y)^2[{d}(x)+{s}(x)]
+\bar{u}(x)+\bar{c}(x)}.
\en
These inclusive asymmetries are mainly sensitive to the $u$ and $d$-quark 
flavor.

\vskip 0.3cm
$\bullet$ {\it semi-inclusive DIS with charged current} \cite{Maul}~~
Consider the charmed meson production in the semi-inclusive DIS:
$\ell^\mp+p\to \nu(\bar{\nu})+D+X$. The main subprocesses are
$s+W^+\to D$ and $\bar{s}+W^-\to \bar{D}$. The single asymmetries are given by
\cite{Maul}
\be
{\cal A}_{LL}^{W^-,D}(x) &\equiv& {d\sigma^{W^-,D}_{\up\down}-d\sigma
^{W^-,D}_{\up\up}\over d\sigma^{W^-,D}_{\up\down}+d\sigma
^{W^-,D}_{\up\up}}=\,-{\Delta\bar{s}(x)+\tan\theta_C^2\Delta\bar{d}(x)\over
\bar{s}(x)+\tan\theta_C^2\bar{d}(x)},   \non \\
{\cal A}_{LL}^{W^+,D}(x) &\equiv& {d\sigma^{W^+,D}_{\up\down}-d\sigma
^{W^+,D}_{\up\up}\over d\sigma^{W^+,D}_{\up\down}+d\sigma
^{W^+,D}_{\up\up}}=\,{\Delta{s}(x)+\tan\theta_C^2\Delta{d}(x)\over
{s}(x)+\tan\theta_C^2{d}(x)},
\en
where $\theta_C$ is a Cabibbo mixing angle. These asymmetries allow to extract 
the strange sea spin distributions $\Delta s(x)$ and $\Delta\bar{s}(x)$
separately. It is of great interest to test if $\Delta\bar{s}(x)$ is 
identical to $\Delta s(x)$.

\vskip 0.3cm
$\bullet$ {\it elastic $\nu N$ scattering} 
\cite{Ellis88,Kap,Garvey,Bass92,Alb}~~
Assuming a negligible $\Delta c$, it was originally argued that the 
axial-vector form factor $G_A(q^2)$ appearing in the matrix element $\la 
N|A_\mu^Z|N\ra$ for $\nu N$ elastic scattering is related to the quark 
polarization by $G_A(0)={1\over 2}(\Delta u-\Delta d-\Delta s)$. Hence, a 
measurement of $G_A(0)$ will determine $\Delta s$ independently. Since 
the limit $q^2=0$ is experimentally unattainable, the $q^2$ dependence
of $G_A(q^2)$ is usually assumed to have a dipole form. The $\nu p$ and
$\bar{\nu}p$ experiments in 1987 \cite{Ahrens} indicated that $\Delta s(0)=
-0.15\pm 0.09$ \cite{Ellis88,Kap}.\footnote{The value of $G_A(0)$ is very 
sensitive to the dipole mass $M_A$. For example, it is shown in \cite{Garvey1}
that $G_A(0)=-0.15\pm 0.07$ is obtained for $M_A=1.049\pm 0.019$ MeV, but
the data also can be fitted with $G_A(0)=0$ and $M_A=1.086\pm 0.015$ MeV.}
It becomes clear now that what measured
in $\nu p\to\nu p$ scattering is the combination $\Delta s-\Delta c$ rather 
than $\Delta s$ itself \cite{Bass92}. First, contrary to the scale-dependent 
$\Delta s$,
the quantity $\Delta s-\Delta c$ is scale independent as it is anomaly 
free. It is possible that $\Delta s$ at a relatively low scale is zero, but
it evolves dramatically from the quark-model scale to the EMC scale
$Q^2_{\rm EMC}=10\,{\rm GeV}^2$ \cite{Jaffe87}. Therefore, the previous
interpretation for $\Delta s(0)$ cannot be extrapolated to $\Delta 
s(Q^2_{\rm EMC})$ directly, but what we can say now is $\Delta s(Q^2)=-(0.15
\pm 0.09)+\Delta c(Q^2)$.
Second, far below charm threshold, charmed quarks stop making 
contributions to DIS, but they still contribute to $G_A$ via the 
triangle diagram. As stressed in \cite{Bass92}, the value of $\Delta c$
defined in $\nu p$ scattering can only be interpretated as $\Delta c$
in DIS well above the charm threshold. A new $\nu N$ scattering experiment 
using LSND (Liquid Scintillator Neutrino
Detector) is currently underway at Los Alamos (see e.g., \cite{Garvey}). 
\vskip 0.3cm
$\bullet$ {\it semi-inclusive $\Lambda$ production in DIS} 
\cite{Lu,Jaffe96b,Ellis96}~~Consider the semi-inclusive decay 
$\ell+\vec{p}\to\ell+\Lambda+X$ with a longitudinally polarized proton 
target. Since in the naive quark model the spin 
of the $\Lambda$ is carried by the strange-quark's spin, it is expected
that the negative strange sea polarization in a polarized proton will be
transferred to the longitudinal $\Lambda$ polarization in the current
fragmentation region. In the simple parton model, the longitudinal
polarization of the $\Lambda$ is given by \cite{Lu}
\be
{\cal P}_\Lambda(x,z,Q^2)=\,{dN^{\Lambda^\up}/dz-dN^{\Lambda^\down}/dz\over
dN^{\Lambda^\up}/dz+dN^{\Lambda^\down}/dz}=\,{\Delta s(x,Q^2)\Delta 
D_s^\Lambda(z,Q^2)\over s(x,Q^2)D_s^\Lambda(z,Q^2)},
\en
where $\Delta D_s^\Lambda=D^{\Lambda^\up}_{s^\up}-D^{\Lambda^\up}_{s^\down}=
D^{\Lambda^\up}_{s^\up}-D^{\Lambda^\down}_{s^\up}$. Very little is known
about $D_s^\Lambda(z)$ and $\Delta D_s^\Lambda(z)$. It is suggested in 
\cite{CHW} a simple parametrization for $D_s^\Lambda(z)$: 
$zD^\Lambda_s(z)=0.5z(1.08-z)^3-0.06
(1-z)^4$. In the absence of experimental data or a detailed theory, the
construction of $\Delta D_s^\Lambda(z)$ is necessarily {\it ad hoc}.
Nevertheless, we expect that the polarization of the outgoing $\Lambda$ 
is equal to that of the strange quark at $z=1$. Beyond the non-relativistic
quark model, the $u$ and $d$ quarks in the $\Lambda$ are also polarized
and will contribute to ${\cal P}_\Lambda$ \cite{Jaffe96b}.
The longitudinal polarization of the $\Lambda$ also can be produced in
the target fragmentation region in
deep-inelastic $\nu N,~\vec{\mu}(\vec{e})\vec{N}(N)$ scatterings and various
underlying mechanisms for ${\cal P}_\Lambda$ are discussed in 
\cite{Ellis96}.\\

\noindent {\bf gluon spin distribution $\Delta G(x)$}

\begin{table}
\centerline{{\small Table III. Various processes which are sensitive to the 
gluon spin distribution.}}
{\small \begin{center}
\begin{tabular}{|c|c|c|} \hline
Process  &  Dominant subprocess & References   \\  \hline
charm or $J/\psi$ leptoproduction &   &    \\
$\vec{\ell}+\vec{p}\to \ell+c+\bar{c}$ & $\gamma^*\,G\to c\,\bar{c}$ & 
\cite{Guill,Morii94} \\ 
$\vec{\ell}+\vec{p}\to \ell+J/\psi+X$ & $\gamma^*\,G\to J/\psi+G$ & \\ \hline
charm or $J/\psi$ photoproduction &  &  \\
$\vec{\gamma}+\vec{p}\to c+\bar{c}$ & $\gamma\,G\to c\,\bar{c}$ & 
\cite{GR,Altar89,Morii94,Frix}  \\ 
$\vec{\gamma}+\vec{p}\to J/\psi+X$ & $\gamma\,G\to J/\psi+G$ & \\ \hline
large-$k_\perp$ two-jet production &  &  \\
$\vec{\ell}+\vec{p}\to {\rm 2~jets}+X$ & $\gamma^*\,G\to q\bar{q}$ & 
\cite{CCM,Man91}\\ \hline
prompt-photon production &  & LO: \cite{CL,Berg,Bourr91,Gull}   \\
$\vec{p}\vec{p}\to \gamma+X$ & $G+q\to\gamma+q$ & NLO: \cite{Cont93} 
\\ \hline
single-jet production & small $x_\perp$: $GG\to GG,\,GG\to q\bar{q}$ & LO: 
\cite{CHL,Bourr91,Strat92,Chia}  \\
$\vec{p}\vec{p}\to {\rm jet}+X$ & intermediate $x_\perp$: $Gq\to Gq$ &   
NLO: \cite{Bern} \\ \hline
two-jet production & small $x_\perp$: $GG\to GG,\,GG\to q\bar{q}$ &  \\
$\vec{p}\vec{p}\to {\rm 2~jets}+X$ & intermediate $x_\perp$: $Gq\to Gq$
&  \cite{CHL,Bourr91,Indu}  \\ \hline
three-jet production & $qG\to qq\bar{q}$ &    \\
$\vec{p}\vec{p}\to {\rm 3~jets}+X$ & $qq\to qqG$ & \cite{Donch91}   \\ \hline
four-jet production &  &  \\
$\vec{p}\vec{p}\to {\rm 4~jets}+X$ & $GG\to GGGG$ & \cite{Fraser}   \\ \hline
two-photon production & $q\bar{q}\to\gamma\gamma$ &  
LO: \cite{Donch92} \\
$\vec{p}\vec{p}\to \gamma\gamma+X$ & NLO: see \cite{Cori} &  
NLO: \cite{Cori} \\ \hline
heavy quark production & $GG\to Q\bar{Q}$ & LO: \cite{Cont90}  \\
$\vec{p}\vec{p}\to Q\bar{Q}+X$ & NLO: see \cite{Kar94} & NLO: \cite{Kar94} 
\\ \hline
charmonium production & $GG\to S$-wave charmonium & 
\cite{Cortes,Cont90,Donch90,Morii94} \\ 
$\vec{p}\vec{p}\to {\rm charmonium}+X$ & $GG\to P$-wave$\,+\,G$  & 
\cite{Rob91} \\ \hline
two-$J/\psi$ production & & \\
$\vec{p}\vec{p}\to J/\psi+J/\psi+X$ &  $GG\to J/\psi+J/\psi$  & \cite{Bara}  
\\ \hline
dimuon production & $q\bar{q}\to \mu^+\mu^-$ &   \\
$p\vec{p}\to\mu^+\mu^-(\vec{\mu}^-)+X$ & $q+G\to q+\mu^+\mu^-$ & 
\cite{Cont91,Carl,Nado94} \\ \hline
\end{tabular}
\end{center} }
\end{table}

\vskip 0.3cm
   Phenomenological signatures of $\Delta G$ can be tested in various ways,
as partly summarized in Table~III. Instead of going through the details for
each process, we will focus on those promising processes which have better
signals, higher event rates and larger asymmetries. Since $\Delta G(Q^2)$ 
increases logarithmically with $Q^2$, it is conceivable that the effects
of gluons will manifest in the polarized $pp$ collider at the RHIC and in
the $ep$ collider at the HERA.

\vskip 0.3cm
$\bullet$ {\it prompt photon production} \cite{CL,Berg,Bourr91,Gull,Cont93}~~
The double-spin asymmetry ${\cal 
A}_{LL}^\gamma$ in the direct photon production at high $p_\perp$ in
longitudinally polarized proton-proton collisions depends strongly on
the polarization of gluons as the Compton subprocess $Gq\to \gamma 
q$ dominates over $q\bar{q}\to\gamma G$ annihilation, reflecting by
the fact that ${\cal A}_{LL}^\gamma$ grows with $x_F$ at fixed $p_\perp$.
This process thus provides a clean, direct, and 
unproblematic possibility for determining $\Delta G(x)$. Such experiments
should be feasible in the near future at the RHIC.

\vskip 0.3cm
$\bullet$ {\it single-jet production} \cite{CHL,Bourr91,Strat92,Chia,Bern}~~
In general the double-spin asymmetry
${\cal A}_{LL}^{\rm jet}$ for a jet production in $pp$ collisions with
a transverse momentum $p_\perp$ is sensitive to the gluon spin
density for $x_\perp=2p_\perp/\sqrt{s}$ not too large. Since the polarized
gluon distribution is large at small $x$, gluon-gluon scattering dominates 
the underlying parton-parton interaction subprocesses at small $x_\perp$.
As the jet momentum increases, quark-gluon scattering becomes more and more
important due to the relatively fast decrease of the gluon spin distribution
with increasing $x$. It is finally governed by quark-quark scattering
at large $x_\perp$. Therefore, a measurement of ${\cal A}_{LL}^{\rm jet}$
in the jet momentum region where the spin asymmetry is dominated by 
$qG$ or $GG$ scattering will furnish important information on $\Delta G(x)$.

\vskip 0.3cm
$\bullet$ {\it hadronic heavy-quark production} 
\cite{Cortes,Cont90,Donch90,Morii94}~~Since the dominant
subprocess for hadronic heavy-quark production in $pp$ collisions is 
$GG\to Q\bar{Q}$, this process depends quadratically on $\Delta G$ and is 
hence very sensitive to the
gluon spin density; it is often considered to be the best and most realistic
test on $\Delta G$.\\

\section{Conclusions}
  The new polarized DIS experiments in recent years have confirmed the
validity of the EMC data and the
controversial conclusions that the observed value of $\Gamma_1^p$, the
first moment of $g_1^p(x)$, is substantially smaller than the 
Ellis-Jaffe conjecture and that only a small fraction of the proton spin
comes from the quarks. However, the proton spin problem now becomes less
severe than before. The new world average is that $\Delta\Sigma\sim 0.30$
and $\Delta s\sim -0.10$ at $Q^2=10\,{\rm GeV}^2$. The Bjorken sum rule
has been tested to an accuracy of 10\% level. Some main conclusions are:

  1). There are two $k_\perp$-factorization schemes of interest: the
chiral-invariant scheme in which the ultraviolet cutoff on the quark spin
distributions respects chiral symmetry and gauge invariance but not the 
axial anomaly, and the gauge-invariant scheme in which the ultraviolet 
regulator satisfies gauge symmetry and the axial anomaly but
breaks chiral symmetry. The usual improved parton model calculation 
corresponds to the chiral-invariant factorization scheme.
There is an anomalous gluonic
contribution to $\Gamma_1^p$ due to the axial
anomaly resided in the box diagram of photon-gluon scattering at 
$k_\perp^2=[(1-x)/4x]Q^2$ with $x\to 0$. As a consequence, $\Delta\Sigma_{\rm
CI}$ is not necessarily small and $\Delta s_{\rm CI}$ is not necessarily
large. For $\Delta G\sim 2.5$ at $Q^2=10\,{\rm GeV}^2$, one has
$\Delta\Sigma_{\rm CI}\sim 0.60$ and $\Delta s_{\rm CI}\sim 0$.

   2). The OPE approach corresponds to the gauge-invariant factorization 
scheme. Hard gluons do not contribute to 
$\Gamma_1$ because the axial anomaly is shifted from the hard photon-gluon 
cross section to the spin-dependent quark distribution. However, it
by no means implies that $\Delta G$ vanishes in a polarized proton.
$\Delta\Sigma_{\rm GI}$ is small because of the negative helicities of sea 
quarks. The chiral-invariant and gauge-invariant factorization schemes are
explicitly shown to be equivalent up to NLO since $\Delta q$'s and 
$\gg_{\rm hard}$'s in these two different schemes 
are related by (4.36) and (4.33), respectively.
As far as the first moment of $g_1(x)$ is concerned, the anomalous gluon
and sea-quark interpretations are thus on the same footing.

   3). Contrary to the gauge-invariant $\Delta q_{\rm GI}$, $\Delta G$ and
chiral-invariant $\Delta q_{\rm CI}$ cannot be expressed as matrix 
elements of local and gauge-invariant operators. Nevertheless, gauge variant
local operator definitions do exist; in the light-front coordinate and
in the light-front gauge $A^+=0$ (or in the infinite momentum frame and in
temporal axial gauge), $\Delta G$ has a local operator
definition given by (4.55) or (4.56), and $\Delta q_{\rm CI}$ by (4.57).
By contrast, they can be also recast as matrix elements of string-like
gauge-invariant but non-local operators.

    4). The U(1) Goldberger-Treiman relation (5.6) in 
terms of the $\eta_0$ remains totally unchanged no matter how one varies
the quark masses and the axial anomaly, while its
two-component expression (5.17) is identified with the connected 
and disconnected insertions [see (5.21)]. We have determined the
physical coupling constants $\getap$ and $\geta$ from the GT relations for
$g_A^0$ and $g_A^8$ and found that $\getap=3.4$ and $\geta=4.7\,$.

    5). For massless sea quarks there are two mechanisms allowing for 
quark helicity flip and producing
sea-quark polarization: the nonperturbative mechanism due to 
instanton-induced interactions (see Sec.~4.4) and the perturbative
way via the axial anomaly (see Sec.~4.3).
The sign of the sea-quark helicity generated by hard gluons via
the latter mechanism is predictable in the framework of perturbative QCD: 
It is negative if the gluon spin component $\Delta G$ is positive. 
The lattice calculation
indicates that sea polarization is almost independent of light quark
flavors; this empirical SU(3)-flavor symmetry implies that it is indeed 
the axial anomaly, which is independent of light quark masses, that
accounts for the bulk of the helicity contribution of sea quarks.

    6). A full and consistent next-to-leading order analysis of the
$g_1(x,Q^2)$ data just became possible recently. We have to work entirely
in the gauge-invariant factorization scheme for the NLO analysis since the
NLO polarized splitting functions are available only in this scheme. While the 
shapes of the spin-dependent valence quark distributions are fairly 
constrained by the data, the sea-quark and gluon spin densities are 
almost completely undetermined. It is thus very important to probe 
$\Delta G(x)$ independently in the hadron-hadron collision processes 
where gluons play a dominant role. The most promising processes are: prompt
photon production, single-jet production and hadronic heavy-quark
production in $pp$ collisions.

    7). As for the spin sum rule ${1\over 2}={1\over 2}\Delta\Sigma+\Delta G
+L_z^q+L_z^G$, the only spin content which is for sure at present is the 
observed value
$\Delta\Sigma\sim 0.30$ at $Q^2=10\,{\rm GeV}^2$. The relativistic quark
model predicts that $\Delta\Sigma=0.75$ and $L_z^q=0.125\,$. Recent lattice
calculations imply that relativistic quark model results are
recovered in the valence approximation. The quark-model's value of 0.75 for
$\Delta\Sigma$ is reduced to the ``canonical" value of $\sim 0.60$ by
negative spin components of cloud quarks, and reduced further to $\sim 0.30$ 
by the negative sea-quark polarization. That is, the deviation of $g_A^0$
from unity expected from the non-relativistic quark model is ascribed to
the negative spin components of cloud and sea quarks and to relativistic
effects. The ``valence" contribution as conventionally referred to is 
actually a combination of cloud-quark and truly valence-quark components. 
It is thus important to estimate the cloud-quark polarization to see
if it is negative in sign and comparable in magnitude to the (one-flavor)
sea-quark helicity. In the asymptotic limit, $J_z^q(\infty)={1\over 
2}\Delta\Sigma(\infty)+L_z^q(\infty)\sim{1\over 4}$ and $J_z^G(\infty)
=\Delta G(\infty)+L_z^G(\infty)\sim {1\over 4}$. If the evolution of
$J_z^q$ and $J_z^G$ is very slow, we will have $L_z^q(10\,{\rm GeV}^2)\sim
0.10$, which is close to the quark-model expectation. The growth of 
$\Delta G$ with $Q^2$ is compensated by the gluon orbital angular momentum,
which also increases like $\ln Q^2$ but with opposite sign.

\vskip 3 cm
\centerline{\bf ACKNOWLEDGMENTS}
\vskip 0.3 cm
    I wish to thank H.L. Yu and H.-n. Li for a careful reading of the 
manuscript. This work was supported in part by the National Science Council 
of ROC under Contract No. NSC85-2112-M-001-010.

%%%%% References %%%%%%%%%%%%%%%%%%%%%%%%%%%%%%%%%%%%%%%%%%%%%%%%%%%%%%%%%%%%%
\renewcommand{\baselinestretch}{1.1}
\newcommand{\bi}{\bibitem}
\end{document}